\documentclass[a4paper,11pt]{article}
\usepackage{jcappub}
\usepackage{aas_macros}

\usepackage{amsmath}
\usepackage{bm}
\usepackage{xcolor}
\usepackage[utf8]{inputenc}
\usepackage{hyperref}
\usepackage{url}
\usepackage{graphicx}
\usepackage{amssymb}
\usepackage[amssymb]{SIunits}
\usepackage{float}  
\usepackage{multirow}
\usepackage{bbold}

\newcommand{\beq} {\begin{equation}}
\newcommand{\eeq} {\end{equation}}
\newcommand{\bal} {\begin{aligned}}
\newcommand{\eal} {\end{aligned}}
\newcommand{\vk} {\bm{k}}

\title{
Multi-tracer intensity mapping: Cross-correlations, Line noise \& Decorrelation
}

\author[a,b]{Emmanuel Schaan}
\author[a,b,c]{and Martin White}
\affiliation[a]{Lawrence Berkeley National Laboratory, One Cyclotron Road, Berkeley, CA 94720, USA}
\affiliation[b]{Berkeley Center for Cosmological Physics, UC Berkeley, CA 94720, USA}
\affiliation[c]{Department of Physics, University of California, Berkeley, CA 94720, USA}
\emailAdd{eschaan@lbl.gov}
\emailAdd{mwhite@berkeley.edu}

\abstract{
Line intensity mapping (LIM) is a rapidly emerging technique for constraining cosmology and galaxy formation using multi-frequency, low angular resolution maps.
Many LIM applications crucially rely on cross-correlations of two line intensity maps, or of intensity maps with galaxy surveys or galaxy/CMB lensing. 
We present a consistent halo model to predict all these cross-correlations and enable joint analyses, in 3D redshift-space and for 2D projected maps.
We extend the conditional luminosity function formalism to the multi-line case, to consistently account for correlated scatter between multiple galaxy line luminosities.
This allows us to model the scale-dependent decorrelation between two line intensity maps,
a key input for foreground rejection and for approaches that estimate auto-spectra from cross-spectra.
This also enables LIM cross-correlations to reveal astrophysical properties of the interstellar medium inacessible with LIM auto-spectra.
We expose the different sources of luminosity scatter or ``line noise'' in LIM, and clarify their effects on the 1-halo and galaxy shot noise terms.
In particular, we show that the effective number density of halos can in some cases exceed that of galaxies, counterintuitively.
Using observational and simulation input, we implement this halo model for the H$\alpha$, [O{{\sc iii}}], Lyman-$\alpha$, CO and [C{\sc ii}] lines.
We encourage observers and simulators to measure galaxy luminosity correlation coefficients for pairs of lines whenever possible.
Our code is publicly available at \url{https://github.com/EmmanuelSchaan/HaloGen/tree/LIM}.
In a companion paper, we use this halo model formalism and code
to highlight the degeneracies between cosmology and astrophysics in LIM, and to compare the LIM observables to galaxy detection for a number of surveys.
}

\begin{document}
\maketitle
\flushbottom

\section{Introduction}

Our ability to measure the large-scale structure of the Universe is improving rapidly, providing a promising means to study a wide variety of topics from theories of the early Universe to the formation and evolution of galaxies within the evolving cosmic web.  The high redshift Universe, in particular, combines a large cosmological volume for precise statistical inference with a more linear density field which can be more reliably modeled and which is better correlated with the initial conditions.

Line intensity mapping (LIM) has emerged as a promising means of efficiently mapping the high $z$ Universe in 3D with a high density of tracers over a large fraction of the sky \cite{Kovetz17,Kovetz19}.
LIM refers to a low resolution, spectroscopic survey to measure integrated flux from unresolved sources on large areas of sky at different frequencies.
LIM surveys capture the largest elements of the cosmic web and map out the distribution of matter in very large cosmological volumes in a fast and efficient manner, with good radial resolution.
In this paper we will focus on LIM in optical/infra-red and sub-mm bands. 
More specifically, we implement a halo model for the following lines:
H$\alpha$ (656.28~nm), Lyman-$\alpha$ (121.6~nm), [O{\sc iii}] (495.9~nm and 500.7~nm), [C{\sc ii}] (158~$\mu$m) and CO 1-0 (2.6~mm).
Although we are leaving aside the field of radio-frequency LIM (for a recent survey of this area, see refs.~\cite{CVDE21cm,Ahmed19}), the formalism presented here applies identically to the post-reionization 21~cm.

While they can be efficient, LIM surveys are not without their challenges.  
Confusion among different lines from different redshifts observed at the same frequency provides a fundamental challenge to tomographic measurements.
To tackle these issues, many methods have been proposed (see \cite{Kovetz17, Pullen13} for a summary).
Continuum interlopers can be controled by masking Fourier modes that vary slowly along the line of sight (LOS), corresponding to smooth spectral energy distributions.
Line interlopers can be reduced in some cases by masking bright voxels, secondary line identification, cross-correlation with a contaminant template (such as a catalog of unwanted sources).
Interlopers at a different redshift can cause a detectable anisotropy in the power spectrum, analogous to the Alcock-Paczynski effect \cite{Lidz16, Gong20}.
In many cases, cross-correlations of different lines \cite{Gong20}, or of a line with a galaxy survey or with CMB lensing, allow one to suppress uncorrelated foregrounds \cite{Furlanetto06, Lidz08, Visbal10, Visbal11, Beane19}.
This way, cross-correlations may be the path towards unbiased astrophysical and cosmological constraints at high redshift.

In this case, an accurate model of auto and cross power spectra is required.
Galaxy luminosities in two different lines are generally not perfectly correlated.
As a result, two intensity maps from the same galaxies in two different lines are not perfectly correlated either.
The level of decorrelation determines the ability to use one of the lines as a proxy for the other, as in foreground cleaning.
Accurately modeling the level of decorrelation is also crucial for estimating auto-spectra from observed cross-spectra, and to use LIM cross-correlations as astrophysical probes.
Such modeling is generally achieved with semi-analytical models \cite{Chung18, Li16, Popping18, Saito20, Hainline20, Chevallard16, Gutkin16},
frequently based on insights from hydrodynamical simulations \cite{Hirschmann17, Moriwaki18, Pallottini19, Katz19}
and spectroscopic surveys
\cite{DeLooze11, Spinoglio11, DeLooze14, Diaz-Santos17, Schreiber17}.
In this paper, we extend the conditional luminosity function (CLF \cite{Yang03}) formalism to the multivariate case, in order to correctly model the correlated scatter between different lines from the same galaxy.
We build a new halo model \cite{Seljak00, Peacock00} upon this multi-line CLF, to predict the auto- and cross-spectra of intensity maps in different lines, including any potential decorrelation.
This formalism consistently predicts cross-correlations of LIM with galaxy surveys, and galaxy and cosmic microwave background lensing (Appendix.~\ref{app:cross_lensing}).

Because the galaxy populations at high redshifts are currently highly uncertain,
predictions based on different observed or simulated galaxy luminosity functions can differ wildly.
In this paper, we thus do not mean to improve the accuracy of existing models for the LIM auto-spectra. Our predicted power spectra may differ significantly from others in the literature relying on different input.
Instead, our goal is to provide a unified formalism, to consistently convert luminosity function inputs into joint predictions of the LIM auto-spectra, cross-spectra, and their cross-correlations with galaxy surveys and galaxy and CMB lensing.

We finish with a brief outline of the paper. 
In \S\ref{sec:formalism}, we derive the general halo model based on the multivariate CLF formalism, to consistently predict the mean intensity and power spectrum of LIM and galaxy clustering in redshift-space and projected maps.
We highlight the importance of scatter in galaxy and halo luminosities, and their effects on the auto and cross-power spectra.
In \S\ref{sec:observational}, we review existing observational constraints on the LIM model derived in \S\ref{sec:formalism} for optical, UV and far IR lines.
Specifically, we focus on the H$\alpha$, [O{\sc iii}], Lyman-$\alpha$, CO and [C{\sc ii}] lines.
Finally, we use the multi-line halo model to quantify the decorrelation between intensity maps in different lines, a key quantity for several methods of line deconfusion.
In the companion paper~\cite{paper2}, we apply our formalism to highlight the degeneracies between astrophysics and cosmology with LIM.
There, we compare the LIM observables to galaxy detection as probes of faint galaxies and tracers of the matter density field.

Throughout this paper, we assume a flat $\Lambda$CDM cosmology from ref.~\cite{PlanckParams16}, with $\Omega_\text{CDM}=0.267$, $\Omega_b = 0.0493$, $H_0 = 67.12\,$km/s/Mpc, $A_S = 2.3\times 10^{-9}$, $n_s=0.9624$, $N_\text{eff}=3.046$ with massless neutrinos.
All distances, volumes and wavevectors are quoted in comoving units, and typically in $h^{-1}$Mpc or $h\,{\rm Mpc}^{-1}$ units unless otherwise specified.
Halo masses refer to virial masses in $h^{-1}M_\odot$, defined as the mass enclosed within the virial radius, where the density is a factor $\Delta_\text{vir crit}(z) = 18\pi^2 + 82\left[\Omega_m(z)-1\right] - 39\left[\Omega_m(z)-1\right]^2$ higher than the critical density $\rho_\text{crit}(z)$ \cite{Bryan98}.
In all the halo model calculations, we make use of the Sheth \& Tormen halo mass distribution function and linear bias \cite{Sheth99}.  We summarize the key symbols used in this paper in Table~\ref{tab:symbol_list}.

\section{Multi-line halo model formalism}
\label{sec:formalism}

In this section, we present our halo model formalism, based on the conditional luminosity function. 
In the companion paper \cite{paper2}, we use this formalism to quantify the detectability of various line power spectra, disentangle the cosmological and astrophysical information, and compare LIM to galaxy detection.

\subsection{Origin of nebular emission lines \& line correlations}
\label{sec:physical_origin}

\begin{figure}[h!]
\centering
\includegraphics[width=0.95\textwidth]{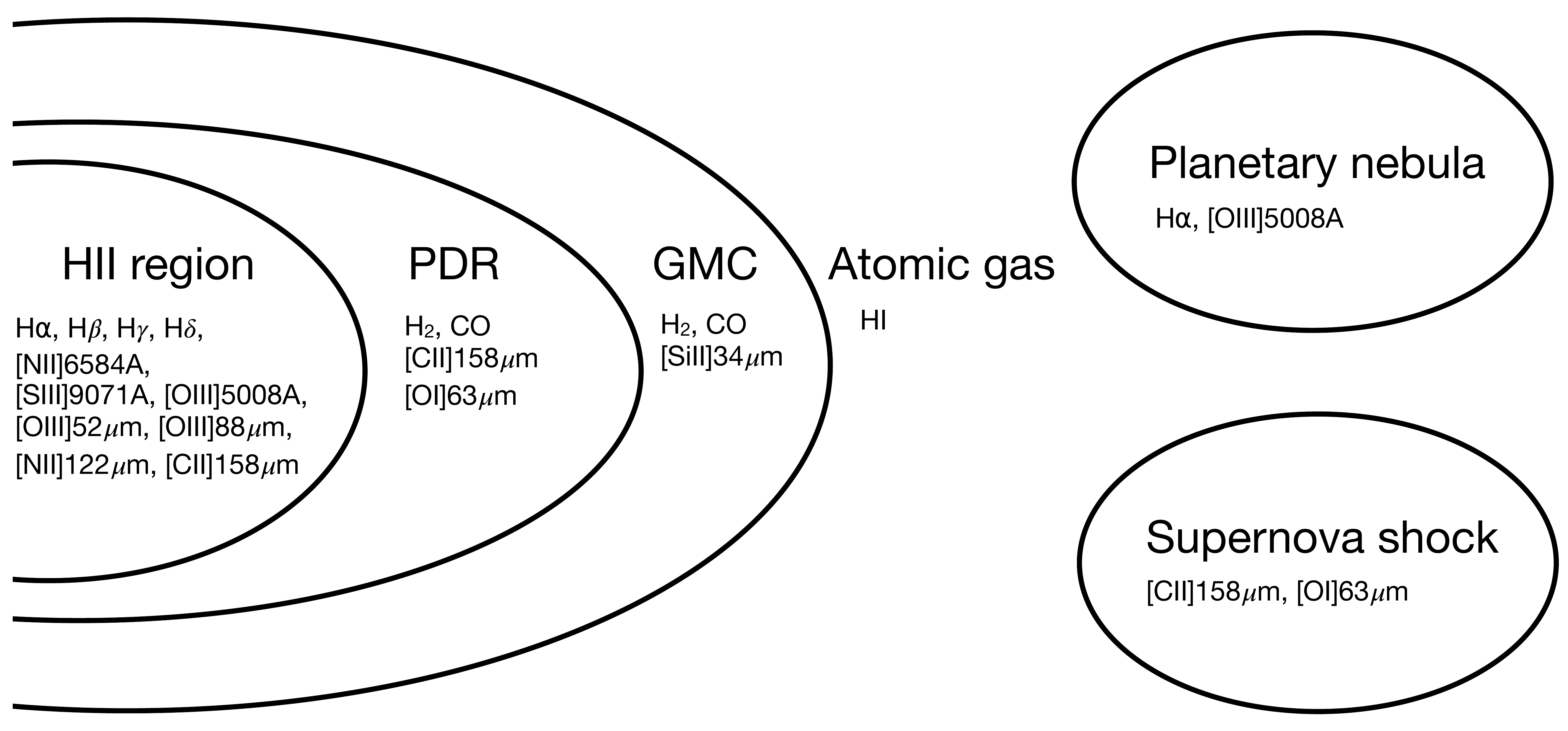}
\caption{
The line emission of a given galaxy is the sum of the contributions from the atomic gas, giant molecular clouds (GMC), photodissociation regions (PDR), H{\sc ii} regions around bright star, as well planetary nebulae and supernova remnants.
The luminosities of lines coming from the same regions inside a galaxy are likely to be highly correlated.
}
\label{fig:illustration_origin_lines}
\end{figure}

The lines of interest to us come from hot ($10^4\,$K) clouds of interstellar gas that are heated, excited and ionized by ultraviolet photons (and the associated photoelectrons) from nearby hot stars or white dwarfs.
These lines have long served as a major source of information about astrophysical phenomena in galaxies, such as star formation rate, metallicity, gas density and temperature \cite{Osterbrock06, Levesque10, Gutkin16, Byler17}.
However, understanding the physical origin of nebular emission lines is also crucial for cosmology.
Indeed, it determines which galaxies and which halos, are most effective at producing a given line.
This in turn determines whether the corresponding line intensity map is a useful tracer of the matter density field, through the value of its bias and shot noise (of galaxies and halos).
The physical origin of nebular emission lines also informs the level of correlation between a galaxy's luminosity in two different lines.
Indeed, two lines produced in the same regions (e.g., H{\sc ii} regions) will likely be more correlated than lines produced in distinct regions (e.g., photodissociation region and supernova remnant).
As we show in this paper, this has implications for, e.g., our ability to undo the contamination from line interlopers, and thus for constraining cosmology from LIM.
For this reason, we briefly summarize the physical origin of nebular emission lines, based on existing thorough reviews \cite{Ferland03,Peimbert17,Osterbrock06,Draine11}, and illustrate it in Fig.~\ref{fig:illustration_origin_lines}.

At nebular densities, collisions are largely unimportant for the H and He lines we see, which are recombination lines.  
In many nebulae, all Lyman line photons are absorbed by other atoms (case B recombination).  
In this scenario the number of optical, Balmer photons (e.g.\ H$\alpha$ and H$\beta$) equals the number of ionizing photons emitted by the stars.  The level populations depend weakly on electron density and temperature so the line ratios are largely fixed by atomic physics, though observed line luminosities are additionally sensitive to absorption and dust attenuation within the nebula and in the interstellar medium.  
The situation is more complex for other lines.  
For example the strength of the collisionally ionized [O{\sc ii}] and [O{\sc iii}] optical lines depends upon metallicity and excitation state of the gas, and is more sensitive to dust extinction.  
The relative strength of the IR CO transitions is sensitive to a nearly degenerate combination of H${}_2$ gas density and temperature, with higher $J$ lines being associated with warmer and denser gas.  
The strength of collisionally excited heavy element lines, relative to hydrogen recombination lines, is determined by the hardness of the ionizing radiation field (see \S7.4 and Eq.~11 in \cite{Ferland03}).
In addition, there can be multiple gas components within a galaxy, each with differing metallicity and dust redenning,  impacting the line ratios of optical and UV lines (e.g.\ H$\alpha$, [O{\sc ii}] and [O{\sc iii}]) \cite{Moustakas06,Shivaei16}, and far IR lines (e.g.\ the various CO transitions) \cite{Daddi15}.  
As expected, metallicity has a strong effect on the relative flux of lines of different atomic species, and dust has a large effect on the relative flux of lines widely separated in frequency.  
Finally, it is expected that gas fraction, density, temperature, pressure, metallicity and the intensity and hardness of the radiation field will evolve strongly with redshift.

\subsection{Multi-line CLF: Jointly modeling multiple line intensities and power spectra}
\label{sec:clf}

One particularly clear way of formalizing the evolution of galaxy properties and their dependence on the host halo mass is through the use of the ``conditional luminosity function'' (CLF; \cite{Yang03}), which we generalize to allow multiple luminosities (for the continuum and each line of interest).
We introduce the multivariate conditional luminosity function $\phi\left( L_1,..., L_n | m \right)$ such that
$\phi\left( L_1, ..., L_n | m \right) dL_1 ... dL_n$ 
is the mean number of galaxies in one halo of mass $m$
with luminosities $L_1, ..., L_n$ in lines $1, ..., n$.
They satisfy:
\beq
\phi(L_1, ..., L_{n-1}|m)
=
\int dL_n\;
\phi(L_1, ..., L_n|m).
\eeq
The multivariate CLF describes the distribution of galaxy line luminosities inside halos. It determines the evolution of the galaxy population with halo mass and redshift.

The usual (unconditional) galaxy luminosity function $\Phi$, frequently found in the literature, can be recovered from the CLF as
\beq
\Phi(L_1, ..., L_n)
=
\int dm\ n(m)
\phi(L_1, ..., L_n | m),
\eeq
such that 
$\Phi(L_1, ..., L_n) dL_1 ... dL_n$ is the mean number density of galaxies with luminosities $L_1, ..., L_n$ in lines $1, ..., n$, regardless of host halo mass.

In this paper, we focus on the mean line intensity and the line intensity auto- and cross-spectra, i.e.\ we discard the information from higher order correlations (e.g.\ the bispectrum \cite{Beane18} or voxel intensity distribution \cite{Breysse17a}).
We give a step-by-step derivation
in App.~\ref{app:pedagogical_derivation} and \ref{app:luminosity_scatter_halo_model}, and simply summarize and discuss the results here.  The case of cross-correlating a line intensity map with a traditional galaxy or quasar survey is discussed in App.~\ref{app:tracer_cross_correlation} while the specialization to angular clustering is discussed in App.~\ref{app:angular_clustering}.
In what follows we shall neglect the distinction between satellites and centrals primarily due to a paucity of data for constraining such a model.  The full case is described in the appendices and we discuss this simplification further later in the text.
Finally, we ignore radiative transfer once the light is emitted by the galaxies but note that this is issue can be particularly delicate and important for the (resonant) Lyman-$\alpha$ line \cite{Behrens18, Byrohl19, Gurung21}.

\begin{table}[h!]
\centering
\begin{tabular}{ |p{3cm} p{11cm}|}
\hline
$n(m, z)$ & Halo mass function\\
$b(m, z)$ & Halo bias\\
$\text{SFR}(m, z)$ & Total star formation rate in a halo of mass $m$ at redshift $z$\\
\hline
\hline
$\phi\left( L_1, ..., L_n | m \right)$ & Multi-line conditional luminosity function\\
$\Phi\left( L_1, ..., L_n \right)$ & Multi-line (unconditional) luminosity function\\
$\phi\left( L_j | m \right)$ & Single-line conditional luminosity function\\
$\Phi\left( L_j \right)$ & Single-line (unconditional) luminosity function\\
$N_\text{gal}(m)$ & Mean number of galaxies in a halo of mass $m$\\
$\bar{n}^\text{gal}$ & Total comoving number density of galaxies\\
$L_j(m)$ & Mean line-$j$ luminosity of a halo of mass $m$\\
$\mathcal{L}_j$ & Total luminosity density per unit comoving volume\\
$L^\text{gal}_j(m)$ & Mean luminosity of a galaxy in a halo of mass $m$\\
$\bar{L}^\text{gal}_j$ & Mean luminosity of a galaxy in the universe\\ 
$\bar{I}_j$ & mean specific intensity in line $j$\\
\hline
\hline
$P_{i,j}(k, \mu, z)$ & Redshift-space cross-spectrum of lines $i$ and $j$\\
\hline
$P_{i,j}^\text{2-halo}(k, \mu, z)$ & 2-halo term for lines $i$ and $j$\\ 
$b_j(k, \mu, z)$ & Scale-dependent intensity bias\\
$F(k, \mu, z)$ & Effective growth rate of structure\\
\hline
$P_{i,j}^\text{1-halo}(k, \mu, z)$ & 1-halo term for lines $i$ and $j$\\ 
$\bar{n}^\text{h eff}_{i,j}(z)$ & Effective mean number density of halos for the 1-halo term\\
$U_{i,j}^2(k, \mu, z)$ & Effective squared halo profile for the 1-halo term.\\
$\sigma^2_{\text{h }i,j}$ & Halo ``line noise'' enhancing the 1-halo term\\
\hline
$P_{i,j}^\text{shot}(z)$ & Shot noise power spectrum for lines $i$ and $j$\\
$\bar{n}^\text{gal eff}_{i,j}$ & Effective mean number density of galaxies for the shot noise term.\\
$\sigma^2_{\text{gal }i,j}$ & Galaxy ``line noise'' enhancing the shot noise term\\
\hline
\end{tabular}
\caption{
List of the main symbols used in this paper.
}
\label{tab:symbol_list}
\end{table}

\subsubsection{Mean intensity}

The mean intensity, $\bar{I}_j$, in line $j$ is determined by the univariate CLF, $\phi\left( L_j | m \right)$.
Indeed, the zero-th moment of the CLF gives the mean number of galaxies in a halo of mass $m$, $N_\text{gal}(m)$, and the total comoving number density of galaxies:
\beq
\left\{
\bal
&N_\text{gal}(m)
=
\int dL_1\ \phi_1(L_1|m)\\
&\bar{n}^\text{gal}
=
\int dm \ n(m) N_\text{gal}(m)\\
\eal
\right.
,
\eeq
where $n(m)$ is the halo mass function, the number of dark matter halos per volume per mass.

The first moment of the CLF gives the mean line $j$ luminosity of a halo of mass $m$, $L_j(m)$, the total luminosity density per unit comoving volume, $\mathcal{L}_j$, the mean luminosity of a galaxy in a halo of mass $m$, $L^\text{gal}_j(m)$, and the mean luminosity of a galaxy in the universe, $\bar{L}^\text{gal}_j$:
\beq
\left\{
\bal
&L_j(m)
=
\int dL_j\ \phi(L_j|m) L_j\\
&\mathcal{L}_j
\equiv
\int dm \ n(m) L_j(m)\\
&L^\text{gal}_j(m)
=
L_j(m) / N_\text{gal}(m)\\
&\bar{L}^\text{gal}_j
=
\mathcal{L}_j / \bar{n}^\text{gal}\\
\eal
\right.
.
\eeq
The mean halo luminosity $L_j(m)$, the key ingredient of many halo models in the literature, can thus be derived from the CLF.

The mean specific intensity in line $j$ is 
\beq
\bar{I}_j
=
\frac{1}{4\pi \nu_j^0}
\frac{c}{H(z)}
\ \int dL_j\ \Phi(L_j) L_j
,
\label{eq:mean_intensity_lf}
\eeq
where $\nu_j^0$ is the rest-frame frequency of line $j$.  More intuitively
\beq
\boxed{
\bar{I}_j
=
\frac{1}{4\pi \nu_j^0}
\frac{c}{H(z)}
\ \mathcal{L}_j,
}
\eeq
which simply converts the total galaxy luminosity per unit volume into an intensity at a given redshift and frequency.

While the mean intensity is determined by the univariate CLF, $\phi\left( L_1 | m \right)$, the power spectrum also requires knowledge of the bivariate CLF, $\phi\left( L_1, L_2 | m \right)$.
The total power spectrum, a function of the comoving wave vector modulus $k$, cosine of the angle to the line of sight $\mu$ and redshift $z$, is made of three contributions:
\begin{equation}
    P_{i,j}(k,\mu,z) = P_{i,j}^\text{2-halo}(k, \mu, z) +
    P_{i,j}^\text{1-halo}(k, \mu, z) +
    P_{i,j}^\text{shot}(z),
\label{eqn:Ptotal}
\end{equation}
which we now describe in detail.

\subsubsection{Power spectrum: 2-halo}

\begin{figure}[h!]
\centering
\includegraphics[width=0.45\textwidth]{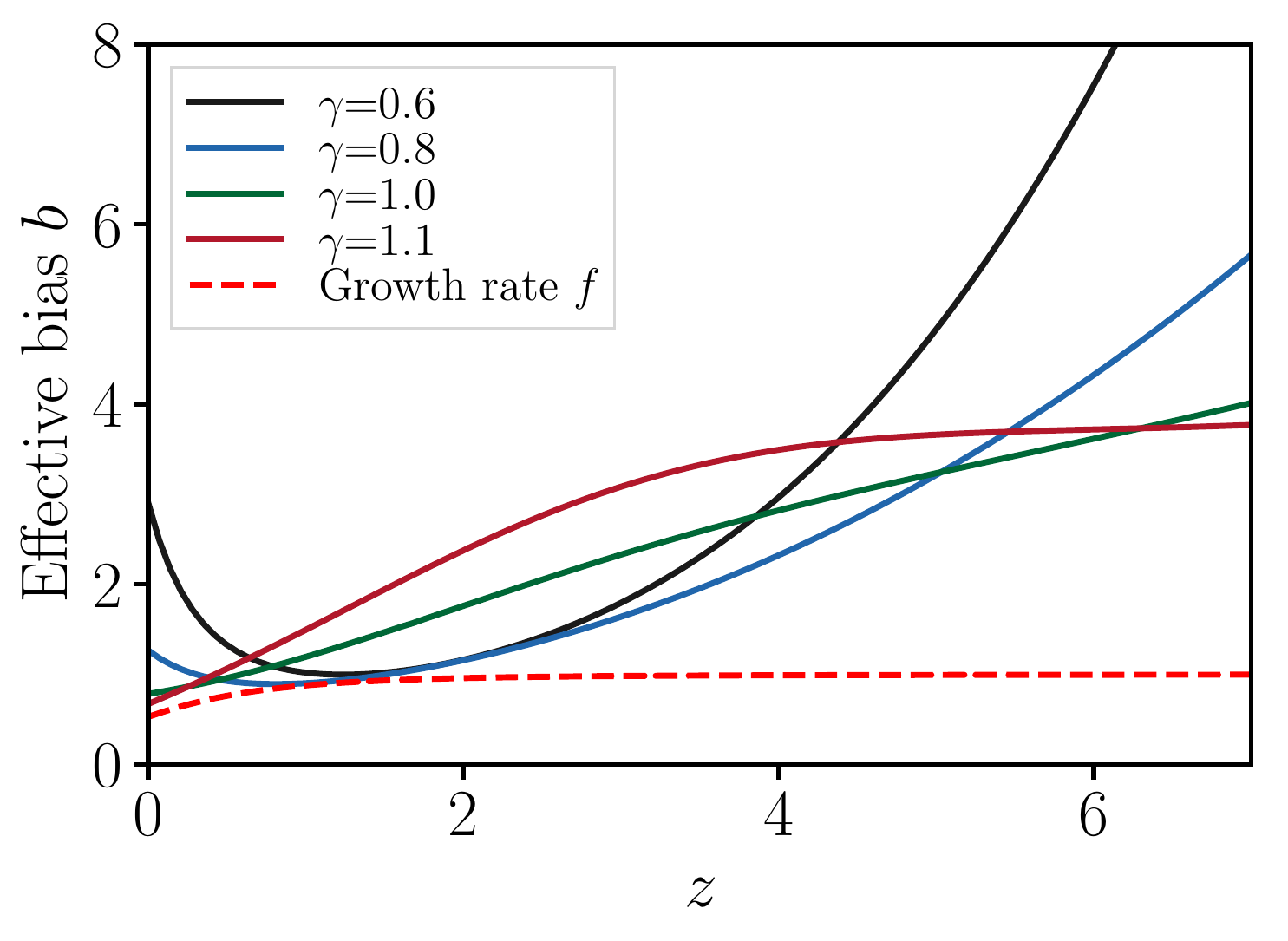}
\includegraphics[width=0.45\textwidth]{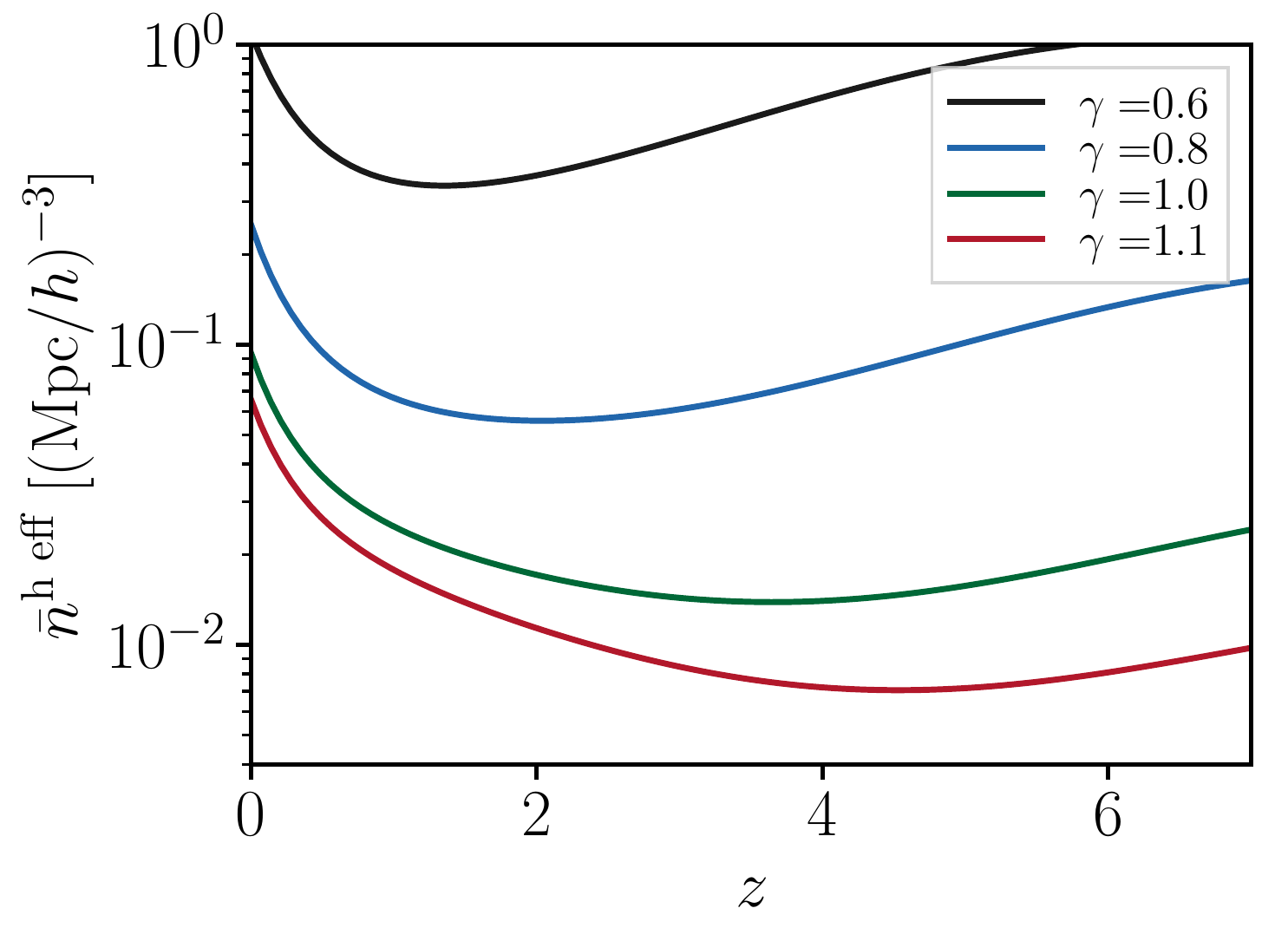}
\caption{
Redshift evolution of the effective halo bias at $k=0$ (left), which determines the amplitude of the 2-halo term.
Redshift evolution of the effective halo number density (Eq.~\ref{eqn:nheff}; right), which determines the amplitude of the 1-halo term.
Different curves correspond to halo luminosities $L(M)\propto \text{SFR}^\gamma$, with $\gamma\simeq 1$ for optical and UV lines, and $\gamma=0.6-1.1$ for far IR lines \cite{Cheng16, Fonseca17}.
}
\label{fig:b_nh}
\end{figure}

On large scales, the number density of halos traces the underlying linear matter density field. 
This gives rise to the 2-halo term in the intensity power spectrum:
\beq
\boxed{
P_{i,j}^\text{2-halo}(k, \mu, z)
=
\bar{I}_i \bar{I}_j
\left[ b_i + F \mu^2 \right]
\left[ b_j + F \mu^2 \right]\; 
P_\text{lin}.
}
\label{eqn:P2halo}
\eeq
The standard assumption of a deterministic, linear, scale-independent but mass-dependent halo bias $b(m)$, combined with a (normalized) halo profile $u(k,m)$ distorted by the finger-of-god (FOG) effect, gives rise to a scale-dependent intensity bias:
\beq
b_j(k, \mu, z)
\equiv \mathcal{L}_j^{-1}
\int
dm\ n(m)\;
L_j(m)
\;
b(m) 
\;
u(k,m)
e^{-k^2 \mu^2\sigma_d^2(m) / 2}
,
\label{eq:effective_bias}
\eeq
where $\sigma_d = \sigma_{v\ \text{1D}}/aH$ is the dispersion of the spurious displacement in redshift space due to the random LOS motion within halos $\sigma_{v\ \text{1D}}$.
Following ref.~\cite{White01}, we approximate the LOS velocity dispersion as that of a singular isothermal sphere of mass $m$ and radius $r_\text{vir}$, i.e.
$\sigma_{v\ \text{1D}}^2 = \mathcal{G} m / 2 r_\text{vir}$.
This simple analytical estimate is a good match to the simulation results reported in ref.~\cite{Evrard08}.
On large scales, where
$u(k,m) e^{-k^2 \mu^2\sigma_d^2 / 2} \longrightarrow 1$,
the intensity bias becomes scale-independent, and converges to the halo-luminosity-weighted averaged halo bias, shown in Fig.~\ref{fig:b_nh}.

In redshift-space, the effective growth rate appears: 
\beq
F(k, \mu, z)
\equiv
f\;
\int dm \; n(m)\;
\left( \frac{m}{\bar{\rho}} \right)
u(k,m) 
e^{-k^2 \mu^2 \sigma_d^2(m) /2}.
\label{eq:effective_growth_rate}
\eeq
Here the weighting is simply by halo mass, rather than halo luminosity, since the Kaiser effect is due to the large scale velocities, which are sourced by the mass.
On large scales, where the FOG-corrected halo profile tends to one, we recover the usual logarithmic growth rate
$F \longrightarrow f\equiv d\ln D / d\ln a$,
as long as the matter density consistency relation 
$\int dm \; n(m)\;
\left(m/\bar{\rho} \right) = 1$
is satisfied\footnote{In practice, for most mass functions, recovering this consistency relation requires integrating down to unrealistically low halo masses. As is usual in the halo model, we ``correct'' this issue by dividing Eq.~\eqref{eq:effective_growth_rate} by $\int dm \; n(m)\;
\left( m/\bar{\rho} \right)$. 
Such a correction is only necessary for Eq.~\eqref{eq:effective_growth_rate}, not Eqs.~(\ref{eq:effective_bias},\ref{eq:p1h}), where the luminosity or squared luminosity weighting provides an effective cutoff at low halo masses. 
The convergence of the mean intensity, bias and 1-halo term at low masses is shown in Fig.~\ref{fig:didl_dshotdl} for $\gamma=1$. We verified it for $\gamma=0.6-1.1$.}.
As shown in Fig.~\ref{fig:b_nh}, the growth rate of structure is comparable to the effective halo bias at the lower redshifts, implying that redshift-space distortions are an important effect.

\subsubsection{Power spectrum: 1-halo}

On intermediate scales, the assumed Poisson distribution of halos results in a halo shot noise, called the 1-halo term:
\beq
P_{i,j}^\text{1-halo}(k, \mu, z)
=
\left(\frac{c}{4\pi H(z)}\right)^2 \frac{1}{\nu_i^0 \nu_j^0}
\int dm \; n(m)
\left| u(k,m) \right|^2
e^{-k^2\mu^2\sigma^2}
L_i(m) L_j(m).
\label{eq:p1h}
\eeq
Although the CLF formalism can describe the distinction between central and satellite galaxies (see \S\ref{app:centrals_satellites}),
we ignore this distinction here and assume that all galaxies follow the NFW profile $u(k,m)$, with $u(k\rightarrow0,m)\rightarrow 1$.
The 1-halo term can be rewritten more transparently as:
\beq
\boxed{
P_{i,j}^\text{1-halo}(k, \mu, z)
=
\bar{I}_i \bar{I}_j
\frac{U_{i,j}^2(k, \mu, z)}{\bar{n}^\text{h eff}_{i,j}}.
}
\eeq
As expected for a halo shot noise, the 1-halo term scales as the inverse mean number density of halos.
The relevant, effective mean number density of halos, $\bar{n}^\text{h eff}_{i,j}$, depends on the lines $i$ and $j$ considered and properly counts the halos by taking into account their luminosities in these lines:
\beq
\bar{n}^\text{h eff}_{i,j}
=
\frac{
\left( \int dm \; n(m) L_i(m)  \right)
\left( \int dm \; n(m) L_j(m)  \right)
}
{\int dm \; n(m) L_i(m) L_j(m)}.
\label{eqn:nheff}
\eeq
This expression can be understood intuitively in several simple cases.
For instance, if line luminosities were independent of halo mass above some cutoff (i.e.\  $L_i(m)$ and $L_j(m)$ do not depend on $m$), then we simply recover
$\bar{n}^\text{h eff}_{i,j} 
=
\int_{\rm min} dm\; n(m)$, 
i.e.\ the total number of halos.
More realistically, if the halo luminosities scale as the star formation rate in a halo of mass $m$, SFR$(m, z)$, we see that the halo shot noise scales as the mean squared halo star formation rate:
\beq
P_{i,j}^\text{1-halo}
\propto
\frac{1}{\bar{n}^\text{h eff}_{i,j}}
\propto
\int dm\; n(m) \text{SFR}^2(m)
.
\eeq
The effective halo number density is shown in Fig.~\ref{fig:b_nh} for cases where $L\propto\mathrm{SFR}^\gamma$ for $\gamma=0.6-1.1$, as is typical for optical, UV and IR lines.

The scale and angular dependence of the 1-halo term is determined by the effective squared halo profile:
\beq
U_{i,j}^2(k, \mu, z)
\equiv
\frac{
\int dm \; n(m)
\left| u(k,m) \right|^2
e^{-k^2\mu^2\sigma^2}
L_i(m) L_j(m)
}
{
\int dm \; n(m)
L_i(m) L_j(m)
}.
\eeq
This is simply the average squared halo profile, weighted by the product of the halo luminosities in lines $1$ and $2$.
On large-enough scales, where the FOG-corrected halo profile goes to unity,
this takes the simple limit
$U_{1,2}^2 \rightarrow 1$, such that the 1-halo term is simply:
\beq
P_{i,j}^\text{1-halo}(k\rightarrow 0, \mu, z)
\rightarrow
\frac{\bar{I}_i \bar{I}_j}{\bar{n}^\text{h eff}_{i,j}}
=
\left(\frac{c}{4\pi H(z)}\right)^2 \frac{1}{\nu_i^0 \nu_j^0}
\int dm\; n(m)
L_i(m)L_j(m),
\eeq
This last equation shows that it is the correlation between halo luminosities in different lines at the same mass that matters to the amplitude of the 1-halo term.
The corresponding 1-halo amplitude in the $k\to 0$ limit is shown as a function of redshift in the right panel of Fig.~\ref{fig:b_nh}.

In the halo model, the 1-halo term generically tends to a non-zero constant on large scales.
This behavior appears unphysical for the matter density, where the low-$k$ power spectrum is expected to approach the linear power spectrum $P_\text{lin}(k)\propto k$ \cite{Cooray02}, and any perturbation conserving mass and momentum on large scales should scale as $\delta P(k) \propto k^4$ \cite{Zeldovich65, Peebles80, Mercolli14}.
This may be corrected by including the effect of halo exclusion \cite{Baldauf13} or compensated halo profiles \cite{Cooray02}, but the complete resolution of this problem remains an open question.
In what follows, we therefore focus only on the small-scale behavior of the 1-halo term, and leave its large-scale behavior unspecified.

\subsubsection{Power spectrum: shot noise}

\begin{figure}[h!]
\centering
\includegraphics[width=0.45\textwidth]{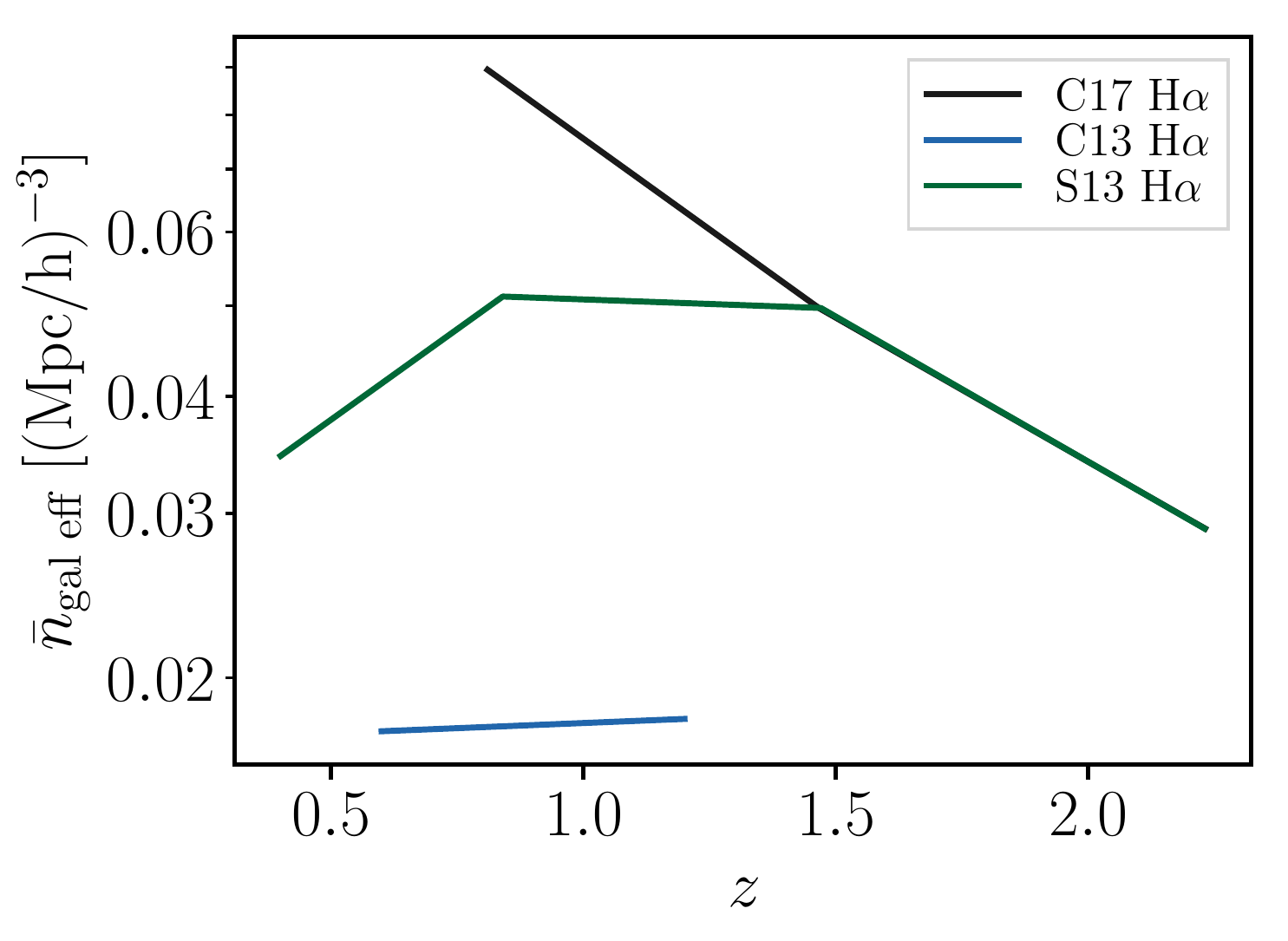}
\includegraphics[width=0.45\textwidth]{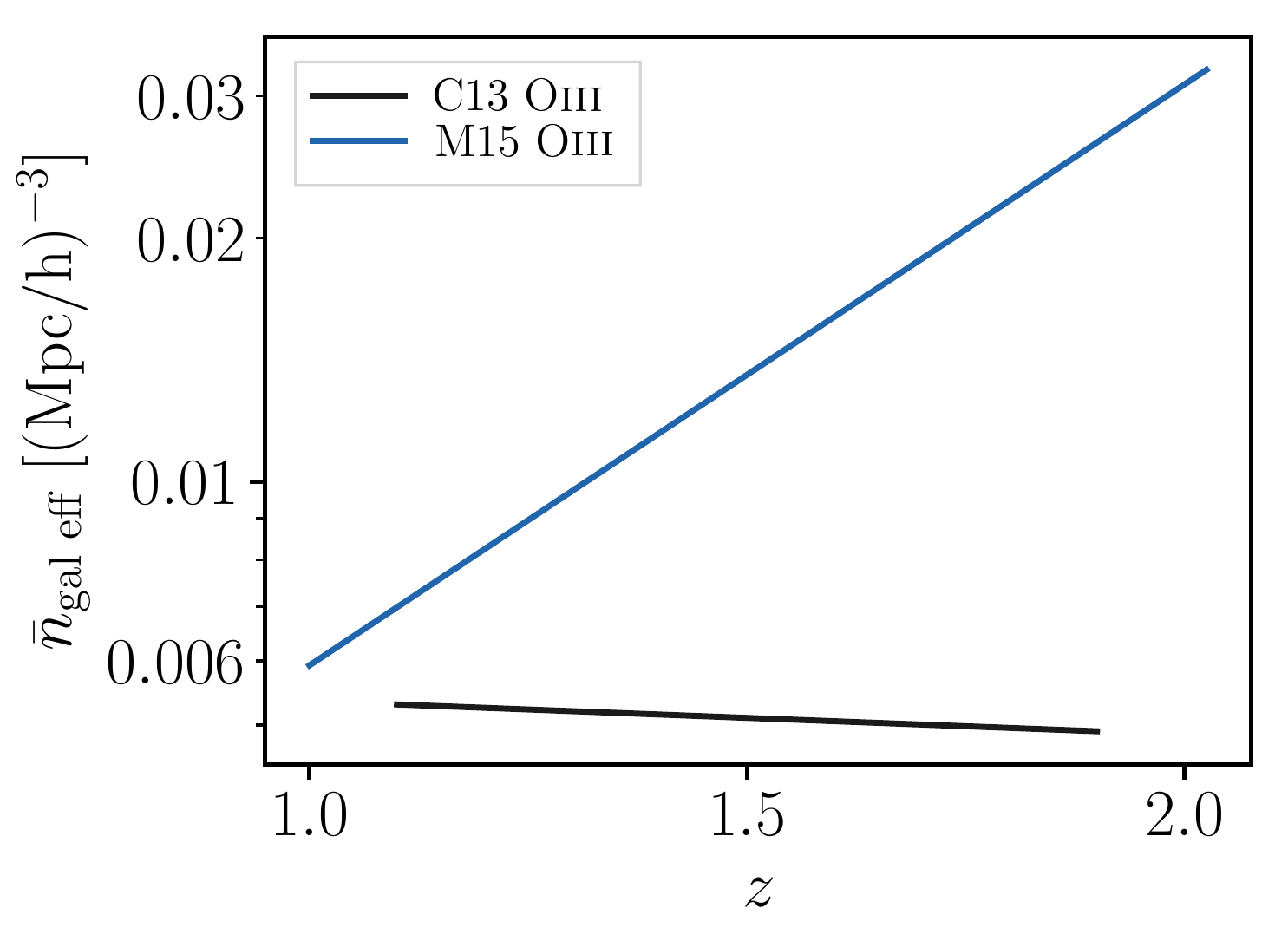}
\caption{
Comparison of the effective galaxy number densities, $\bar{n}_{\rm gal,eff}$ (Eq.~\ref{eqn:ngaleff}), as a function of redshift, for H$\alpha$ (left) and [O{\sc iii}] (right), as predicted from several observed luminosity functions (S13 \cite{Sobral13}; C13 \cite{Colbert13}; M15 \cite{Mehta15}; C17 \cite{Cochrane17}).
The various references imply very different $\bar{n}_{\rm gal,eff}$, indicating that a measurement of the galaxy shot noise would greatly increase our knowledge of the galaxy luminosity functions.
In all cases, the effective number density of galaxies is similar to that of halos (Fig.~\ref{fig:b_nh}), indicating that the shot noise and 1-halo terms are comparable in amplitude for H$\alpha$ and [O{\sc iii}].
}
\label{fig:ngaleff_z}
\end{figure}

On the smallest scales, the Poisson fluctuations in the number of galaxies in a given halo produce a galaxy shot noise, which enhances the intensity power spectrum:
\beq
P_{i,j}^\text{shot}(z)
=
\left(\frac{c}{4\pi H(z)}\right)^2 \frac{1}{\nu_i^0 \nu_j^0}
\int dL_i dL_j\ \Phi(L_i, L_j) L_i L_j.
\eeq
This galaxy shot noise term is the only one which truly requires the bivariate CLF, i.e.\ a knowledge of the correlation between the luminosity in different lines from the same galaxy.
This can be rephrased, similarly to the 1-halo term:
\beq
\boxed{
P_{i,j}^\text{shot}(z)
=
\frac{\bar{I}_i \bar{I}_j}{\bar{n}^\text{gal eff}_{i,j}}
,
}
\eeq
where the effective number density of galaxies properly takes into account the contribution from each galaxy to the intensity:
\beq
\bar{n}^\text{gal eff}_{i,j}
=
\frac{
\left( \int dL_i\ \Phi(L_i) L_i \right)
\left( \int dL_j\ \Phi(L_j) L_j  \right)
}
{
\int dL_i dL_j\ \Phi(L_i, L_j) L_i L_j
}
.
\label{eqn:ngaleff}
\eeq
Because galaxies are point sources for our purposes, this power spectrum is scale-independent.
Its amplitude scales as the inverse mean number density of galaxies, as expected for a galaxy shot noise.
In Fig.~\ref{fig:ngaleff_z}, we predict the effective galaxy number density from various luminosity functions in the literature, described in \S\ref{subsec:lf}.  Different observational determinations of the luminosity functions lead to quite different $\bar{n}_{\rm gal,eff}(z)$, though in all cases $\bar{n}_{\rm gal,eff}$ is comparable to the effective density of halos.  This suggests that observations of small-scale clustering in LIM surveys could provide valuable information about the form and evolution of the underlying luminosity functions.

Looking at the mean intensity and the various power spectrum terms, we see that all the information is encoded in the multi-line CLF.
The mean intensity is entirely determined by the first moment of the (unconditional) luminosity function, i.e. the mean luminosity density in the Universe $\mathcal{L}_j$.
The 2-halo and 1-halo power spectra are determined by the first moment of the single-line CLF, i.e.\ $L_j(m)$.
Finally, the shot noise requires the second moment of the two-line (unconditional) LF $\int dL_i dL_j\ \Phi(L_i, L_j) L_i L_j$.
The LIM literature often uses the quantities $\mathcal{L}_j$, $L_j(m)$ and $\int dL_i\ \Phi(L_i) L_i^2$ as their starting point.
Our formalism thus connects to the literature, extends it to cross-correlations, and shows how these quantities all derive from the multi-line CLF.

\begin{figure}[h!]
\centering
\includegraphics[width=0.45\textwidth]{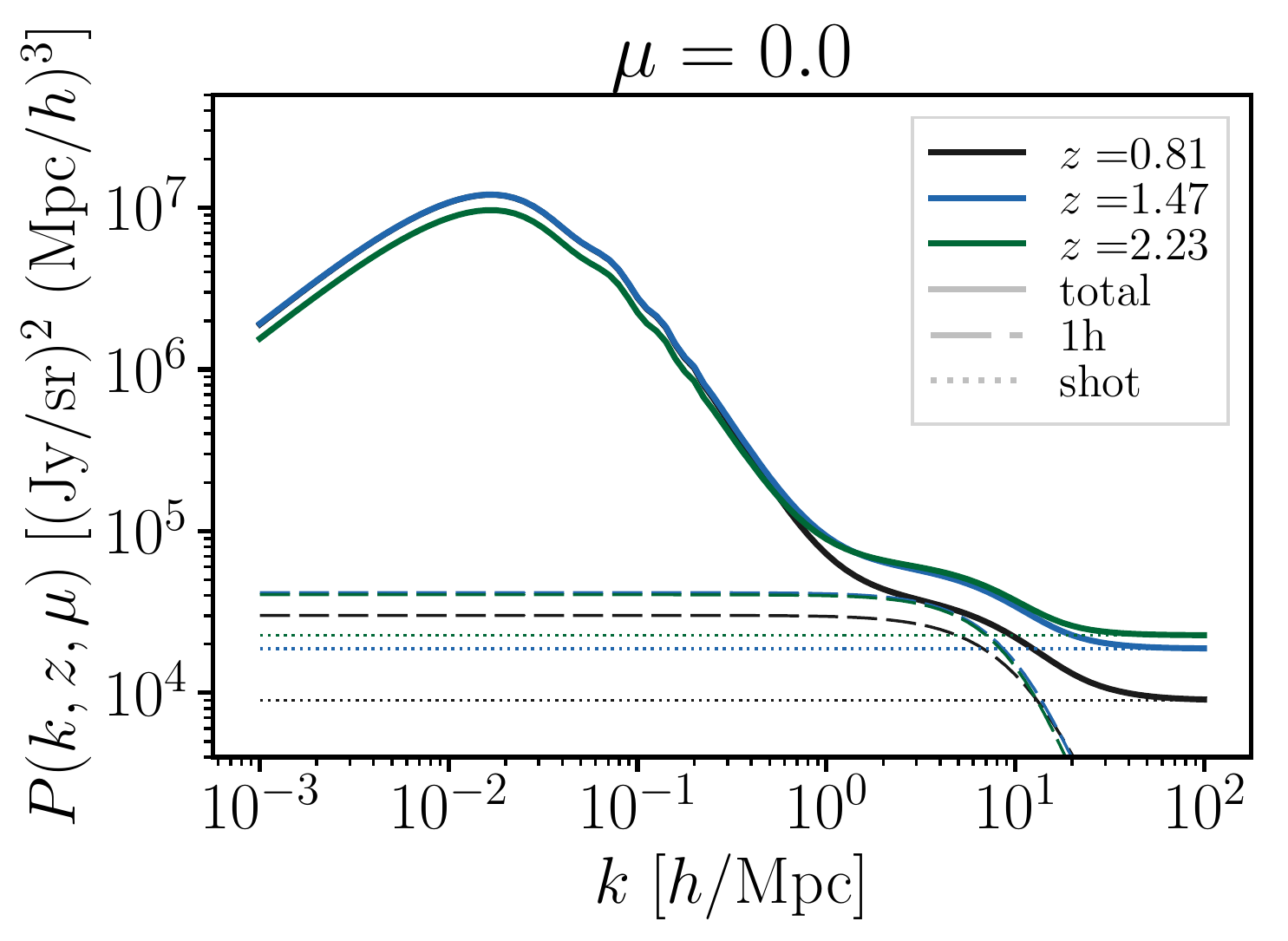}
\includegraphics[width=0.45\textwidth]{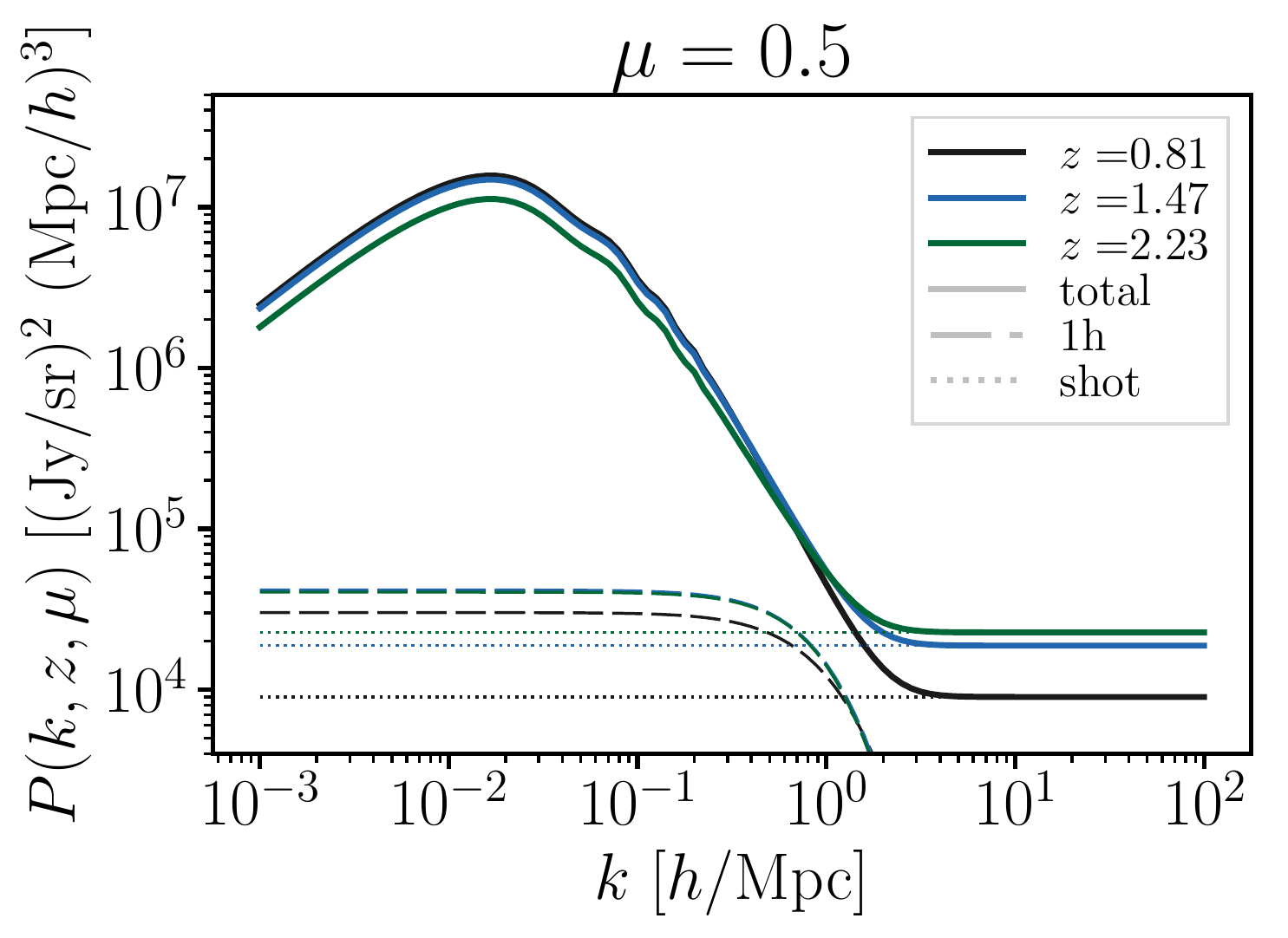}
\caption{
The total power spectrum (solid), 1-halo contribution (dashed) and shot noise (dotted) for the H$\alpha$ power spectrum as a function of redshift (line colors), for $\mu=0$ (left) and $\mu=0.5$ (right).
This assumes the H$\alpha$ luminosity function from ref.~\cite{Cochrane17}.
}
\label{fig:pk_terms}
\end{figure}

We combine the different components to form the observed power spectrum, in redshift space, in figure \ref{fig:pk_terms}.  
For illustration figure \ref{fig:pk_terms} shows the H$\alpha$ power spectrum for $\mu=0$ (i.e.\ the real-space or tranverse power spectrum) and for $\mu=0.5$ at $z=0.81$, 1.47 and 2.23 with the different contributions to the total indicated by different line types. 
The power spectrum evolves only slowly with redshift over the interval shown, since the evolution of the matter clustering is largely compensated by the evolution of the halo-luminosity connection.  
One can see that the two halo term dominates on large scales (low $k$) with the 1-halo term dominating at intermediate scales with the shot-noise dominating at very small scales (high $k$).  For modes close to along the line of sight direction ($\mu\approx 1$) the 2-halo and 1-halo terms are more strongly damped near $k\simeq 1\,h\,{\rm Mpc}^{-1}$ while for transverse modes the 1-halo term has power to significantly higher $k$.
Comparing the $\mu=0$ and $\mu=0.5$, the FOGs act as a switch, turning the 1-halo term on across the LOS and off along the LOS.
Inaccuracies in our modeling of the FOG only change the scale at which this switch occurs, but not its effect.

\begin{figure}[h!]
\centering
\includegraphics[width=0.45\textwidth]{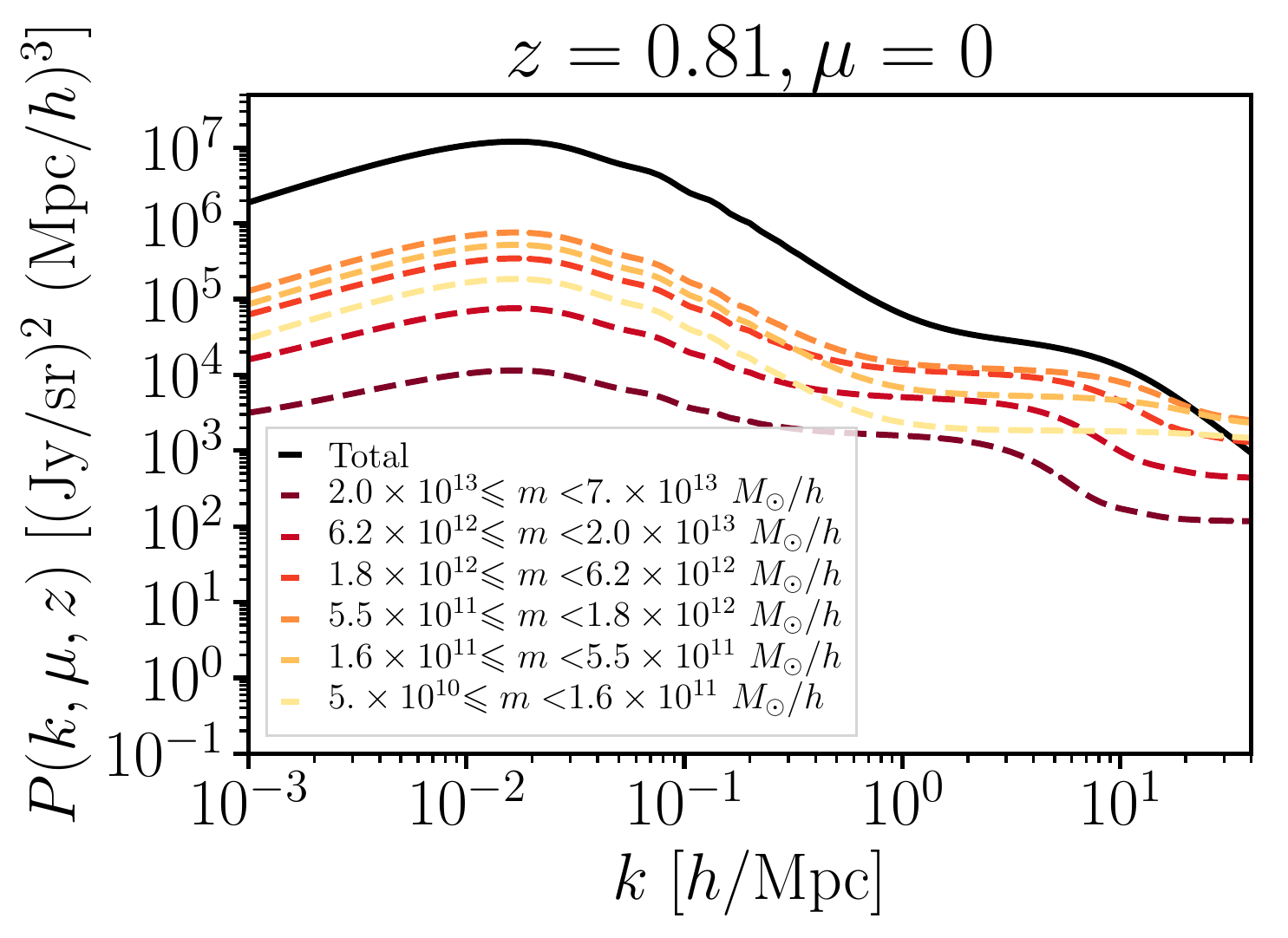}
\includegraphics[width=0.45\textwidth]{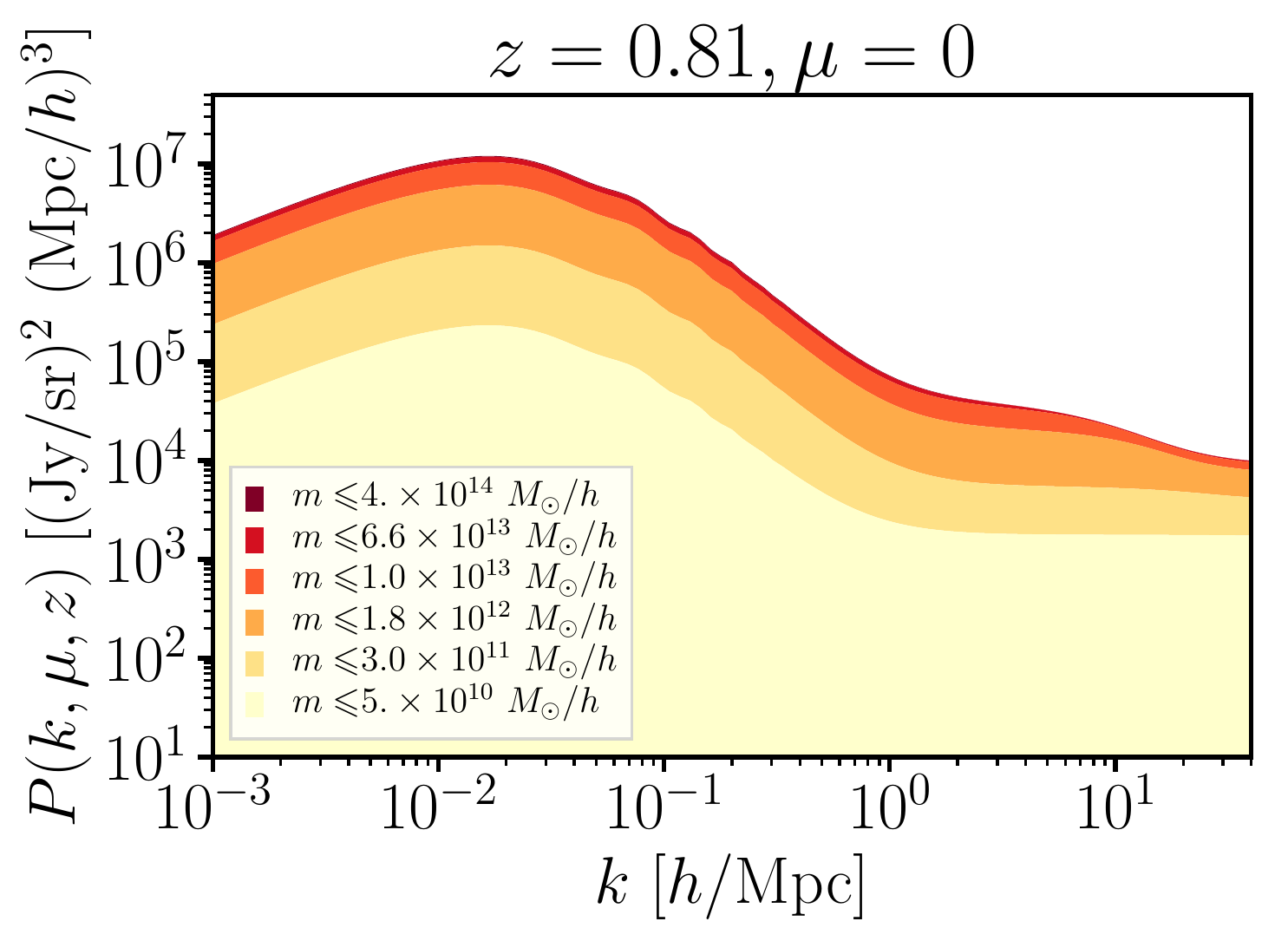}
\includegraphics[width=0.45\textwidth]{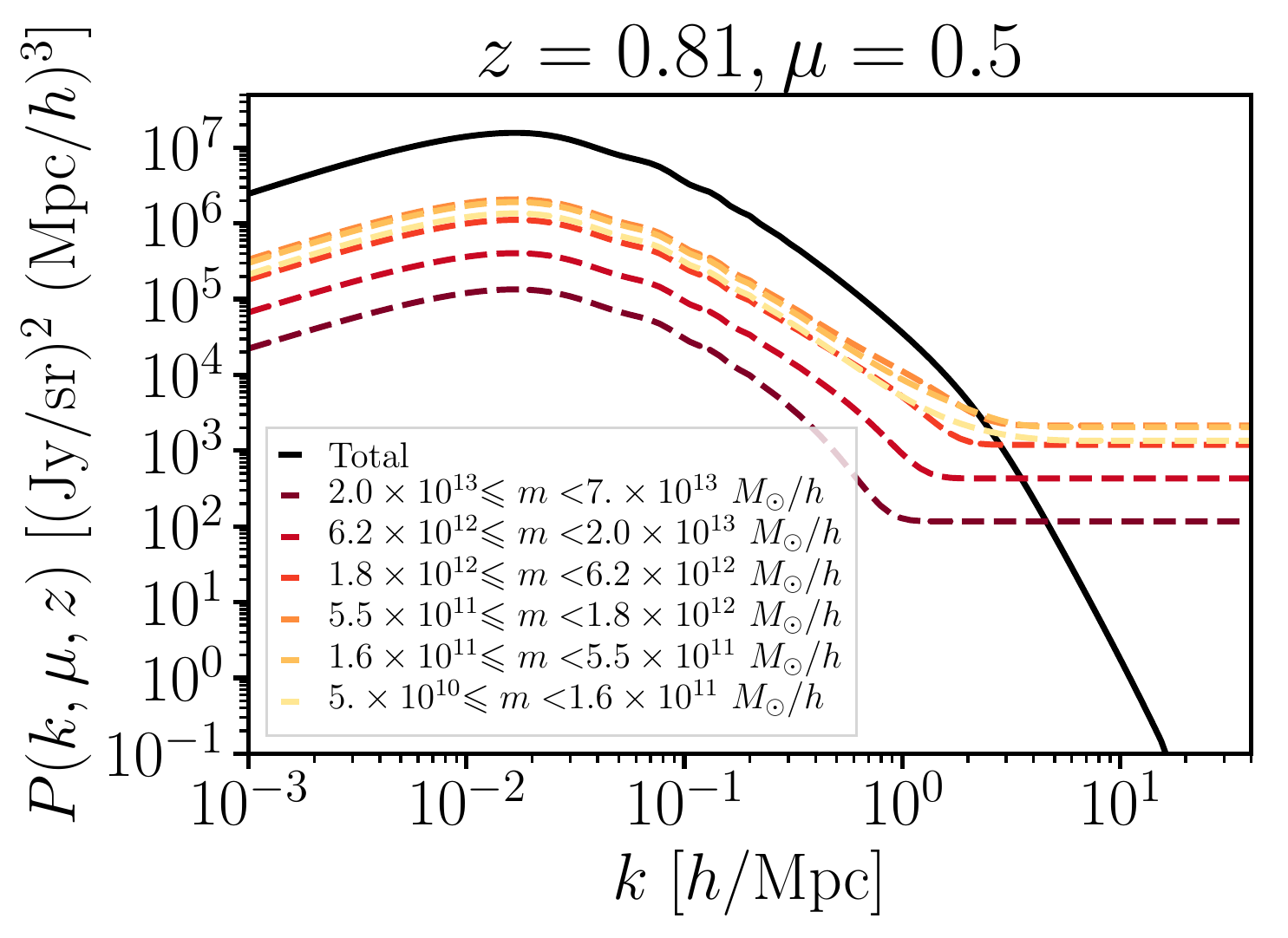}
\includegraphics[width=0.45\textwidth]{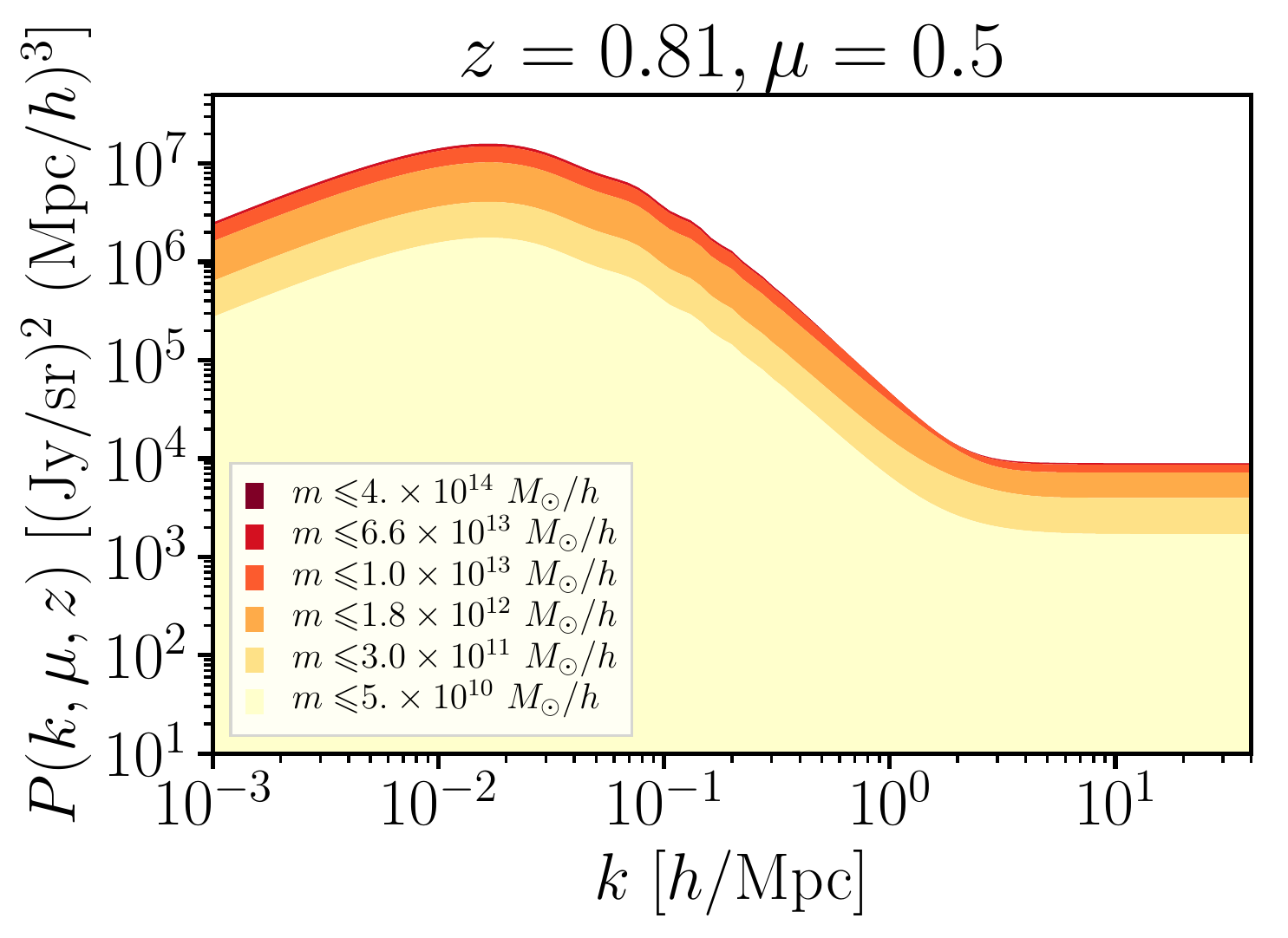}
\caption{
Contribution from different halo mass to the redshift-space power spectrum of H$\alpha$.
The power spectrum is shown for wave vectors across the LOS ($\mu=0$, top) and for wave vectors $60^\circ$ away from the LOS ($\mu=0.5$, bottom).
The mass contributions are shown for each mass interval (left) and cumulatively (right).
This assumes the H$\alpha$ luminosity function from ref.~\cite{Cochrane17}.
}
\label{fig:pk_mass_contributions}
\end{figure}

The contribution of different halo masses to the total H$\alpha$ power spectrum is shown in figure \ref{fig:pk_mass_contributions}, again for $\mu=0$ and $\mu=0.5$ at $z=0.81$.  
The left column shows the differential contribution while the right column shows the cumulative contribution.  
For this line, at this redshift, we see the bulk of the power spectrum is contributed by halos in the range $10^{11}-10^{12}\,h^{-1}M_\odot$. 
This is expected based on the relationship between H$\alpha$ luminosity and star-formation rate, the relationship between $\dot{M}_\star$ and $M_\star$ and the stellar-mass--halo-mass relation which we discuss further in \S\ref{sec:observational}.  
While different halo mass ranges do not contribute identically to the 2-halo, 1-halo and shot-noise terms the correspondence is in fact quite good as can be seen by comparing the ordering of the lines at low and high $k$ in the left column of figure \ref{fig:pk_mass_contributions}.  In this respect having access to the $\mu$-dependence of the clustering signal provides significant additional information as can be seen by comparing the upper and lower panels in the left column of figure \ref{fig:pk_mass_contributions} near $k\simeq 1\,h\,{\rm Mpc}^{-1}$.

\subsubsection{Origin of galaxy and halo shot noise}

The halo shot noise, which corresponds to the 1-halo term in the halo model, is determined by the effective number density of halos $\bar{n}^{\rm h,eff}$.
According to Eq.~\ref{eqn:nheff}, this is the integral of the mass function conditioned by the manner in which the mean line luminosity in a halo depends upon halo mass.  In particular we expect star formation to be ineffective in low mass halos, making these halos essentially invisible to the LIM experiment and raising the 1-halo term.  A measurement of the amplitude of the 1-halo term provides an integral constraint on the way in which luminosity is produced in halos as a function of halo mass.

By contrast the galaxy shot noise term, determined by $\bar{n}^{\rm gal,eff}$, is a sum of the fluctuations in galaxy line luminosity within halos of fixed mass and so is a probe of the stochasticity of the luminosity generation mechanism. One way to think of this stochasticity is in terms of ``line noise'',  a concept we introduce below.

Finally, the existence of these two shot noise terms (1-halo term and galaxy shot noise) results from the assumptions that the number of galaxies in a given halo is Poisson distributed.
This assumption is broken in a toy model where every halo contains either zero or one galaxy. 
In this case, the 1-halo and galaxy shot noise terms are identical, and should not be double counted.
However, our assumption holds in the more realistic case where the halo luminosity is dominated by one galaxy, and most halos only contain zero or one galaxy, but some contain more, following a Poisson statistics.

\subsection{Enhanced 1-halo term and shot noise from halo and galaxy ``line noises''}
\label{sec:line_noise}

The above formalism presents two interesting differences with the more familiar, galaxy-based halo model, leading to modifications to the power spectrum which we shall refer to as ``line noise''.  These noise sources can be understood by looking at the expressions for $\bar{n}_{ij}^{\rm h,eff}$ and $\bar{n}_{ij}^{\rm gal,eff}$ which enter the 1-halo and shot-noise power spectra, respectively. In both cases, the line noise reduces the effective halo and galaxy number densities compared to the naive expectations, thus enhancing the 1-halo and shot noise power spectra.

As is apparent from Eq.~(\ref{eqn:nheff}), if the mass dependence of the mean halo luminosity in line $i$ is significantly different from that in line $j$, $\bar{n}_{ij}^{\rm h,eff}$ can differ significantly from our naive expectations.
Let us formally introduce a ``halo line noise'' $\sigma^2_{\text{h } i,j}$ as the fractional covariance of halo luminosities in lines $i$ and $j$ across all halos:
\beq
\sigma^2_{\text{h }i,j}
\equiv \frac{\langle L_i(m)L_j(m)\rangle_h - \langle L_i(m)\rangle_h \langle L_j(m)\rangle_h}{\langle L_i(m)\rangle_h \langle L_j(m)\rangle_h} 
\eeq
where the expectation value of any function X over all halos is defined as\footnote{A subtlety with this formulation is that both $\sigma_{\text{h }i,j}^2$ and $\bar{n}_h = \int dm\ n(m)$ are ``cutoff dependent'', i.e.\ dependent on the arbitrary choice of the lower bound in the mass integrals. The expression for the 1-halo power spectrum remains correct for any sufficiently small choice of lower mass bound though, and the intuitive meaning of halo line noise remains.}:
\beq
\langle X \rangle_h
\equiv
\frac{1}{\bar{n}_h}
\int dm\ n(m)\ X(m)
\quad\text{with}\quad
\bar{n}_h = \int dm\ n(m).
\eeq
One can then rephrase the 1-halo power spectrum as:
\beq
P^\text{1-halo}_{i,j}(k\rightarrow0, \mu, z)
\rightarrow 
\frac{\bar{I}_i \bar{I}_j}{\bar{n}^\text{h eff}_{i,j}}
=
\frac{\bar{I}_i \bar{I}_j}{\bar{n}_h}
\left[
1 + \sigma^2_{\text{h }i,j}
\right].
\eeq
This intuitive expression shows that any variance in the halo line luminosities reduces the effective number density of halos and enhances the 1-halo term.

A similar situation arises in the galaxy shot-noise term.  This is the only term that depends upon the bivariate CLF, and so is sensitive to galaxy-line decorrelations.  As Eq.~(\ref{eqn:ngaleff}) makes clear, $\bar{n}_{ij}^{\rm gal,eff}$ depends upon how $\langle L_iL_j\rangle$ compares to $\langle L_i\rangle\langle L_j\rangle$ with the averages being over the CLF and halo mass function as shown explicitly in Eq.~(\ref{eqn:ngaleff}).  To bring out this point more clearly, and to connect to Appendix \ref{app:luminosity_scatter_halo_model}, let us introduce ``galaxy line noise'', $\sigma^2_{\text{gal }i,j}$, as the fractional covariance of lines $i$ and $j$ luminosities across all galaxies:
\beq
\sigma^2_{\text{gal }i,j}
\equiv
\frac{\langle L_iL_j\rangle_\text{gal} - \langle L_i\rangle_\text{gal} \langle L_j\rangle_\text{gal}}{\langle L_i\rangle_\text{gal} \langle L_j\rangle_\text{gal}}
\eeq
where the expectation value over all galaxies is defined as:
\beq
\langle X \rangle_\text{gal}
\equiv
\frac{1}{\bar{n}_\text{gal}}
\int dL_i dL_j\ \Phi(L_i, L_j) X,
\eeq
allowing the shot noise power spectrum to be written as:
\beq
P_{i,j}^\text{shot}
=
\frac{\bar{I}_1 \bar{I}_2}{\bar{n}^\text{gal eff}_{i,j}}
=
\frac{\bar{I}_1 \bar{I}_2}{\bar{n}_\text{gal}}
\left[
1 + \sigma^2_{\text{gal }i,j}
\right]
.
\eeq
The ``galaxy line noise'' is thus indeed an additional noise term, which enhances the galaxy shot noise.
Ref.~\cite{Chen20} recently used the a similar formalism to show the effect of line noise on the galaxy shot noise. Here, we have shown that the same formulation applies to the 1-halo term as well.

We quantify the importance of the halo and galaxy line noises below, and discuss the potential decorrelation they produce between the two intensity maps in line $i$ and $j$ in \S\ref{sec:decorrelation}.

\subsection{Special cases and ansatz for the CLF}

In order to gain some intuition for these noise sources, and for $\bar{n}_{ij}^{\rm h,eff}$ and $\bar{n}_{ij}^{\rm gal,eff}$, let us consider some special forms of the CLF and see how they affect the amplitudes of the 2-halo, 1-halo and shot-noise terms.

\paragraph{Equal galaxy luminosities}

If all galaxies in the Universe have the same luminosity, i.e.\ if the CLF is
\beq
\phi\left( L_1, ..., L_n | m \right)
\equiv
\prod_{i=i}^n
\delta^D\left( L_i - \bar{L}_i \right)
\;
N_\text{gal}(m),
\eeq
then we recover the standard HOD description of the galaxy clustering (with $N_\text{gal}$ including both the central and satellite galaxies, which we are not differentiating between).
In particular, as expected, the galaxy line noise $\sigma^2_{\text{gal }i,j}$ is zero in this case, and the effective galaxy number density is simply $\bar{n}^\text{gal eff}_{i,j} = \bar{n}_\text{gal}$, without any luminosity weighting.

However, intuitively, there is a non-zero halo line noise, since the halo luminosity $L_i(m) = \bar{L_i} N_\text{gal}(m)$ varies with mass. Formally:
\beq
\sigma^2_{\text{h } i,j}
=
\frac{\langle N_\text{gal}^2(m) \rangle_h}{\langle N_\text{gal}(m) \rangle_h^2}
-1.
\eeq
Further, if $N_\text{gal}(m) \propto m$ with a large proportionality constant such that the galaxy shot noise is negligible, we simply recover the matter power spectrum (in redshift space), up to a multiplicative factor.

\paragraph{Our ansatz for the CLF}

In what follows, we will assume the following form for the CLF of the lines of interest:
\beq
\phi\left( L_1, ..., L_n | m \right)
\equiv
\frac{\Phi\left( L_1, ..., L_n \right)}
{\int dL_1 ... dL_n\ \Phi\left( L_1, ..., L_n \right)}
N_\text{gal} \left( m \right)
.
\eeq
In effect, we have separated the dependences in halo mass and galaxy luminosity, a reasonable simplification given the scarcity of observational and simulation constraints on the CLF.
The CLF $\phi$, on the left-hand side, is expressed as a function of the usual (unconditional) luminosity function $\Phi$ on the right-hand side. The ratio can be interpreted as the joint probability distribution function for the luminosities $L_1, ..., L_n$ in the lines $1, ..., n$.
However, this model is sufficient to encode the potential correlation between different line luminosities.
As we describe in \S\ref{subsubsec:mehta15}, ref.\ \cite{Mehta15} provides a fitting function for the joint galaxy luminosities in H$\alpha$ and [O{\sc iii}].
This bivariate LF is particularly valuable, as it allows us to compute the correlation coefficient between these two lines, a crucial input to our ansatz for predicting the LIM cross-power spectra. 
The halo mass dependence is completely encoded in the mean number of galaxies $N_\text{gal} \left( m \right)$ per halo.
In particular, the mean halo luminosity is simply
$L_i(m, z) = \bar{L}_i N_\text{gal}(m,z)$.
This effectively assumes that the galaxy populations in all halos are identical, except for the number of galaxies per halo.
It also means that the halo line ratios $L_i(m, z) / L_j(m,z)$ and galaxy line ratios $\bar{L}^\text{gal}_i(z) / \bar{L}^\text{gal}_j(z)$ are independent of mass.
This also ignores the distinction between central and satellite galaxies inside halos, in order to reduce the number of free parameters in the model, given the sparsity of observational constraints. 
However, in Appendix.~\ref{app:centrals_satellites}, we show that the CLF formalism naturally describes this distinction, and we derive the general expressions there.
In practice, we model the number of galaxies per halo as:
\beq
N_\text{gal}(m,z)
\equiv
\frac{\bar{n}_\text{gal}(z)\ \text{SFR}^\gamma(m,z)}
{\int dm '\ n(m', z) \ \text{SFR}^\gamma(m',z)}
.
\eeq
This corresponds to the mass scaling $N_\text{gal}\left( m \right) \propto \text{SFR}^\gamma(m)$,
customary in the LIM literature\footnote{Taking $L\propto$SFR$^\gamma$ is a common assumption, also made in refs.~\cite{Cheng16,Lidz16,Gong17,Gong20} as just a few recent examples. However, recent simulation work \cite{Lupi20, Yang20} suggests that such a simple model may not be an accurate extrapolation at the lower halo masses.}.
For the halo mass function and linear bias, we implement the fitting function from ref.~\cite{Sheth99}.

To evaluate the star formation rate in a halo of given mass and redshift, we rely on the mean halo-to-stellar mass relation from ref.~\cite{Moster13} (their Eqs.~2, 11-14 and Table 1) and the mean relation between stellar mass and star formation rate \cite{Speagle14}. 
Refs.~\cite{Moster13, Speagle14} find a lognormal scatter of 0.15 dex (respectively 0.2 dex) in the ref.~\cite{Moster13} (respectively ref.~\cite{Speagle14}) relation.
However, as we show in Appendix~\ref{app:luminosity_scatter_halo_model}, the mean intensity, 1-halo and 2-halo terms only depend on the mean $\text{SFR}(m)$ relation, not its scatter.
While this is not true for the galaxy shot noise, this term is computed directly from the line LF, rather than the $\text{SFR}(m)$ relation.
In conclusion, we do not need to include the scatter in the $\text{SFR}(m)$ relation to accurately predict the mean intensity, 1-halo, 2-halo and galaxy shot noise terms in our formalism.
Note that this approach only yields the SFR of the central galaxy in the halo, whereas one should in principle include the sum of the SFRs of all galaxies within halos. 
However, as we show in \S\ref{sec:halo_masses}, the LIM observables are dominated by low-mass halos ($m\sim 10^{12} M_\odot/h$), in which the number and luminosities of satellite galaxies is expected to be small.
Indeed, ref.~\cite{Jimenez20} shows that the mean number of satellites is only a few percent of the mean number of centrals in these low mass halos (see their Fig.~3).
The mass integral is simply $\int dm\ n(m)\ N_\text{gal}(m) = \bar{n}_\text{gal}(z)$, as expected.
Here, the halo line noise is entirely due to the mass dependence of the halo star formation rate:
\beq
\sigma^2_{\text{h } i,j}
\sim
\frac{\langle \text{SFR}^{\gamma_i+\gamma_j}(m) \rangle}
{\langle \text{SFR}^{\gamma_i}(m) \rangle \langle \text{SFR}^{\gamma_j}(m) \rangle}
-1.
\eeq
In practice the mass average above is ill-defined in the absence of a mass cutoff, since the mean number density of halos of all masses diverges.  For this reason we have used a $\sim$ in the expression above, indicating that it is only a formal relation, but we believe it is still useful in highlighting the importance of the relative scatter in the halo SFR.

With our CLF ansatz, the galaxy line noise is simply given by the covariance of the individual galaxy luminosities:
\beq
P^\text{shot}_{i,j}
=
\frac{\bar{I}_i \bar{I}_j} {\bar{n}_\text{gal}}
\left[
1 + \sigma^2_{\text{gal } i,j} 
\right]
\quad\text{with}\quad
\sigma^2_{\text{gal } i,j} 
= 
\frac{\text{Cov}\left[ L_i, L_j \right]}{\bar{L}_i \bar{L}_j}.
\eeq

In summary, with this ansatz, the halo model expressions simplify to:
\beq
\boxed{
\left\{
\bal
&\bar{I}_j
=
\frac{1}{4\pi \nu_j^0}
\frac{c}{H(z)}
\ \int dL_j\ \Phi(L_j) L_j,\\
&P_{i,j}^\text{2-halo}(k, \mu, z)
=
\bar{I}_i \bar{I}_j
\left( b_i + F \mu^2 \right)
\left( b_j + F \mu^2 \right)\; 
P_\text{lin}\\
&\text{with}\quad b_j(k, \mu, z)
=
\frac{
\int
dm\ n(m)\;
\text{SFR}^{\gamma_j}(m)
\;
b(m) 
\;
u(k,m)
e^{-k^2 \mu^2\sigma_d^2(m) / 2}}
{\int
dm\ n(m)\;
\text{SFR}^{\gamma_j}(m)}\\
&\text{and}\quad F(k, \mu, z)
=
f\;
\int dm \; n(m)\;
\left( \frac{m}{\bar{\rho}} \right)
u(k,m) 
e^{-k^2 \mu^2 \sigma_d^2(m) /2},\\
&P_{i,j}^\text{1-halo}(k, \mu, z)
=
\left(\frac{c}{4\pi H(z)}\right)^2 
\frac{\bar{I}_i \bar{I}_j}{\nu_i^0 \nu_j^0}\
\frac{\int dm \; n(m)
\text{SFR}^{\gamma_i + \gamma_j}(m)
\left| u(k,m) \right|^2
e^{-k^2\mu^2\sigma^2}}
{\int dm \; n(m)
\text{SFR}^{\gamma_i + \gamma_j}(m)},\\
&P_{i,j}^\text{shot}(z)
=
\left(\frac{c}{4\pi H(z)}\right)^2 \frac{1}{\nu_i^0 \nu_j^0}
\int dL_i dL_j\ \Phi (L_i, L_j) L_i L_j.\\
\eal
\right.
}
\eeq
In what follows, we gather observational and simulation constraints on $N_\text{gal}(m)$ and $\phi(L_i, L_j)$, 
in order to evaluate the LIM observables.

\subsection{Relative size of the 1-halo and shot noise terms}
\label{sec:galaxy_halo_consistency}

We have seen that the 1-halo and galaxy shot noise terms scale are
$I^2/\bar{n}^\text{h eff}$ and $I^2/\bar{n}^\text{gal eff}$ 
on large scales, respectively.
Since halos can host several galaxies, one might expect the effective number density of galaxies to be higher than that of halos.
One would thus expect the 1-halo term to be larger than the galaxy shot noise.
In this section, we show that this is not necessarily the case, and clarify the regime in which this occurs.

From our Appendix~\ref{app:luminosity_scatter_halo_model}, we can write schematically the 1-halo and shot noise terms as:
\beq
P
=
\bar{n}_h \left[
\underbrace{\bar{N}_g^2 \langle L_g \rangle^2}
_{\text{1-halo}}
+
\underbrace{\bar{N}_g \langle L_g^2 \rangle}
_{\text{shot noise}}
\right],
\label{eq:halo_model_schematic}
\eeq
where $P$ is the LIM power spectrum, $\bar{n}_h$ the mean number density of halos, $\bar{N}_g$ the mean number of galaxies per halo, and $L_g$ an individual (random) galaxy luminosity.
So the comparison is:
\beq
\text{1-halo} \geq \text{shot noise}
\quad\Leftrightarrow\quad
\bar{N}_g \geq 
\frac{\langle L_g^2 \rangle}{\langle L_g \rangle^2}
=
1 + \sigma_g^2.
\eeq
The 1-halo term is thus larger than the shot noise in the regime where halos contain many galaxies, and the fluctuations in galaxy luminosity are small.

The case of galaxy clustering, instead of intensity mapping, is obtained by setting $L_g=1$ for each galaxy, such that galaxies are number-weighted, as opposed to luminosity-weighted.
The comparison between 1-halo term and shot noise is thus simply:
\beq
\text{1-halo} \geq \text{shot noise}
\quad\Leftrightarrow\quad
\bar{N}_g \geq 
1.
\eeq
Thus, for clustering, the 1-halo term is larger than the shot noise if halos host on average more than one galaxy.
In practice, none of these inequalities have to be satisfied. Thus the 1-halo term may be larger or smaller than the shot noise.
In terms of the CLF, the comparison can be expressed as:
\beq
\bal
\text{1-halo} &\geq \text{shot noise}
\quad\Leftrightarrow\quad\\
&\int dm\ n(m)\ \left( \int dL\ \Phi(L|m) L \right)^2
\geq
\int dm\ n(m)\ \int dL\ \Phi(L|m) L^2.\\
\eal
\eeq
In the particular case of our CLF ansatz, this becomes:
\beq
\text{1-halo} \geq \text{shot noise}
\quad\Leftrightarrow\quad
\frac{\int dm \ n(m)\ \text{SFR}^{2\gamma}(m)}
{\left( \int dm \ n(m)\ \text{SFR}^\gamma(m)  \right)^2}
\geq
\frac{\int dL_i\ \phi(L_i) L_i^2}
{\left( \int dL_i \; \phi(L_i) L_i  \right)^2}.
\eeq
Given the observed LF and the $\text{SFR}(m)$ relation at the corresponding redshift, this inequality may or may not be true.
We illustrate both cases in \S\ref{subsec:lf} (see Fig.~\ref{fig:p3d}).

\subsection{Halo masses probed}
\label{sec:halo_masses}

A question of particular interest is how LIM compares to galaxy catalogs, in terms of the halo masses probed.
For instance, galaxies selected from their emission lines are known to reside in lower mass halos than luminous red galaxies \cite{Jimenez20, Favole16, Gonzalez18}, and preferentially in sheets and filaments \cite{Gonzalez20} where their star formation has not been quenched.

Before continuing it is worth reopening the issue of satellite vs.\ central galaxies, a distinction we have neglected in the formalism in the main text since we are unable to meaningfully constrain the more complex model at this point.  For many of our inferences, neglecting this distinction is a reasonable approximation for LIM, since much of the emission-weighted signal comes from quite low mass halos (e.g.\ Fig.~\ref{fig:pk_mass_contributions}) where the satellite contribution is expected to be small.  Further, as we move to higher redshift (where LIM becomes increasingly competitive) the number of high mass halos continues to drop but the manner in which galaxies occupy halos evolves much less dramatically \cite{Moster13} so the relative contribution of satellites is reduced.  Neglecting the satellite-central distinction is however likely to impact our estimates for the FOG effect.  For this reason we regard the estimates presented here as less reliable than many of our other inferences, and we would require more observational constraints in order to be able to properly refine them.  Conversely LIM measurements could provide the much-needed constraints on a model including such a central-satellite split if they were able to tightly constrain the 2D power spectrum in the region where FoG effects dominate.

\begin{figure}[h!]
\centering
\includegraphics[width=\textwidth]{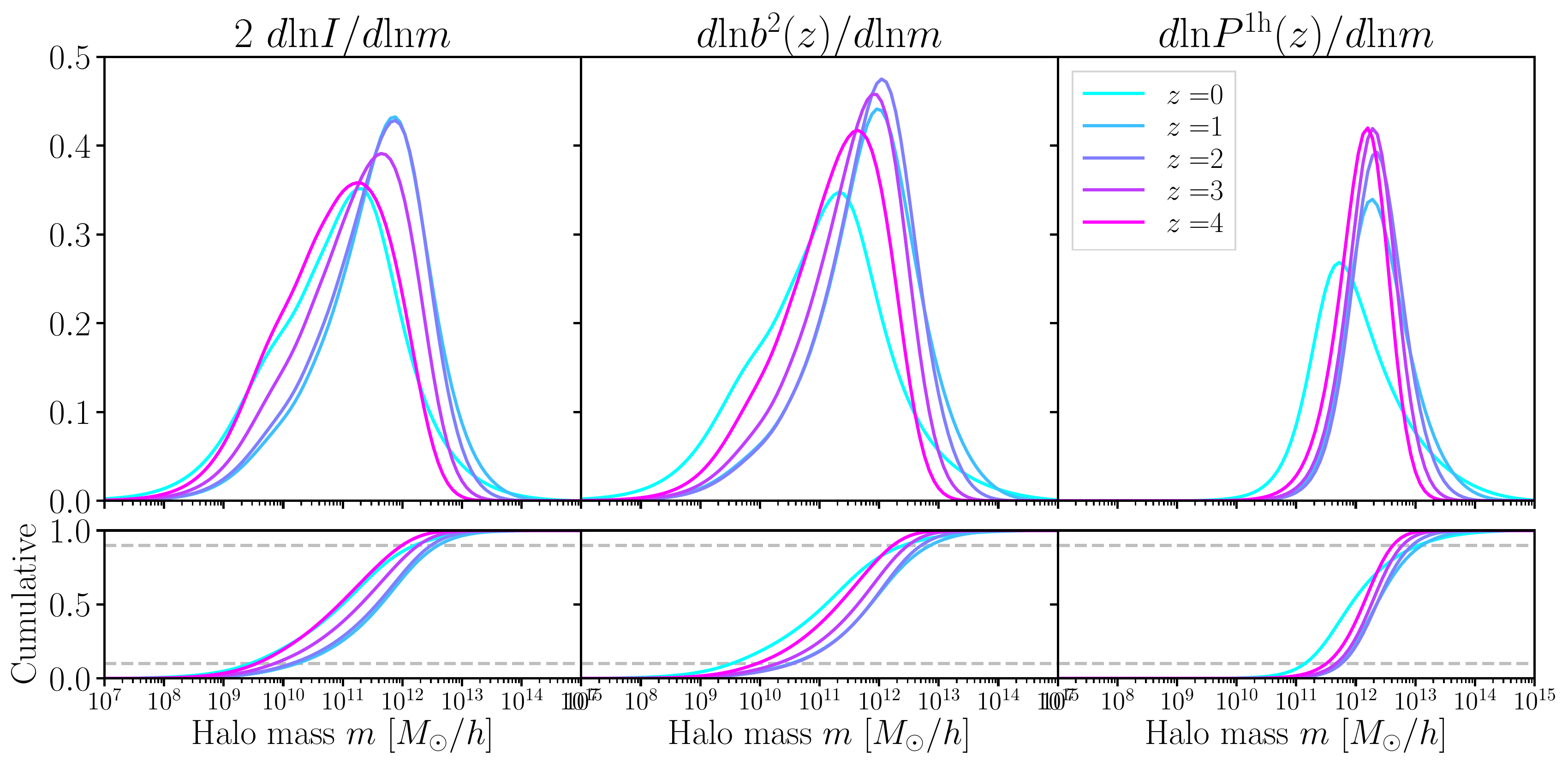}
\caption{
Contribution of each halo mass to the mean intensity (left), the 2-halo term (middle, via the squared effective halo bias) and 1-halo terms (right) at various redshifts.
Here, we assumed that the halo luminosity scales as $\text{SFR}$, i.e. $\gamma=1$, as is the case for most UV and optical lines.
The mean intensity and 2-halo term probe similarly low halo masses, whereas the 1-halo term is sensitive to more massive halos.
The grey dashed lines in the cumulative plots (bottom panels) indicate the 10th and 90th percentiles.
}
\label{fig:dlnalldlnm}
\end{figure}

As shown in Fig.~\ref{fig:dlnalldlnm}, the mean intensity and the 2-halo term (through the squared bias) receive contribution from a broad range of halo masses. 
Masses within $10^{10}-10^{13}\,h^{-1}M_\odot$ contribute more than 80\% of these quantities.
On the other hand, the 1-halo term receives contributions from a narrower and heavier range of mass:
80\% of it comes from $10^{12}-10^{13}\,h^{-1}M_\odot$ halos.
This is expected, since the 1-halo term weighs halos by their squared luminosity, thus up-weighting higher mass halos.
In Fig.~\ref{fig:dlnalldlnm}, the smooth cutoff at low masses is due to the $\text{SFR}(m)$ relation, whereas the sharper cutoff at high masses is due to the exponential cutoff in the halo mass function.
Galaxies selected by luminosity, stellar mass or line emission tend to only be found in halos with mass $m\geq 10^{11}\,h^{-1}M_\odot$ \cite{Jimenez20}, especially at higher $z$.
Through the mean intensity and 2-halo term, LIM may thus enable probing slightly lower mass halos, down to $10^{10}\,h^{-1}M_\odot$.

\subsection{Galaxy luminosities probed}
\label{sec:galaxy_luminosities_probed}

\begin{figure}[h!]
\centering
\includegraphics[width=\textwidth]{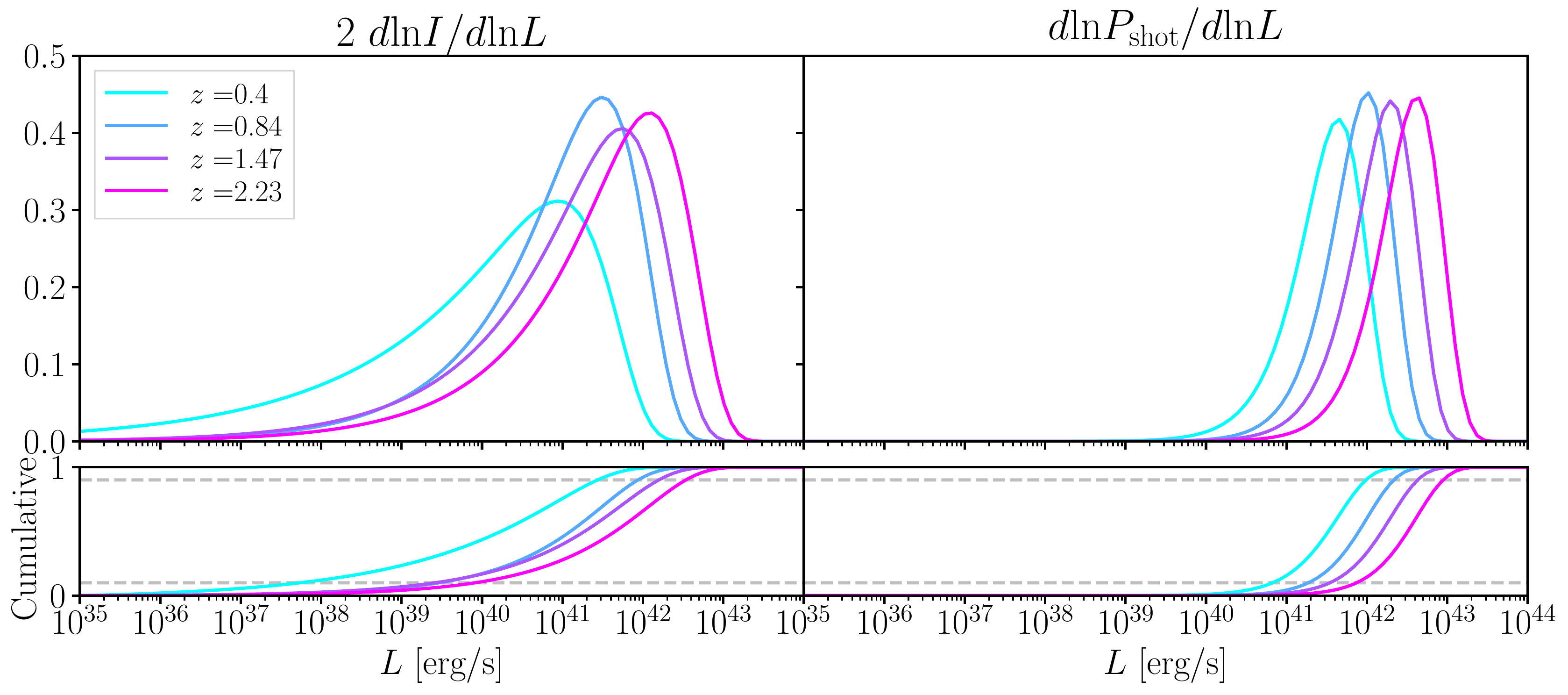}
\caption{
Contribution from each galaxy luminosity to the H$\alpha$ mean intensity (left) and shot noise power spectrum (right), using LFs from ref.~\cite{Sobral13}.
The mean intensity is sensitive to much lower luminosities than the shot noise.
The grey dashed lines in the cumulative plots (bottom panels) indicate the 10th and 90th percentiles.
}
\label{fig:didl_dshotdl}
\end{figure}

A key astrophysics goal of LIM is to learn about the distribution of luminosities of the galaxies which make up the intensity map.
Our halo model predicts the relative weight of each galaxy luminosity bin to the LIM observables, as shown in Fig.~\ref{fig:didl_dshotdl}.

\section{LIM auto-spectra: single-line luminosity functions}
\label{sec:observational}

The key quantity in our halo model is the galaxy CLF.
The single-line CLF entirely determines the LIM auto-spectrum and the multi-line CLF predicts the LIM cross-spectrum.
These CLFs can be measured in galaxy surveys or from simulations. 
In this section, we present several such constraints on the CLF.

\subsection{Observational \& simulation constraints}
\label{subsec:lf}

As we show, modeling the LIM power spectrum requires knowledge not only of the mean, but also of the scatter in the galaxy luminosities.
Line luminosities can vary by large factors from galaxy to galaxy, driven in large part by the scatter in star formation rate \cite{Osterbrock06, Levesque10, Gutkin16, Byler17}.
This scatter in line luminosity enhances the shot noise power spectrum.

Below we model the H$\alpha$, [O{\sc iii}], Ly-$\alpha$, [C{\sc ii}] and CO line luminosities using empirical fits to observed LFs.
Ref.~\cite{Cochrane17} presents Schechter model fits to the H$\alpha$ LF (table 2).
These fits are based on the High-redshift(Z) Emission Line Survey (HiZELS) survey \cite{Sobral13}, including 
additional and near-infrared narrow-band data from the Canada France Hawaii Telescope \cite{Sobral15}.
They cover three redshift bins over $z=0.8-2.2$, and are corrected for the contamination from [N{\sc ii}] lines at $654.8\,$nm and $658.4\,$nm.
These LFs also include a correction for dust attenuation, assuming a uniform 1 mag of dust extinction.
We undo this correction in order to obtain the observed, dust attenuated, luminosity distribution.
Ref.~\cite{Colbert13} presents Schechter model fits to the observed H$\alpha$ and [O{\sc iii}] LFs from the WFC3 Infrared Spectroscopic Parallel Survey (WISP) program.
The H$\alpha$ LF fits describe two redshift bins over $z=0.3-1.5$. 
The LF of the higher redshift bin is consistent with ref.~\cite{Sobral13}, while that of the lower redshift bin is found to be slightly lower.
The [O{\sc iii}] LFs is fit in four redshift bins spanning $z=0.7-2.3$.
Using data from the WISP program, ref.~\cite{Mehta15} further estimates the joint bivariate LF in H$\alpha$ and [O{\sc iii}], providing a fitting function in one redshift bin covering $z=0.8-1.2$. 
This is the observed, dust attenuate LF relevant to our LIM forecast.
These measured H$\alpha$ LFs have been used as a reference for a number of numerical and semi-analytical models, e.g.\ refs.~\cite{Merson18, Merson19, Zhai19, Zhai20, Schreiber17, Gong17}.
Using HiZELS, ref.~\cite{Khostovan15} measure the H$\beta$+[O{\sc iii}] and [O{\sc ii}] LFs in five redshift bins over $z=0.8-5$. 
They compare their Schechter fits to existing measurements, including refs.\ \cite{Sobral13, Colbert13}.  
At $z=1.47$, their H$\beta$+[O{\sc iii}] LF is consistent with the [O{\sc iii}] LF from \cite{Colbert13} at the bright end, where the [O{\sc iii}]-emitting galaxies dominate over the H$\beta$-emitting population, but they diverge at the faint end, where the H$\beta$ galaxies dominate.
At the same redshift, their [O{\sc iii}] LF is consistent with \cite{Sobral13}.
We implement the Schechter fits to the [C{\sc ii}] and CO LFs predicted by the semi-analytical model of \cite{Popping16} over $z=0-6$.
The [C{\sc ii}] LF was also measured in the local Universe by \cite{Hemmati17}.
Finally, we implement the Ly-$\alpha$ LF from \cite{Cassata11} over $z=2-5$.

\begin{figure}[h!]
\centering
\includegraphics[width=0.32\textwidth]{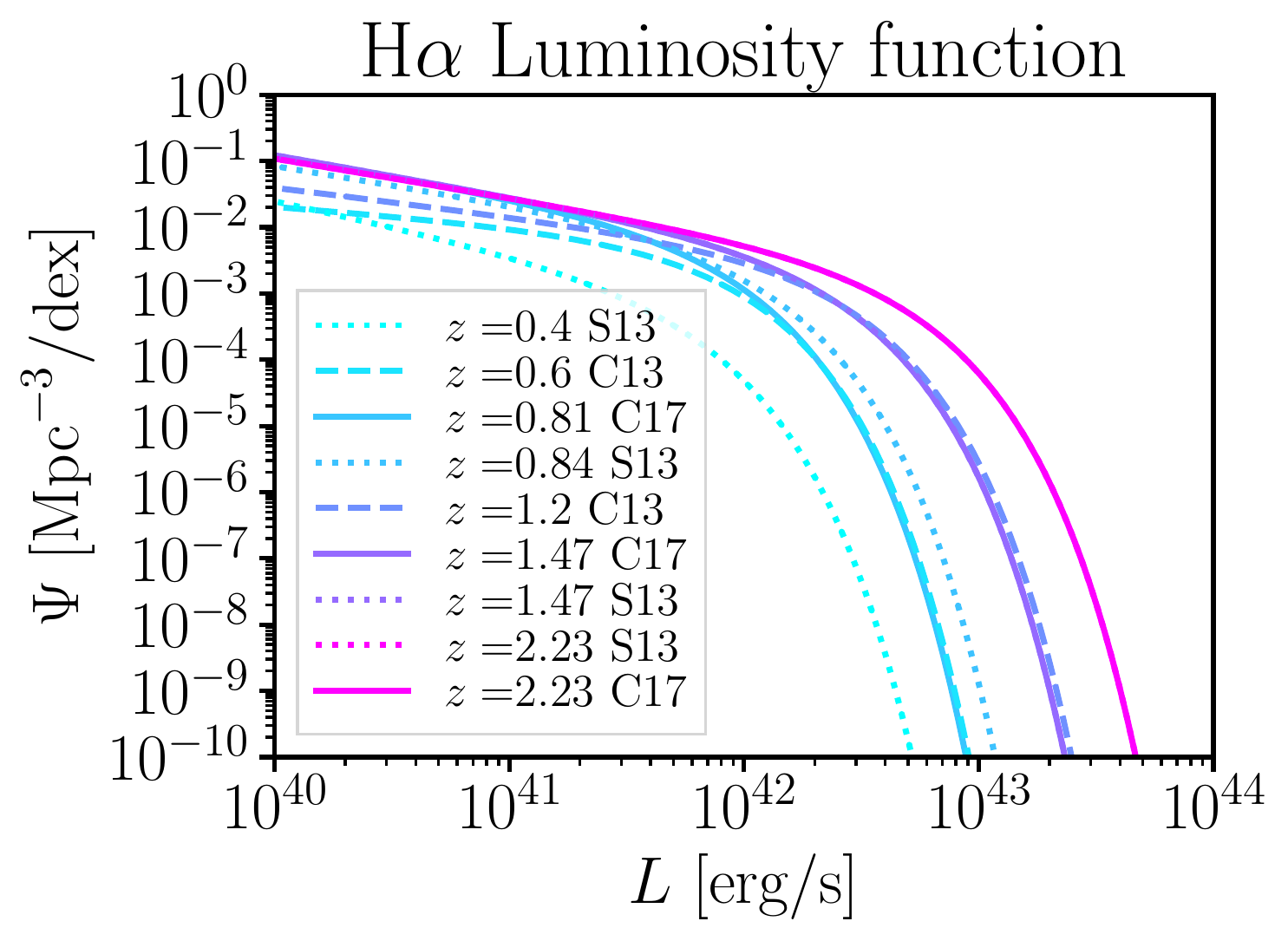}
\includegraphics[width=0.32\textwidth]{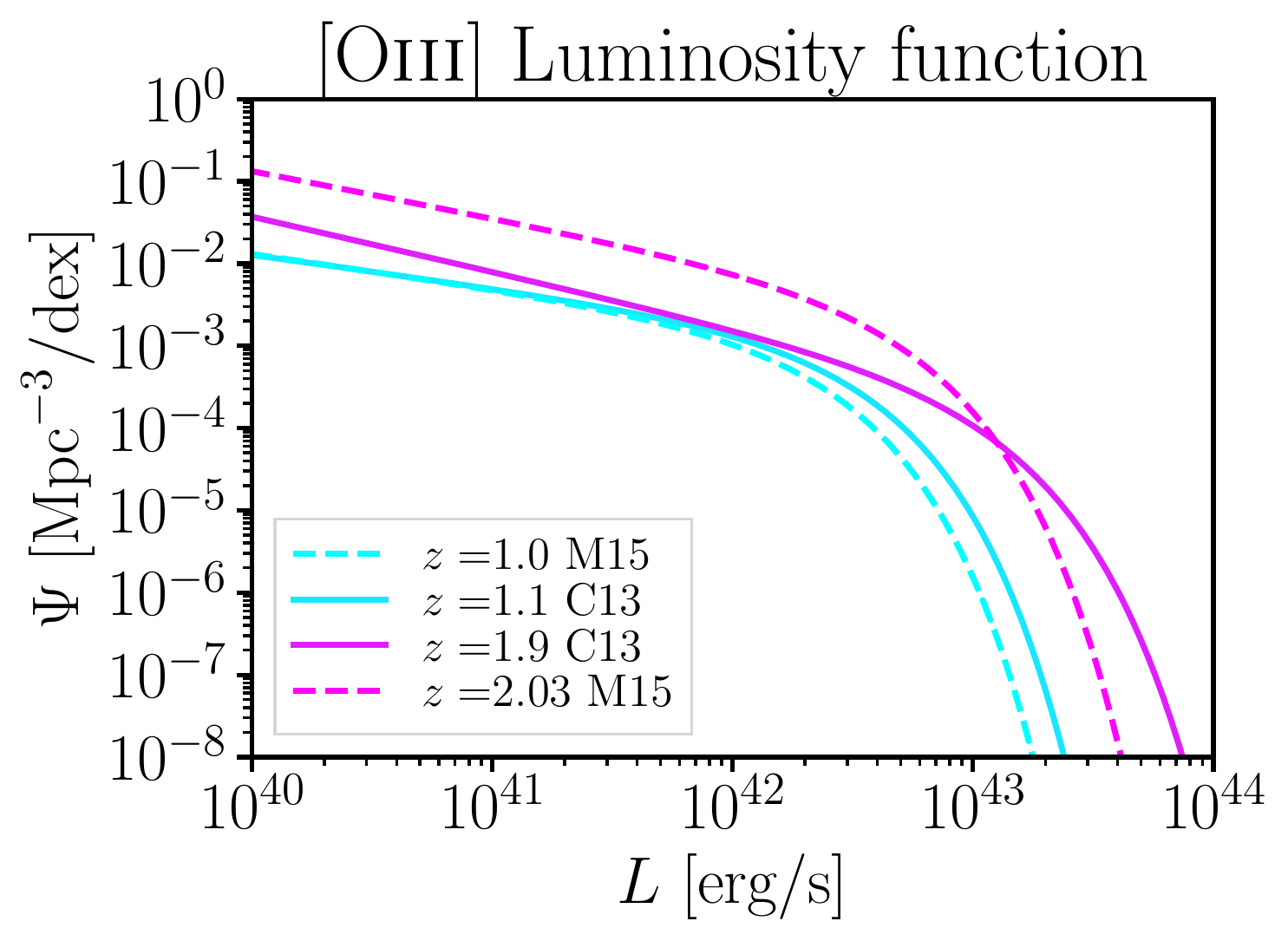}
\includegraphics[width=0.32\textwidth]{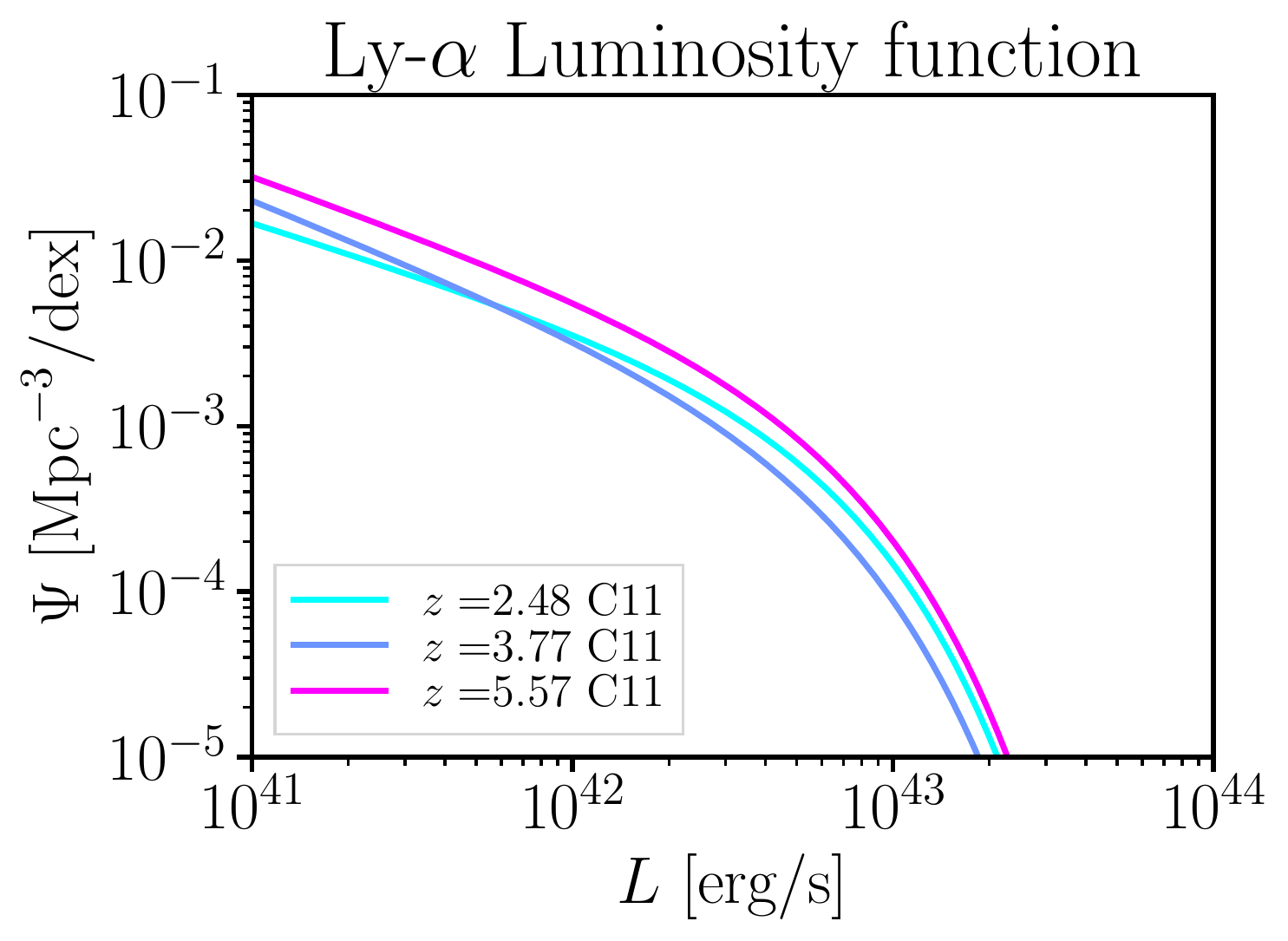}
\includegraphics[width=0.45\textwidth]{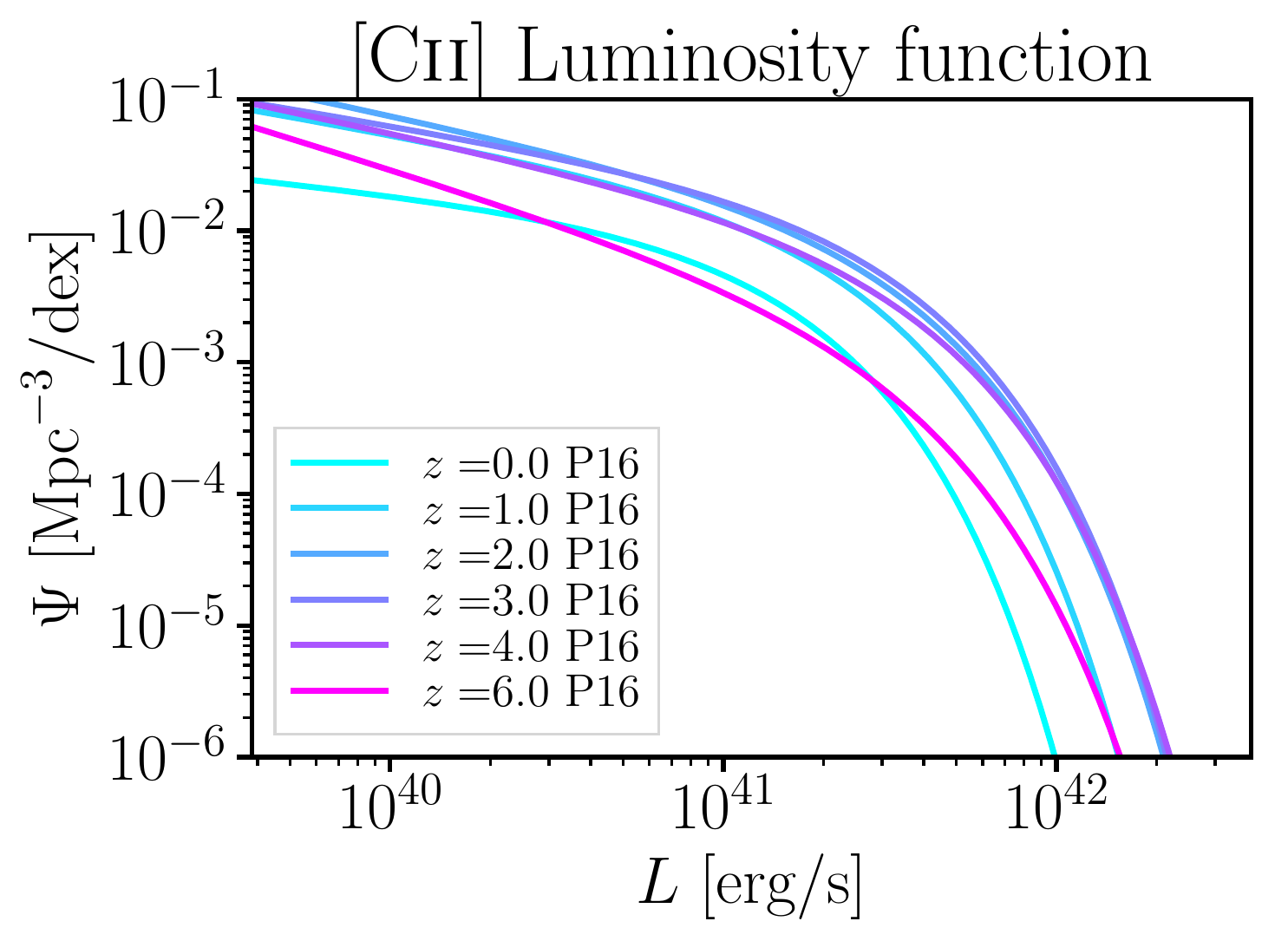}
\includegraphics[width=0.45\textwidth]{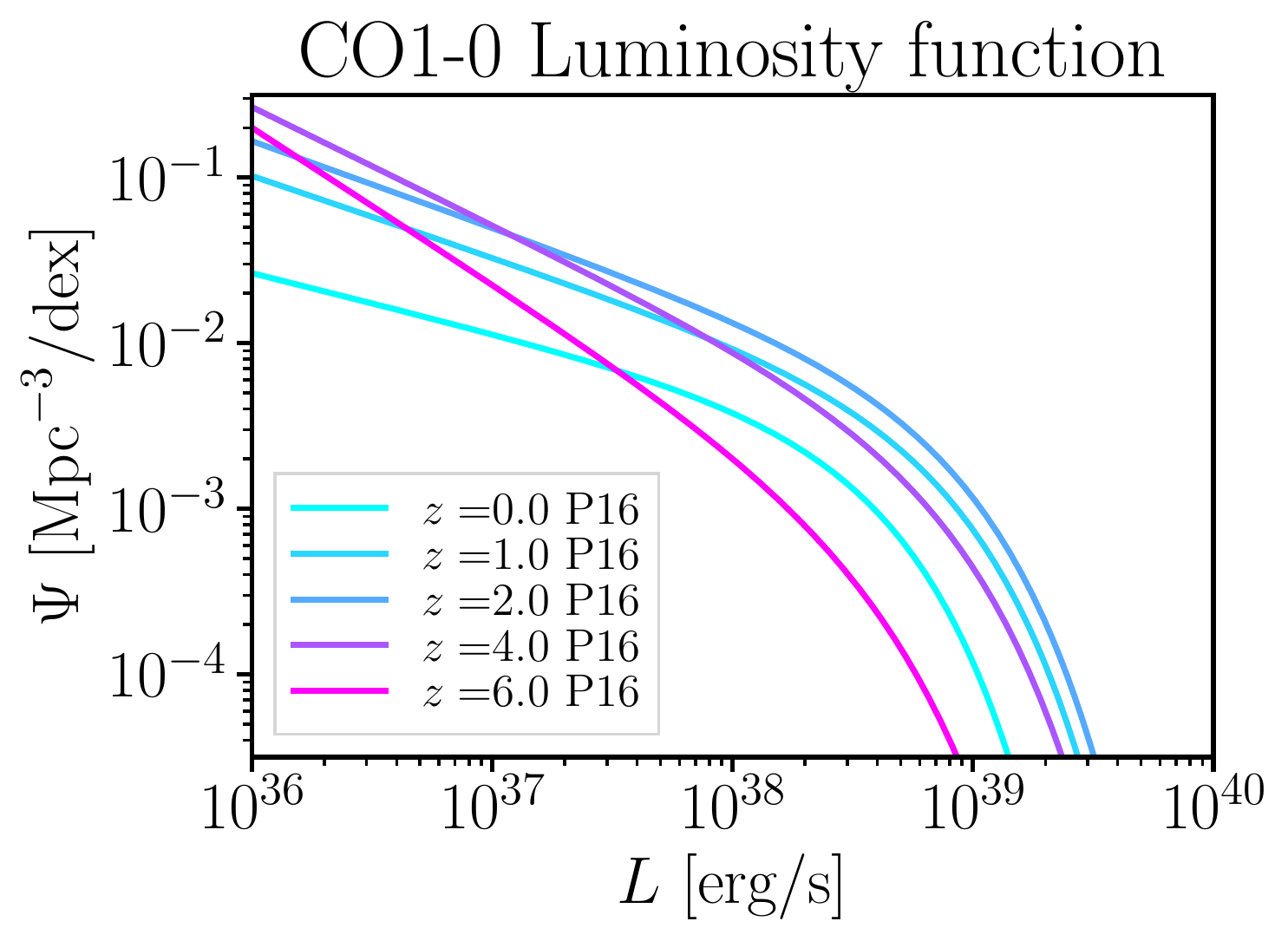}
\caption{
Galaxy luminosity functions $\Psi(L) \equiv dN/d\log_{10}L\,dV$ for H$\alpha$, [O{\sc iii}], Ly-$\alpha$, [C{\sc ii}] and CO1-0 from the various references (S13 \cite{Sobral13}, C13 \cite{Colbert13}, M15 \cite{Mehta15}, C17 \cite{Cochrane17}, P16 \cite{Popping16} and C11 \cite{Cassata11}).
The H$\alpha$ LFs from C17 and S13 are identical at $z=1.47$ and $2.23$.
}
\label{fig:lf}
\end{figure}

The various LFs are shown in Fig.~\ref{fig:lf}.  
There is some disagreement in the H$\alpha$ luminosity function at $z\simeq 0.8$ but the individual estimates agree very well at higher redshift.  
The [O{\sc iii}] luminosity function shows more diversity, especially at the bright end and at higher redshift ($z\simeq 1.9$).

\begin{figure}[h!]
\centering
\includegraphics[width=0.32\textwidth]{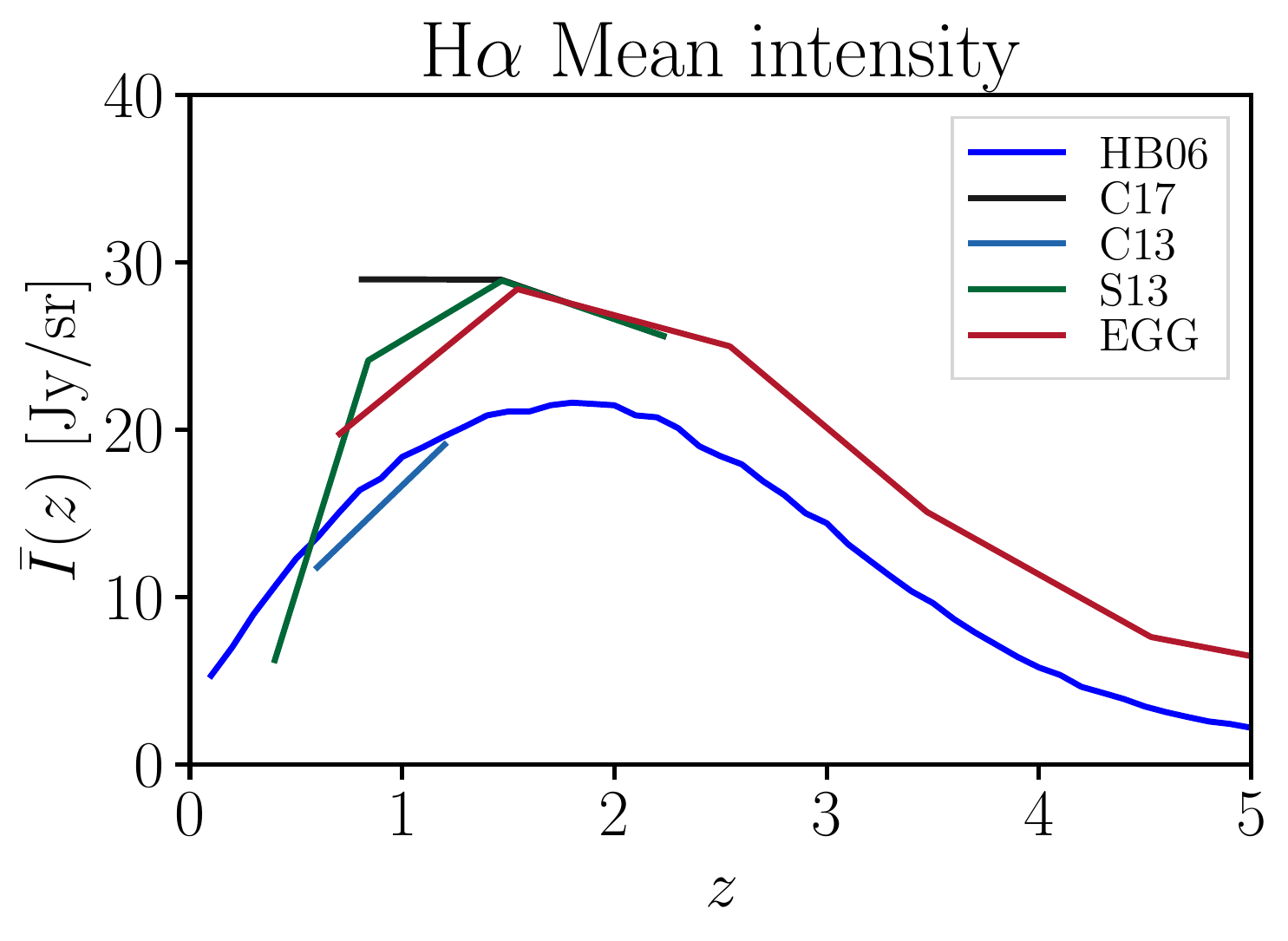}
\includegraphics[width=0.32\textwidth]{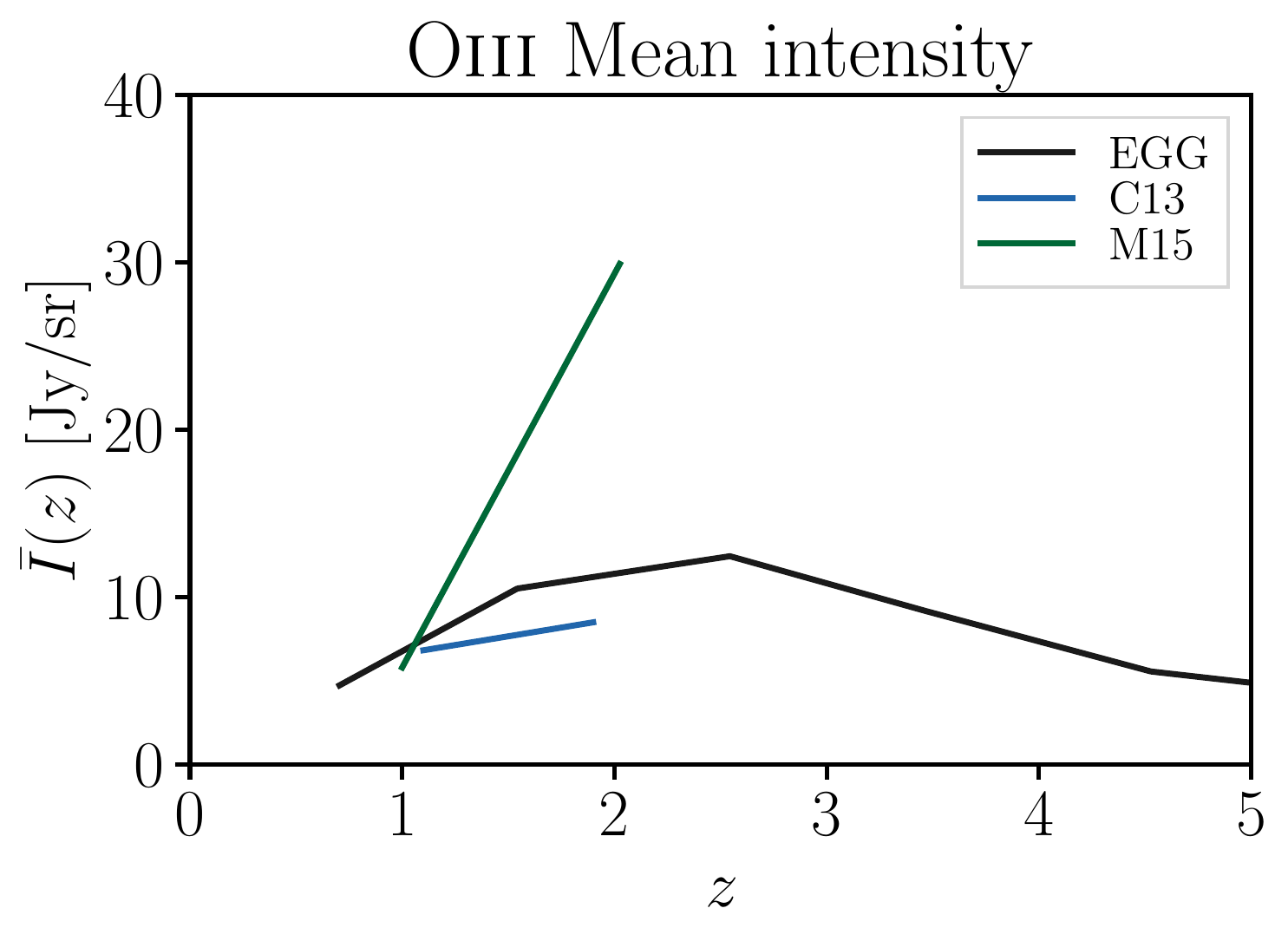}
\includegraphics[width=0.32\textwidth]{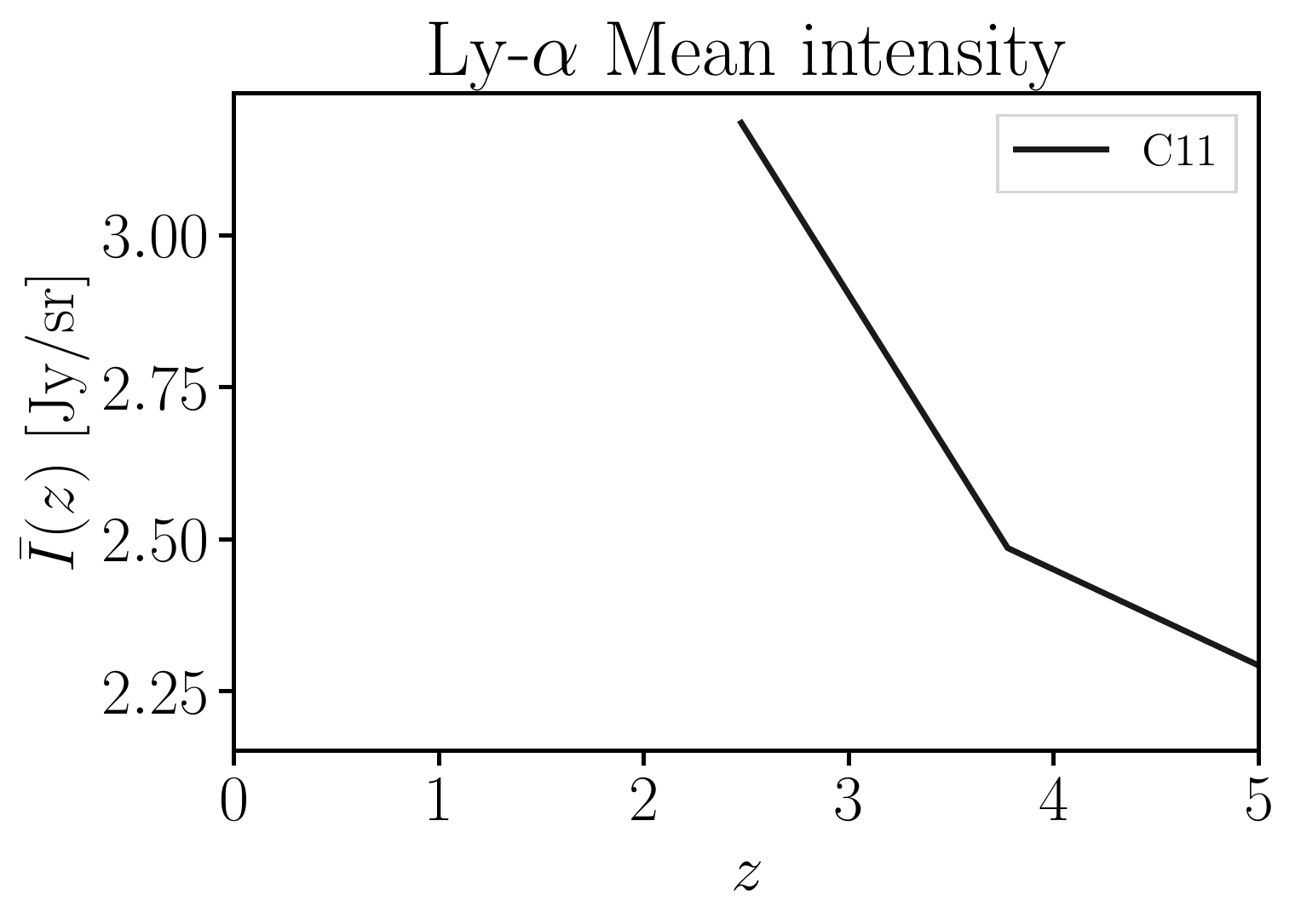}
\includegraphics[width=0.45\textwidth]{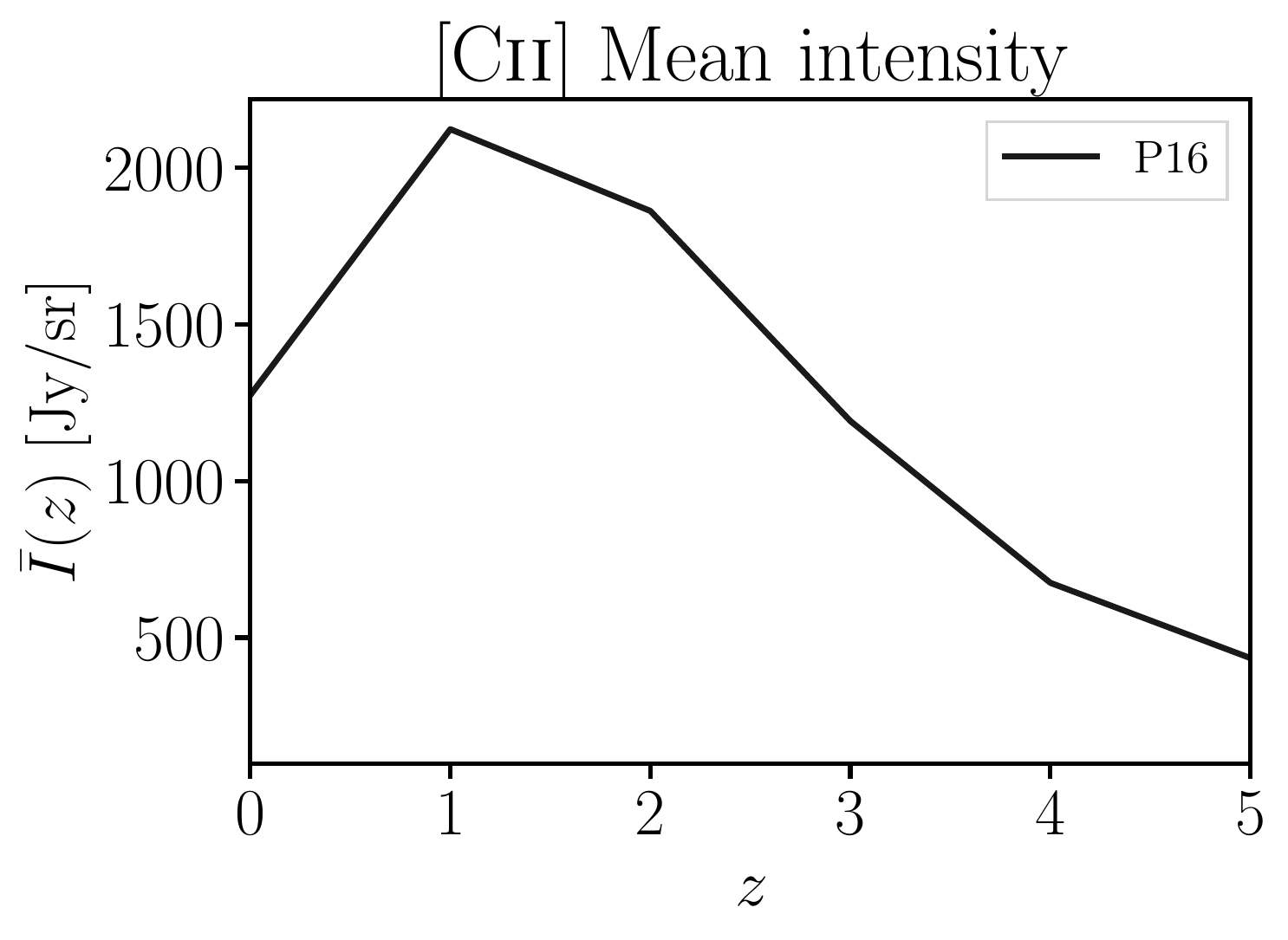}
\includegraphics[width=0.45\textwidth]{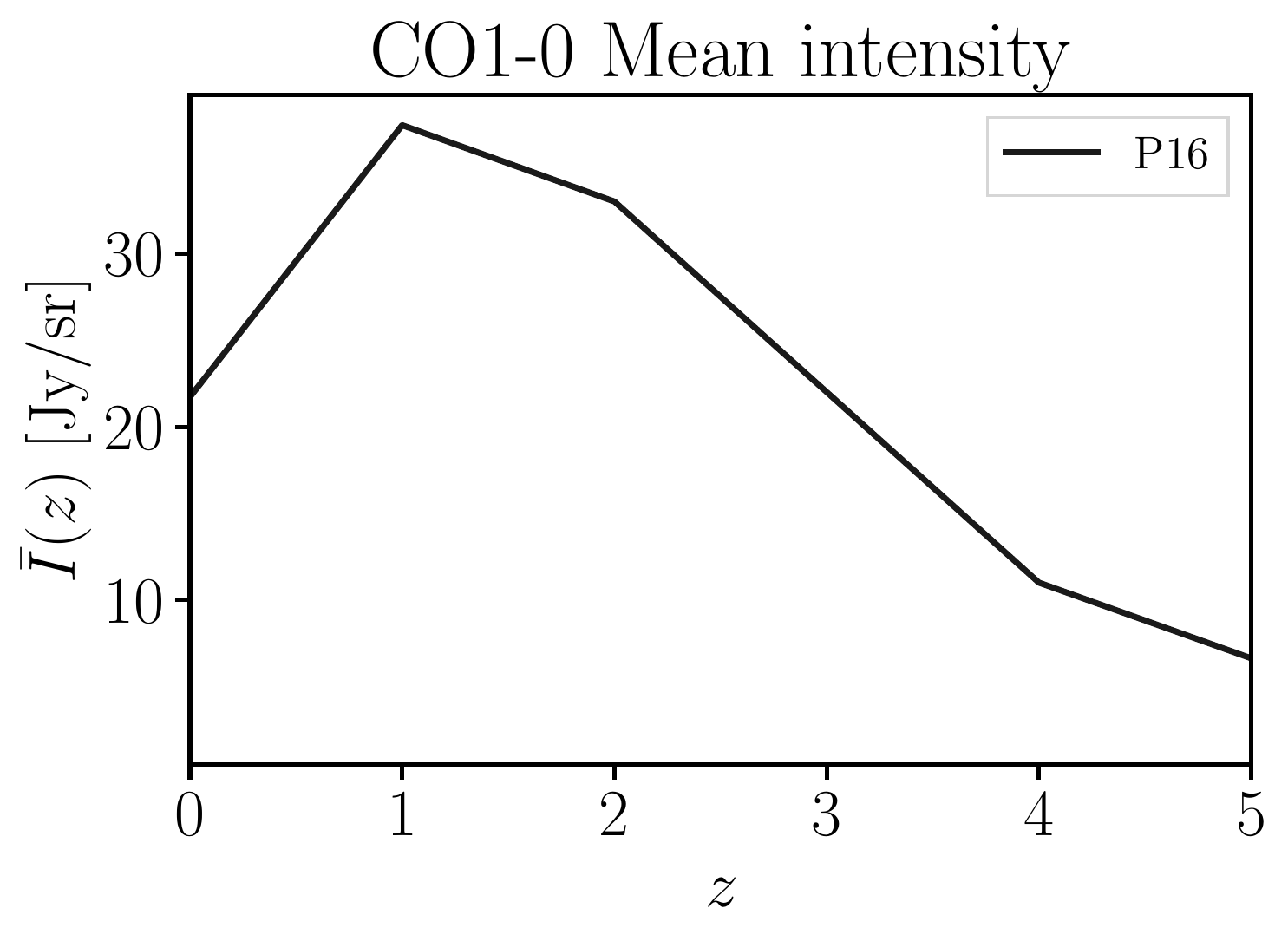}
\caption{
Comparison of the mean intensity as a function of redshift, for H$\alpha$, [O{\sc iii}], Ly-$\alpha$, [C{\sc ii}] and CO1-0, as predicted from the various observed luminosity functions (see Fig.~\ref{fig:lf}) as well as the EGG simulator \cite{Schreiber17}.
This first moment of the galaxy luminosity function also determines the overall normalization of the LIM power spectrum.
The relative sizes of the mean intensities in the various lines can be understood by comparing their LFs (Fig.~\ref{fig:lf}) and rest-frame frequencies.
}
\label{fig:mean_intensity}
\end{figure}

\begin{figure}[h!]
\centering
\includegraphics[width=0.32\textwidth]{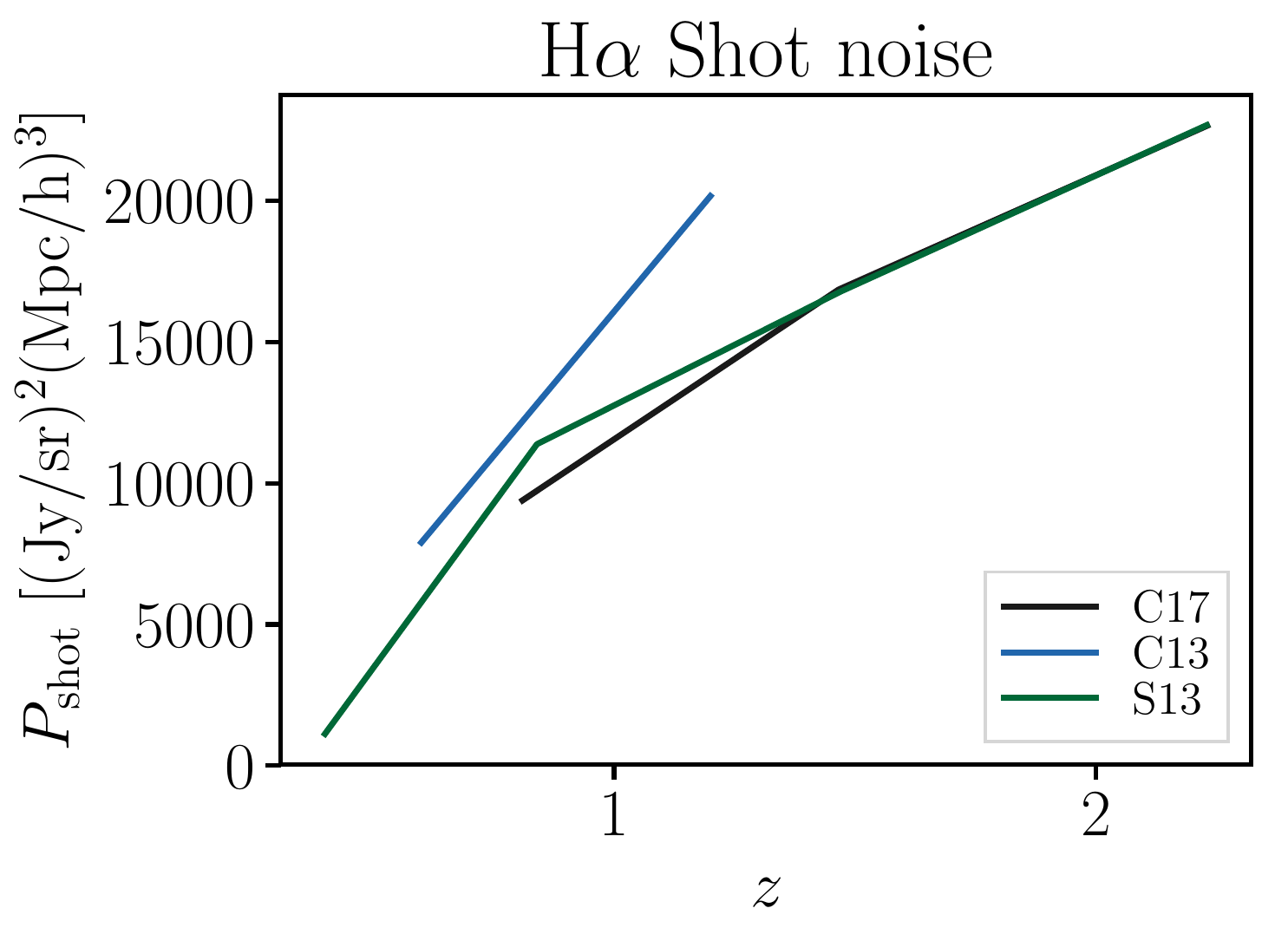}
\includegraphics[width=0.32\textwidth]{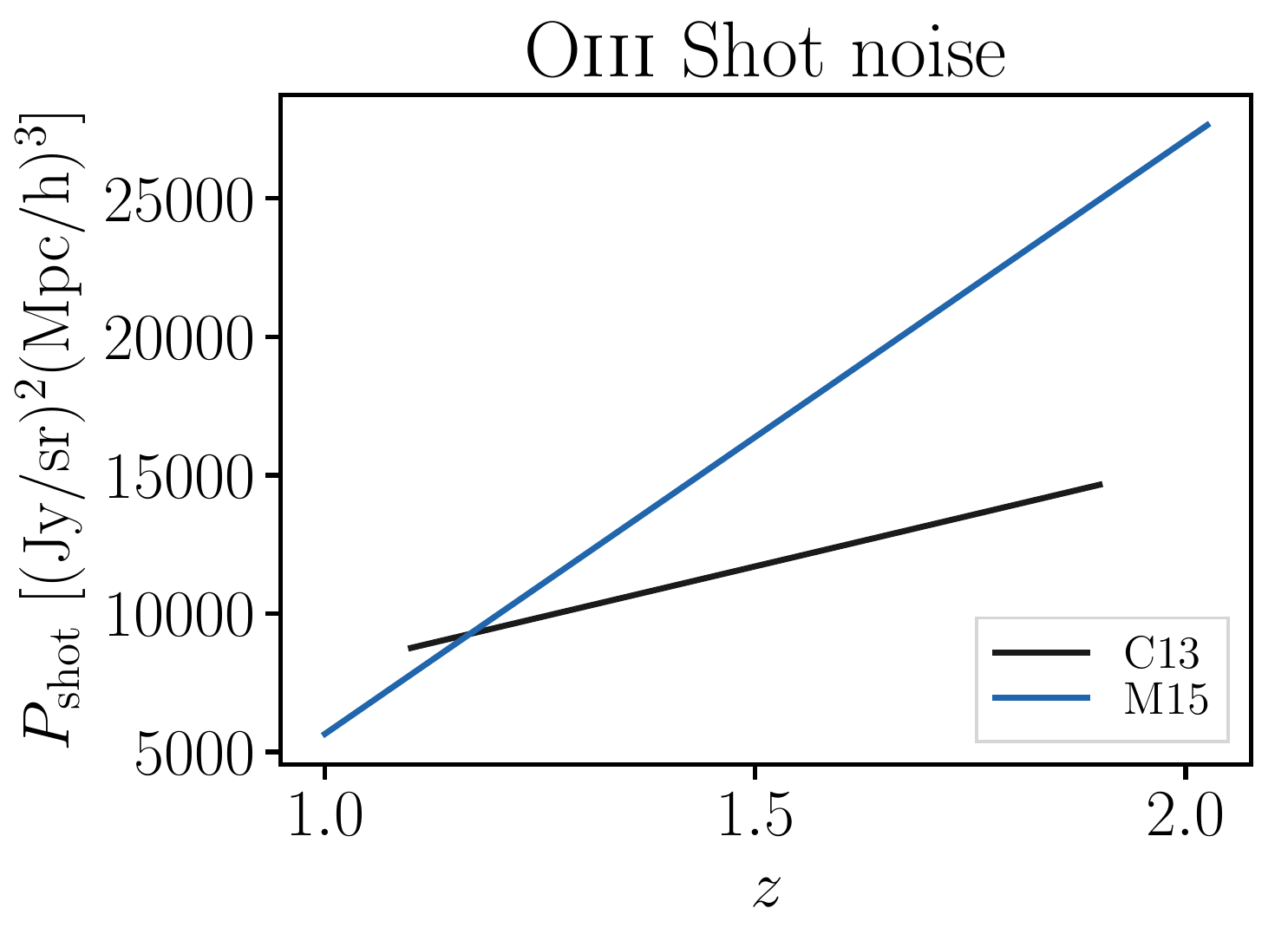}
\includegraphics[width=0.32\textwidth]{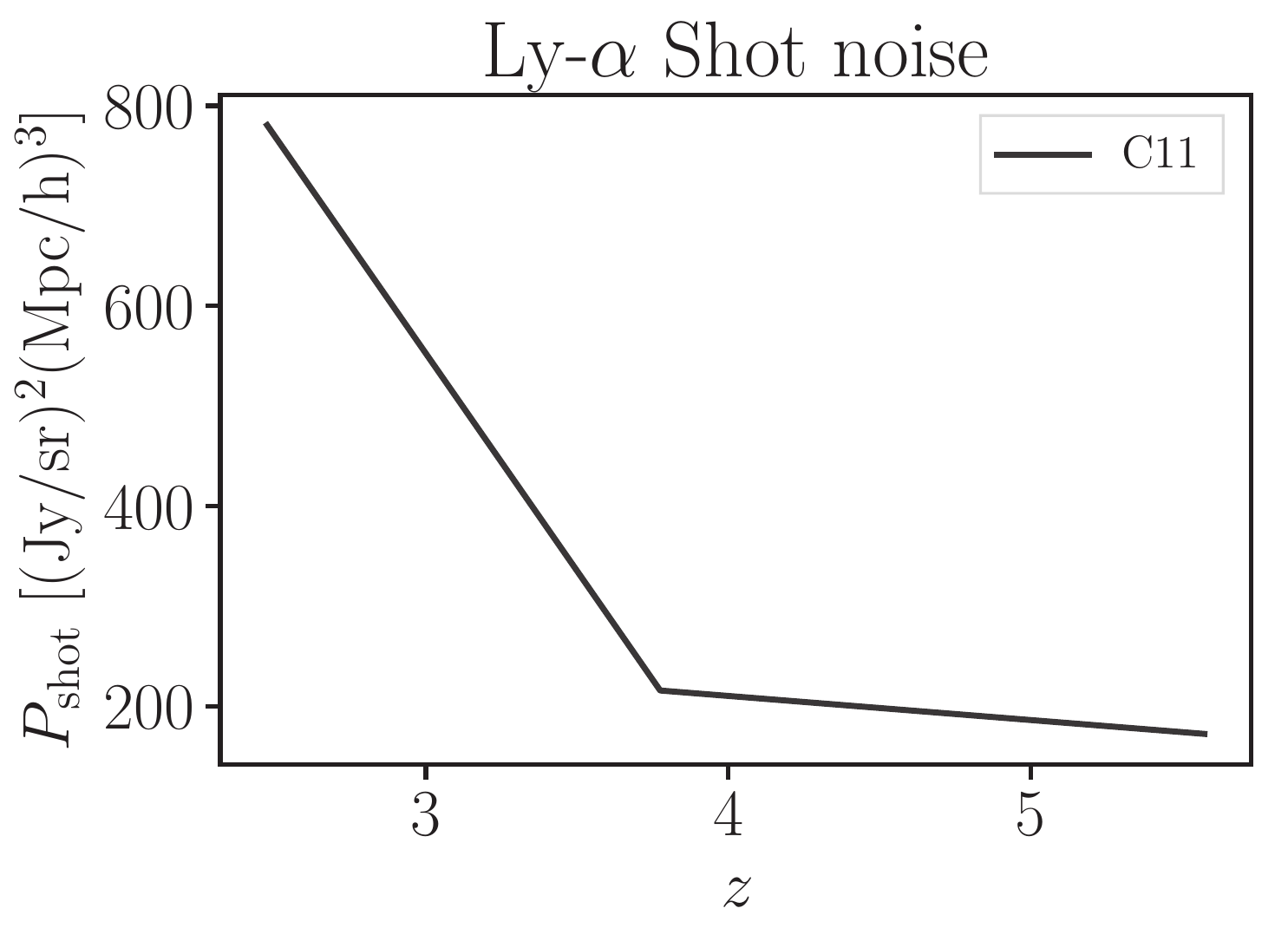}
\includegraphics[width=0.45\textwidth]{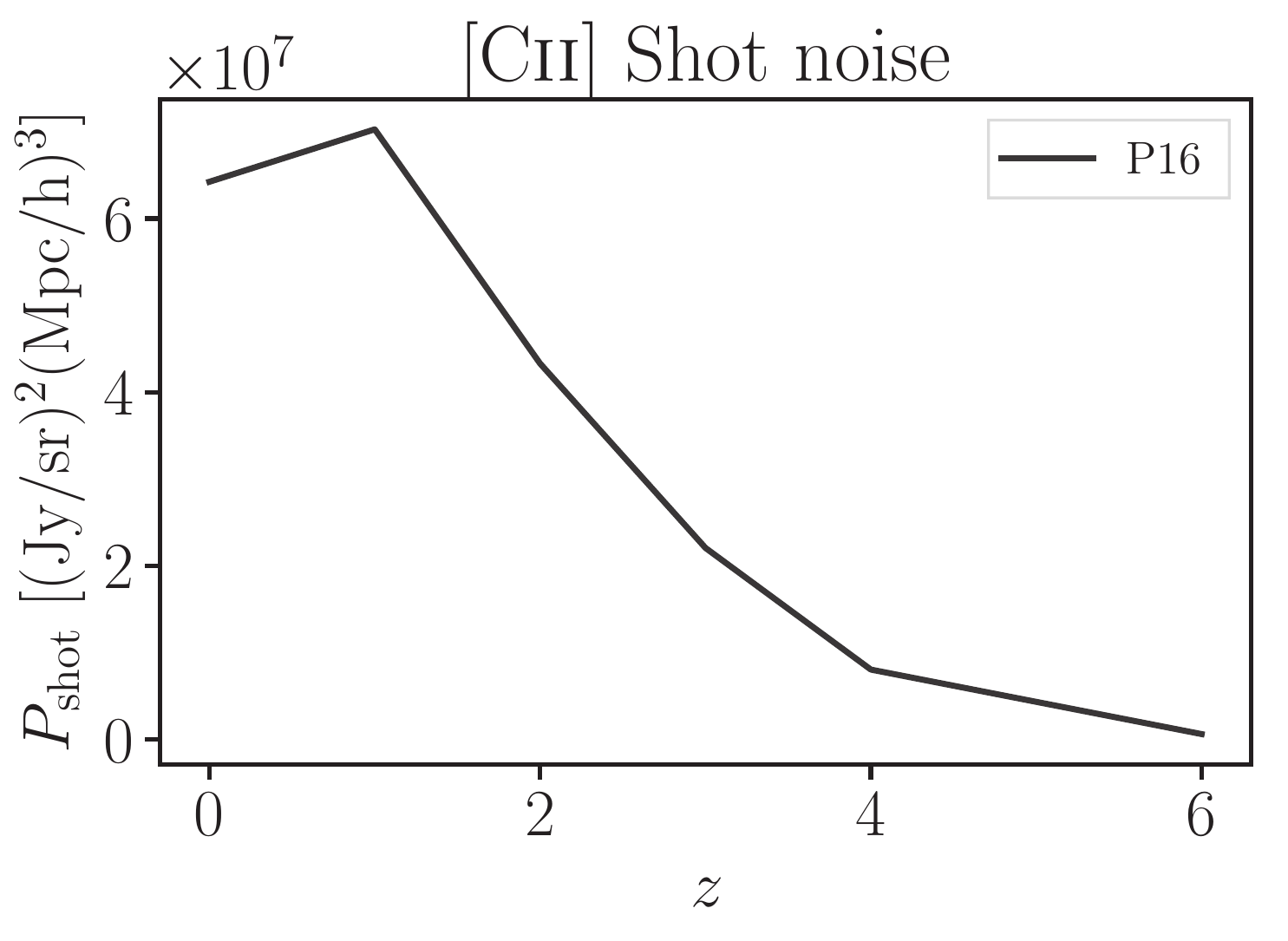}
\includegraphics[width=0.45\textwidth]{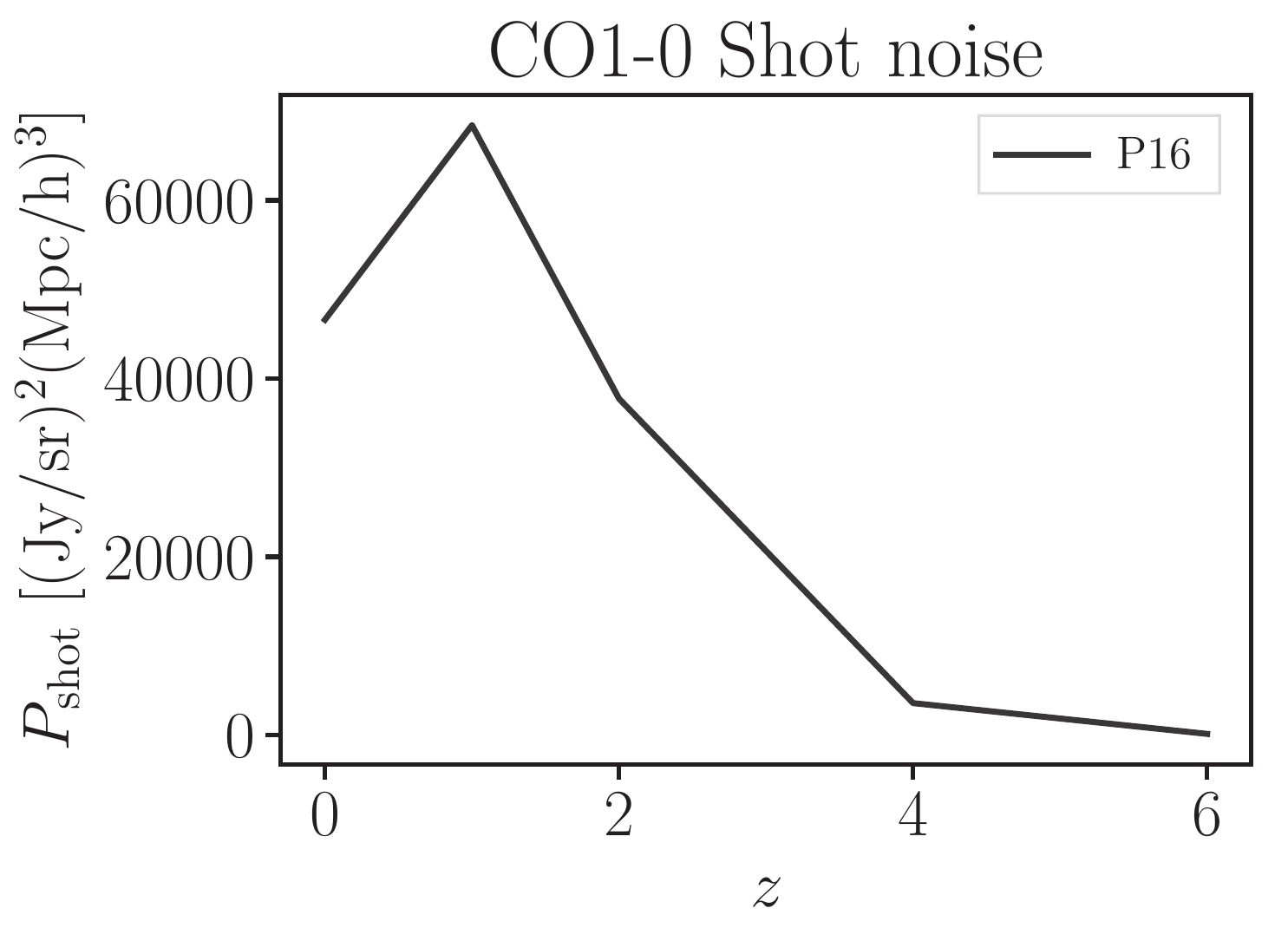}
\caption{
Shot noise power spectrum as a function of redshift, for H$\alpha$, [O{\sc iii}], Ly-$\alpha$, [C{\sc ii}] and CO1-0, as predicted from the various observed luminosity functions (see Fig.~\ref{fig:lf}).
The shot-noise power is determined by the second moment of the galaxy luminosity function.
}
\label{fig:pshot_z}
\end{figure}

We compute the first two moments of these LFs, needed to compute the LIM mean intensity and power spectrum.
The first moment gives the mean intensity (Fig.~\ref{fig:mean_intensity}), which also normalizes the LIM power spectrum.  As might have been expected from the disagreement in the luminosity functions themselves, the amplitude and evolution of the mean intensity is not well constrained at present, especially for [O{\sc iii}].
The relative sizes of the mean intensities in different lines can be understood by comparing their LFs and rest-frame frequencies, since
$\bar{I} \propto \left(\int dL\ \Phi(L) L \right)/ \nu^0$ (Eq.~\eqref{eq:mean_intensity_lf}).
For instance, the mean intensity in CO and optical lines are similar, even though galaxy CO luminosities are typically $\sim 10,000$ fainter.
This is compensated by the CO rest-frame frequency at $\sim 100$ GHz, about $\sim 10,000$ smaller than for optical lines.
On the other hand, the mean intensity is [C{\sc ii}] is much larger than in optical lines.
This can be understood since the galaxy luminosities in [C{\sc ii}] are only somewhat fainter than in the optical lines, and the $\sim$ THz frequency of the [C{\sc ii}] line is $\sim 1,000$ smaller than for the PHz optical lines.
The second moment determines the shot noise term (Fig.~\ref{fig:pshot_z}), through the effective number density of galaxies (Fig.~\ref{fig:ngaleff_z} and Eq.~\ref{eqn:ngaleff}).  The level of agreement is qualitatively similar to that displayed by the first moment, even though the luminosity weightings are of course different in the two.
The wide discrepancies between the predicted mean intensities (Fig.\ \ref{fig:mean_intensity}) and shot noises (Fig.\ \ref{fig:pshot_z}) from the various measured luminosity functions is striking.

In Fig.~\ref{fig:mean_intensity}, \ref{fig:pshot_z}, \ref{fig:p3d}, we only show the redshift ranges where luminosity functions from actual observations are available. We do not show the predictions from simulations such as EGG.

\subsection{LIM auto-spectra: theoretical uncertainty}
\label{subsec:auto_spectra}

\begin{figure}[h!]
\centering
\includegraphics[width=0.29\textwidth]{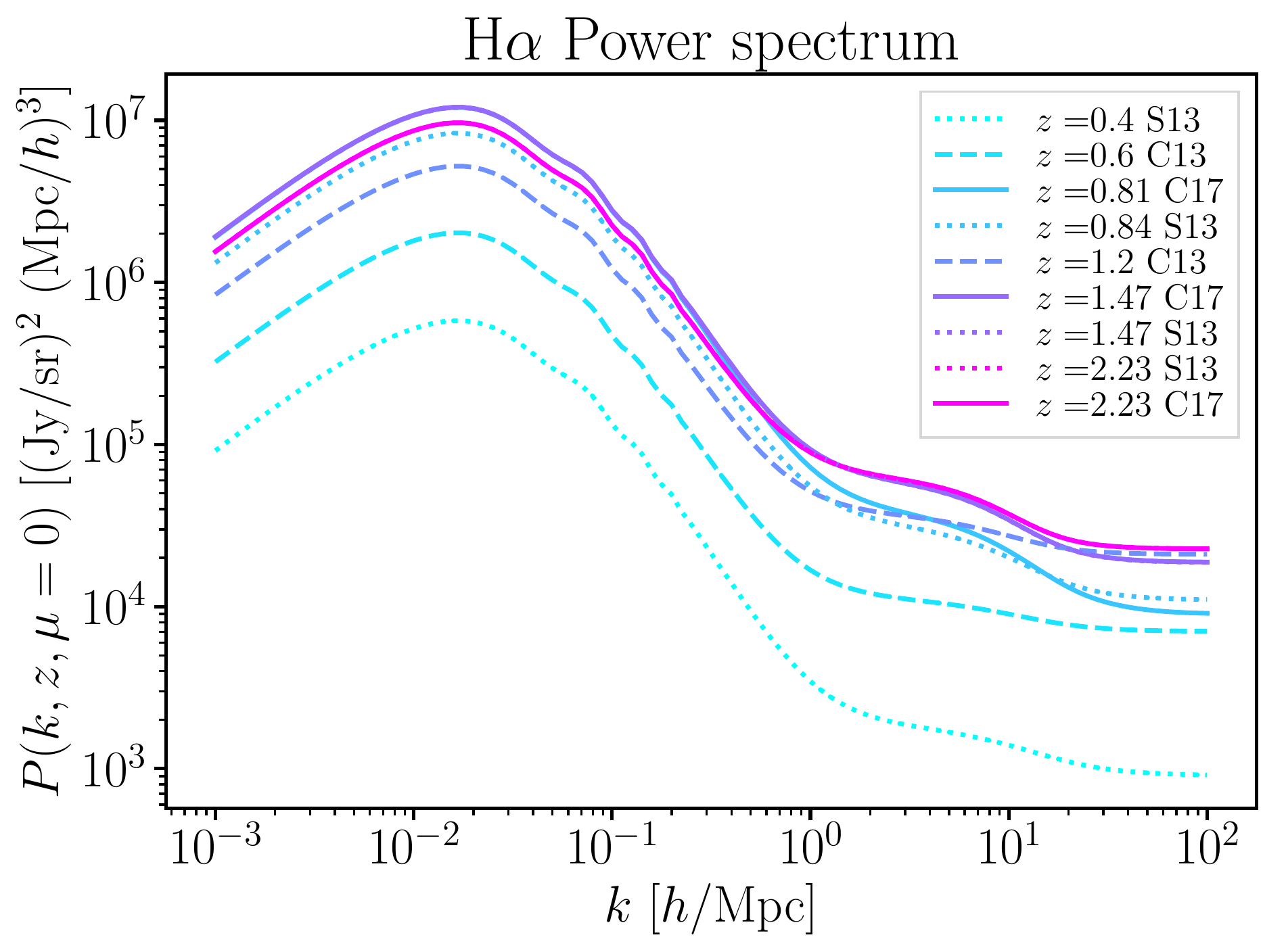}
\includegraphics[width=0.31\textwidth]{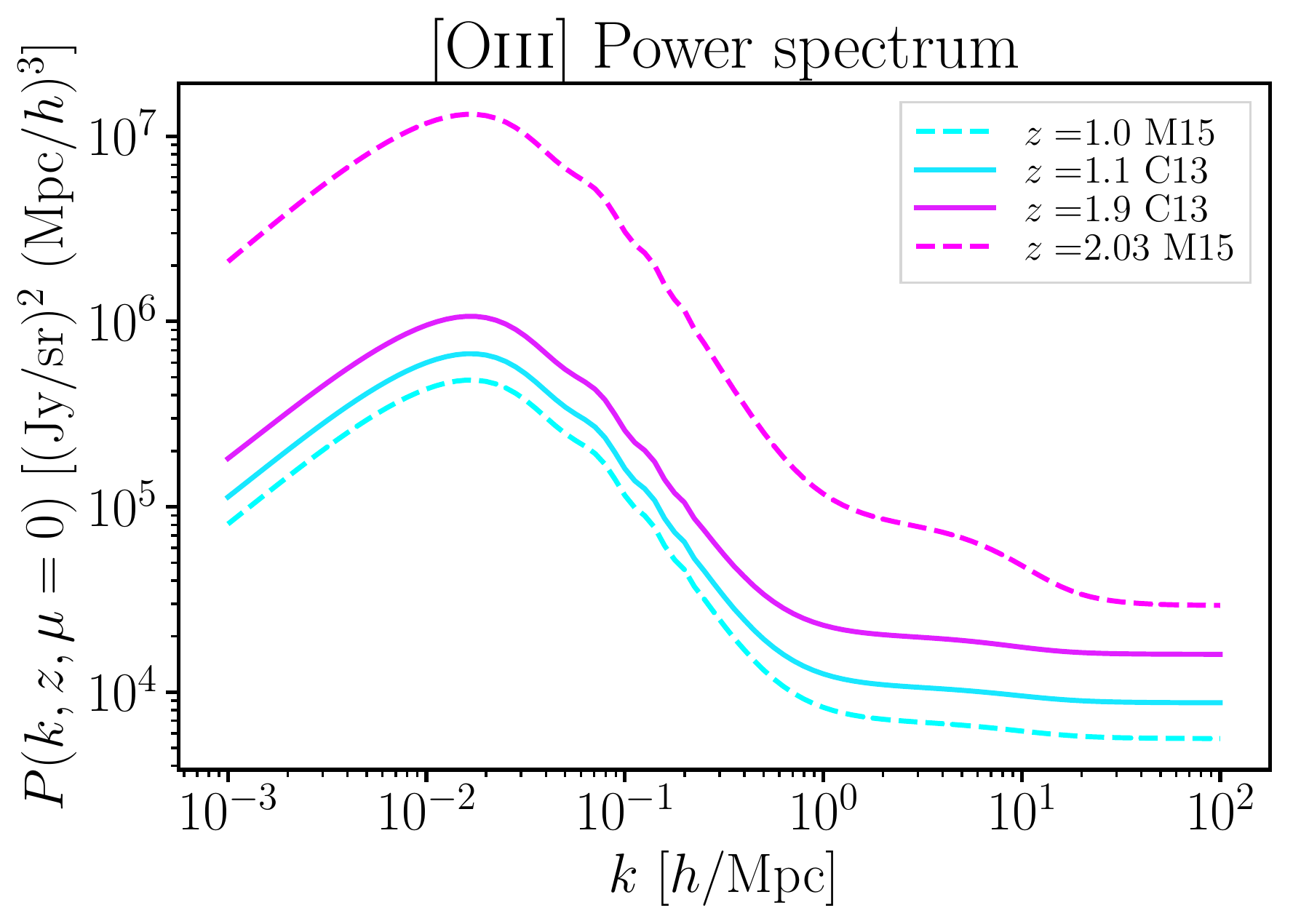}
\includegraphics[width=0.33\textwidth]{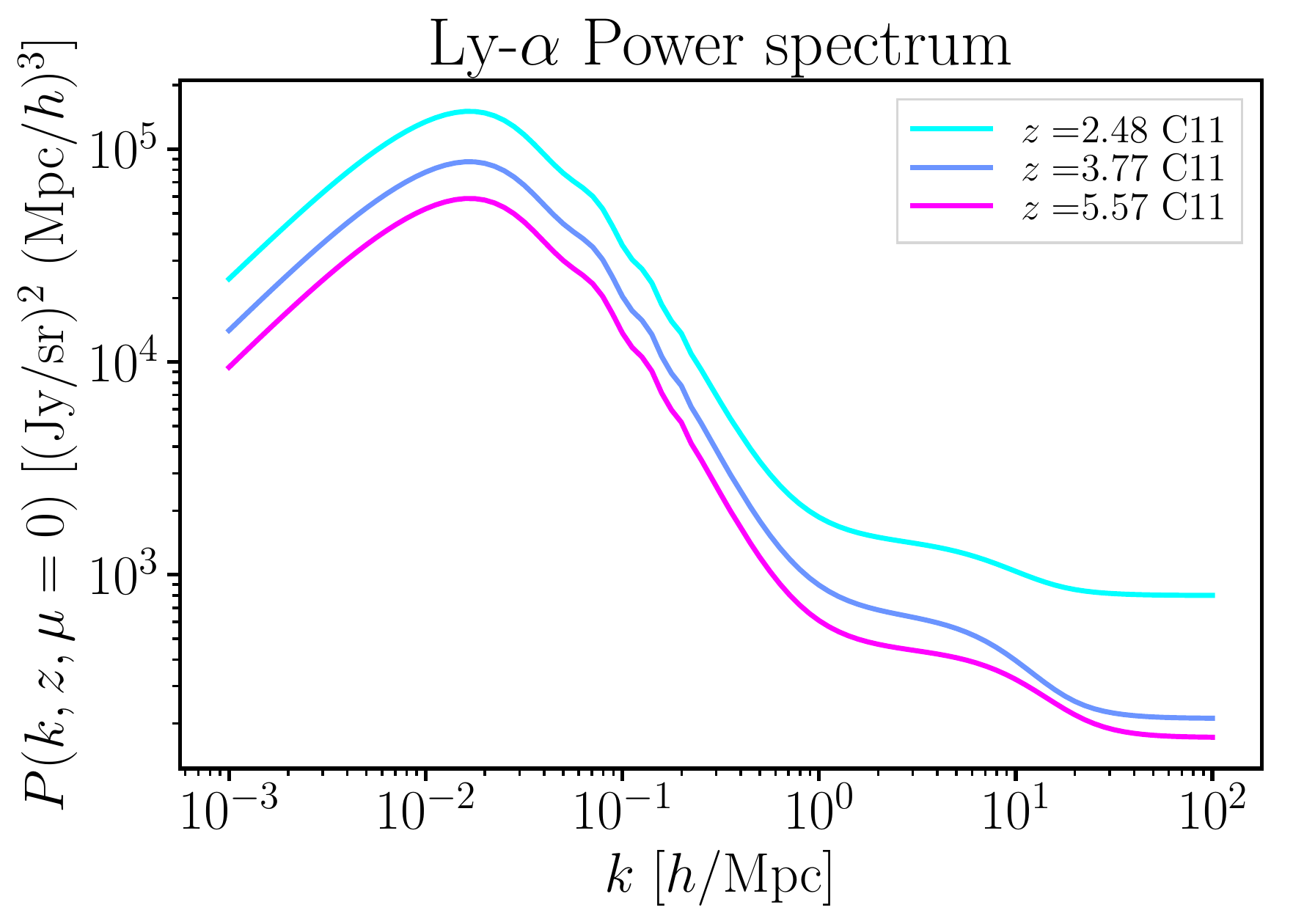}
\includegraphics[width=0.45\textwidth]{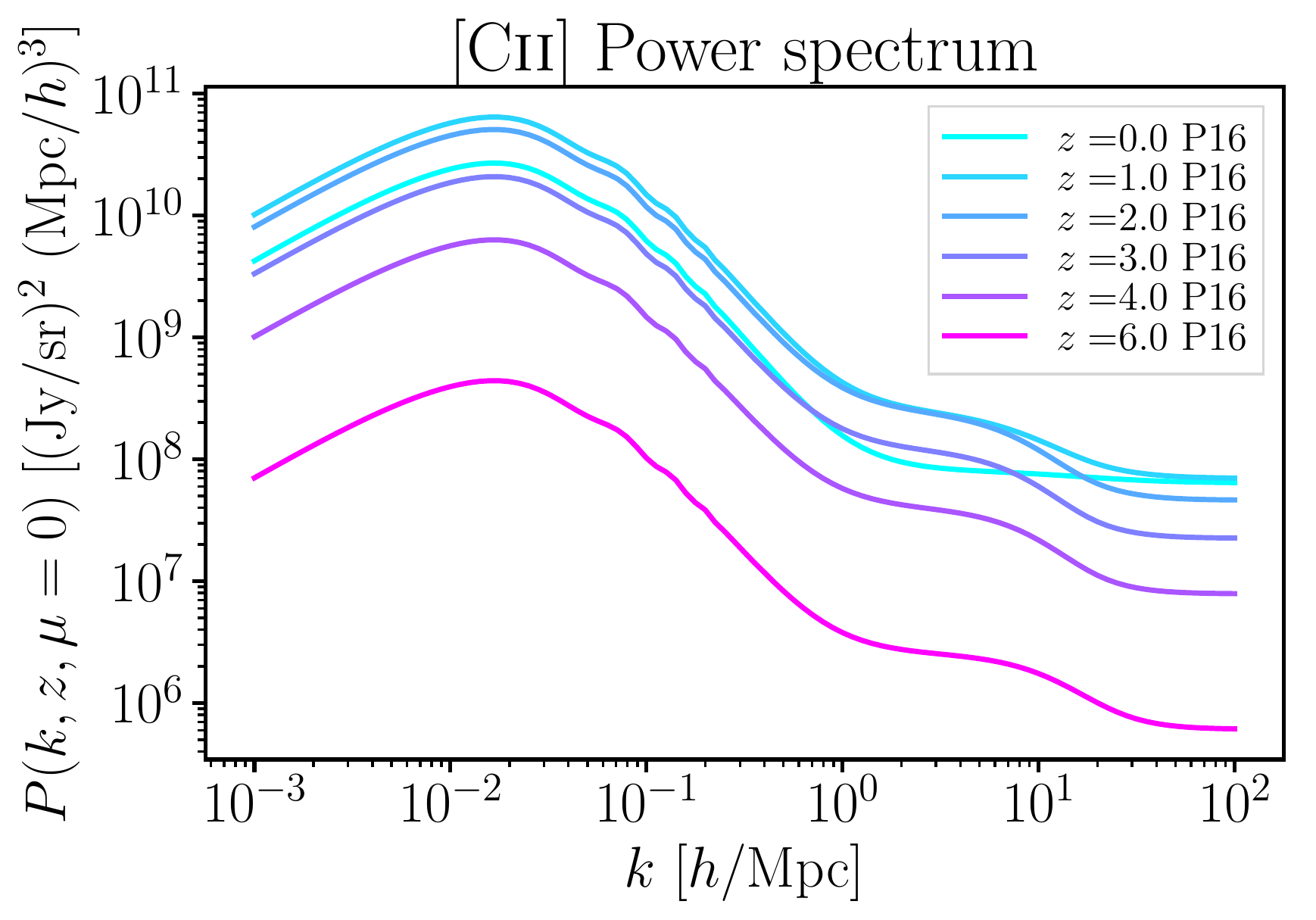}
\includegraphics[width=0.43\textwidth]{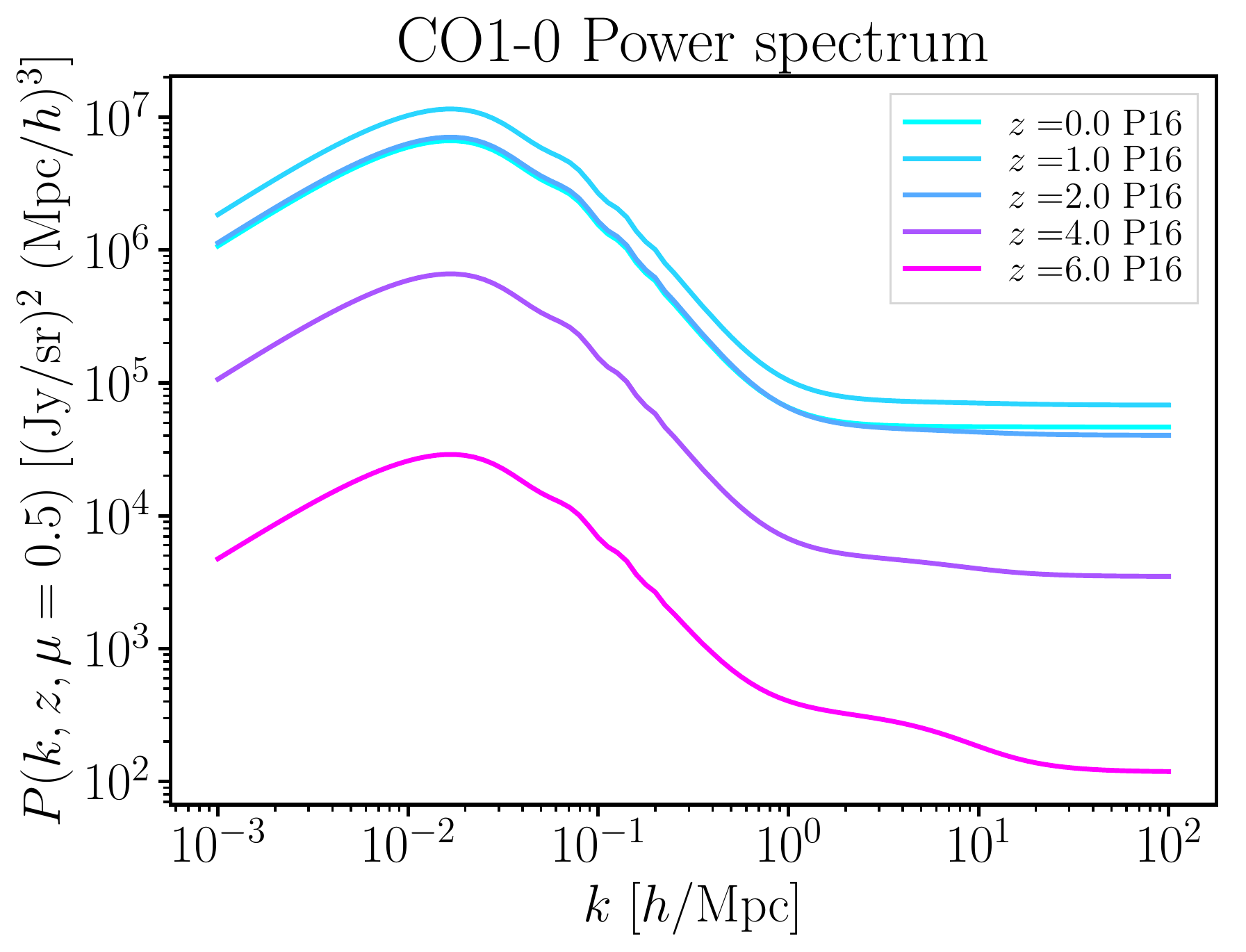}
\caption{
Power spectra of the H$\alpha$, O{\sc iii}, Ly-$\alpha$, [C{\sc ii}] and CO lines, predicted from the various observed luminosity functions (S13 \cite{Sobral13}, C13 \cite{Colbert13}, M15 \cite{Mehta15}, C17 \cite{Cochrane17}, C11 \cite{Cassata11} and P16 \cite{Popping16}).
We show the model predictions out to extremely small scales ($k\sim 100 h/$Mpc) to illustrate that the 2-halo term dominates on the largest scales, the 1-halo term on intermediate scales, and the shot noise on the smallest scales.
However, our model should not be relied upon on such small scales.
}
\label{fig:p3d}
\end{figure}

We show the predicted LIM power spectra from these LFs in Fig.~\ref{fig:p3d}.
We show a wide range of scales, where each of the 2-halo, 1-halo and shot noise terms successively dominate.
However, the model should certainly not be trusted out to these extremely small scales.
Comparing the power spectra of the various lines, the relative size of the 2-halo and 1-halo term is determined by the index $\gamma$ in the relation between mass and SFR.
The overall normalization and its redshift evolution are further determined by the redshift evolution of the line luminosity function.
Overall, the 1-halo and 2-halo terms appear to generally increase out to $z=1-2$, then decrease at higher redshift.
Because the redshift evolutions of the 1-halo and 2-halo terms are not identical, this corresponds to a complex redshift and scale-dependent bias.
Finally, we note that the relative size of the 1-halo and shot noise terms depends crucially on our assumption that the galaxies are NFW-distributed within halos.
If all galaxies were instead centrals, the 1-halo term would be much reduced, subdominant compared to the shot noise.
This may be the case for instance for CO and [C{\sc ii}] at high redshift.

We compare these power spectra to previous predictions in the literature.
Many useful references for comparison exist, including \cite{Sun19} for many of these lines, and \cite{Serra16, Yue19, Chung20, Concerto20, Padmanabhan18, Padmanabhan19} for CO and [C{\sc ii}]. 
For the H$\alpha$ power spectrum, we compare with Fig.~25 in ref.\ \cite{Dore14}. There, the $z=2$ power spectrum matches our prediction in the 1-halo and shot noise regimes, to about a factor two.
Comparing with Fig.~5 in ref.\ \cite{Gong17} at $z=1-2$, the 2-halo terms agree again to a factor of a few, but the shot noise in our analysis is larger by factors $10-100$.
For [O{\sc iii}], comparing with Fig.~25 in ref.\ \cite{Dore14} shows a 2-halo term at $z=2$ consistent with ours from the LF of ref.\ \cite{Mehta15}, and larger than those from the LF of ref.\ \cite{Colbert13}.
We compare our Ly-$\alpha$ power spectrum to that in Fig.~6 of ref.\ \cite{Silva13}. 
Although the redshifts are slightly different ($z=7$ instead of $z=5.57$), the 2-halo terms agree within a factor about two, whereas our shot noise is larger by a a factor 100.
This is probably explained in part by the difference in redshifts, and in part by the theoretical uncertainty, estimated to be a factor 100 in Fig.~25 of ref.\ \cite{Dore14} in this range of redshifts.
For the CO power spectrum, we compare to ref.~\cite{Li16}.
In their Fig.~6 and 10, they show a range of predictions from the literature and from various relations between SFR and CO luminosity. These span a factor of 1000 at $z=2.4-2.8$.
Our prediction is within this range: in the upper end for the 2-halo term and in the middle for the shot noise term.
For the [C{\sc ii}] power spectrum, we compare our prediction to Fig.~16 in ref.\  \cite{Concerto20} at $z=6$. 
This figure shows a factor of about 1000 of theoretical uncertainty.
Our predicted 2-halo term is a factor two lower than their lowest prediction, and our 1-halo term is more than 10 times smaller than their lowest estimate.
Given the large theoretical uncertainty, this is not completely unexpected.
It may be explained by our choice of the LF of ref.\ \cite{Popping16}.
Indeed, refs.~\cite{Popping16, Popping18} mention that the simulated [C{\sc ii}] in\cite{Popping16} is lower than most semi-analytical models.

Thus, in general, the model predictions from different observed LFs or semi-analytical models differ significantly.
This highlights the extreme theoretical uncertainty on the LIM power spectra, especially at high redshift where the LFs are not well known.
This implies that measuring the LIM mean intensity and power spectrum will provide an important further constraint on the galaxy luminosity function, and the combination of the measurements should allow us to better constrain the underlying LFs.

\section{LIM cross-spectra \& decorrelation: multi-line luminosity functions}
\label{sec:cross_spectra}

\subsection{Observational \& simulation constraints}
\label{sec:line_correlation}

Quantifying the correlation between the luminosities in different lines is crucial to correctly predict their cross-spectra in the shot noise regime, and in particular to assess the degree of correlation of the two intensity maps.
As described in \S\ref{sec:physical_origin}, several physical parameters can affect the line luminosity ratios \cite{Osterbrock06}.
Metallicity, dust redenning and the ionization state of the gas determine the line ratios of optical and UV lines, such as H$\alpha$, [O{\sc ii}] and [O{\sc iii}] \cite{Moustakas06},
and far IR lines such as the various CO transitions \cite{Daddi15}.
Naturally, line ratios involving high excitation lines such as [O{\sc iii}] or higher J CO lines are most sensitive to the ionization state \cite{Daddi15}. This ionization state is determined by the local interstellar radiation field, which itself traces the local star formation rate density.
In practice, accurately fitting the spectral line energy distribution of galaxies can require including multiple gas components inside the galaxy, with different ionization states.
On the other hand, ratios of lines from different species, such as [O{\sc ii}]/H$\alpha$ are more sensitive to metallicity \cite{Moustakas06}.
The impact of dust reddening on line ratios should depend mostly on the separation in frequency of the two lines.
Since metallicity, dust properties and ionization state vary from galaxy to galaxy and even inside a single galaxy, they will cause a partial decorrelation of the luminosities in the various lines.
This degree of decorrelation determines the the amplitude of the line cross-power spectrum in the shot noise regime.

Observationally, measurements of the multi-line LFs, i.e.\ joint measurements of the luminosities in several lines for the same galaxy sample, are needed to assess this degree of correlation.

\subsubsection{Mehta model}
\label{subsubsec:mehta15}

Using data from the Hubble WFC3 Infrared Spectroscopic Parallel Survey (WISP), ref.~\cite{Mehta15} provides a fitting function for the joint galaxy luminosities in H$\alpha$ and [O{\sc iii}] over the redshift range $z=0.8-1.2$.
They model the [O{\sc iii}] luminosity function with a realistic Schechter form, and model the probability distribution for the H$\alpha$ of a galaxy, given its [O{\sc iii}] luminosity, as lognormal.
From this fitting function, we compute the correlation coefficient between the line luminosities in H$\alpha$ and [O{\sc iii}], finding $r=0.65$ in that redshift bin.
Such bivariate LF measurements are crucially needed in order to predict the LIM cross-power spectra.

\subsubsection{GOALS survey}

We make use of the publicly available catalog of line fluxes from the Great Observatories All-sky Luminous Infrared Galaxy Survey (GOALS) \cite{Diaz-Santos17}.
This catalog contains [C{\sc ii}] $158\mu$m, [N{\sc ii}] $122\mu$m, [O{\sc i}] $63\mu$m and [O{\sc iii}] $88\mu$m luminosities for 240 galaxies,
observed with the Herschel Photodetector Array Camera and Spectrometer (PACS).
These lines are some of the most important atomic fine-structure lines in the far IR/sub-mm range of $50-200\mu$m.
In this sample, 239 galaxies have measured [C{\sc ii}] line fluxes,
77 galaxies have [N{\sc ii}],
239 have [O{\sc i}]
and 161 have [O{\sc iii}].

\begin{figure}[h!]
\centering
\includegraphics[width=0.95\textwidth]{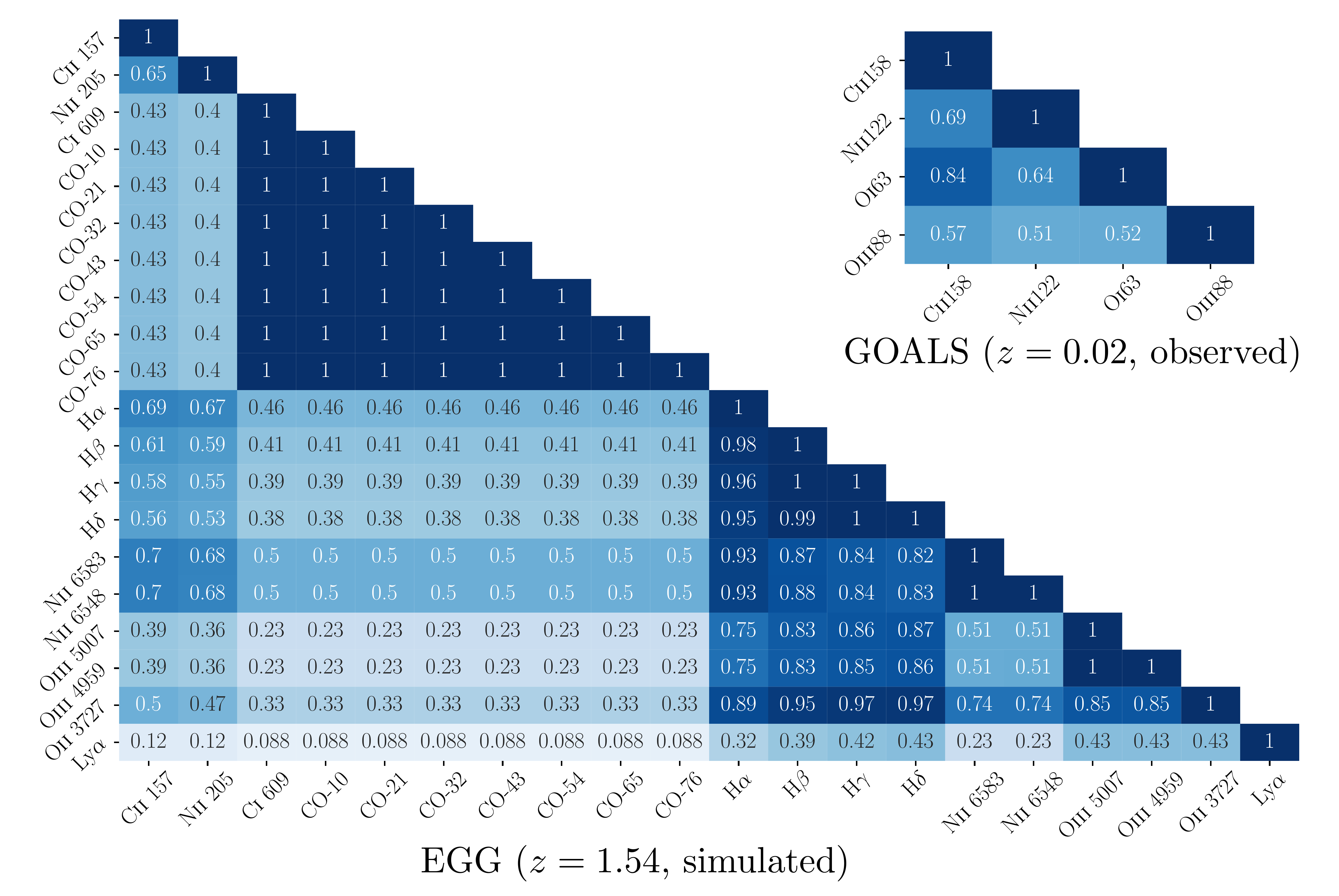}
\caption{
Galaxies emits light in various different lines.
This figure shows the correlation coefficient between the line luminosities of a given galaxy in two different lines.
These correlation coefficients are a function of redshift only, and were estimated
from the GOALS survey at $z=0.02$ (top right) and the EGG simulator at $z=1.54$ (bottom left).
The EGG correlation pattern is qualitatively similar out to $z=4$.
These correlation coefficients at the individual galaxy level are a key input to determine the scale-dependent correlation coefficient between two line intensity maps.
}
\label{fig:rij_goals_egg}
\end{figure}

These galaxies include\footnote{See \url{http://goals.ipac.caltech.edu/} for more details.} approximately 180 Luminous Infrared Galaxies (LIRGs) with total IR luminosity $L_\text{IR} \in [10^{11} L_\odot, 10^{12} L_\odot]$
and more than 20 Ultra-Luminous Infrared Galaxies (ULIRGs)
with $L_\text{IR} \geq 10^{12} L_\odot$.
They were selected from the IRAS Revised Bright Galaxy Sample (RBGS; \cite{Sanders03}) with the aim of 
``provid[ing] an unbiased picture of the processes responsible for enhanced infrared emission in the local Universe, and [to be] excellent analogs for comparisons with infrared and sub-millimeter selected galaxies at high-redshift.''
The RBGS sample includes 629 of the brightest sources at $60\,\mu$m, with flux densities above $5.24\,$Jy and Galactic latitude above $5^\circ$.
``The LIRGs and ULIRGs targeted in GOALS span the full range of nuclear spectral types (type-1 and type-2 AGN, LINERs, and starbursts) and interaction stages (major mergers, minor mergers, and isolated galaxies).''
The sample has a median redshift $z_\text{med}=0.0215$, and spans $z_\text{min}=0.0030$ to $z_\text{max}=0.0918$. 
From this data, we estimate the line covariances and their correlation coefficients, shown in Fig.~\ref{fig:rij_goals_egg}.
We neglect the scatter from measurement error, which was found to be smaller \cite{Diaz-Santos17}.

\subsubsection{EGG simulator}

In order to estimate the degree of correlation we might expect between different lines, we also make use of the ``empirical galaxy generator'' (EGG\footnote{http://cschreib.github.io/egg/})  \cite{Schreiber17}.
EGG works by sampling galaxy stellar masses and redshifts from observed stellar mass functions for quiescent and star forming galaxies.
For each galaxy, the stellar mass, redshift and star forming or quiescent nature then determine the star formation rate, the UVJ colors and the panchromatic broadband of the galaxy, through empirical prescriptions.
The following emission lines are then simulated: 
C{\sc ii} $157\,\mu$m, N{\sc ii} $205\,\mu$m, C{\sc i} $609\,\mu$m, CO1-0 $2.6\,$mm, CO2-1 $1.3\,$mm, CO3-2 $867\,\mu$m, CO4-3 $650\,\mu$m, CO5-4 $520\,\mu$m, CO6-5 $434\,\mu$m, CO7-6 $372\,\mu$m, H$\alpha$ $656.3\,$nm, H$\beta$ $486.1\,$nm, H$\gamma$ $434.1\,$nm, H$\delta$ $410.2\,$nm, N{\sc ii} $658.4\,$nm, N{\sc ii} $654.8\,$nm, O{\sc iii} $500.7\,$nm, O{\sc iii} $495.9\,$nm, O{\sc ii} $372.7\,$nm, Ly$\alpha$ $121.6\,$nm.
This includes all the lines that are relevant for LIM with SPHEREx, specifically:
H$\alpha$ $656.3\,$nm, H$\beta$ $486.1\,$nm, [O{\sc iii}] $500.7\,$nm and Ly$\alpha$ $121.6\,$nm.

The emission line modeling is described in ref.~\cite{SchreiberInPrep}.
The H$\alpha$ luminosity is assumed to scale with the star formation rate of the galaxy, with a fixed random scatter in the Kennicutt relation \cite{1998ApJ...498..541K}.
The other lines in the Balmer series (H$\beta$, H$\gamma$, H$\delta$) are then inferred from H$\alpha$ using fixed line ratios from case-B recombination, without scatter.
Luminosities in [O{\sc iii}] and [N{\sc ii}] are then inferred from the H$\alpha$ and H$\beta$ luminosities using empirical relations, with a mean and scatter dependent on the galaxy metallicity.
These relations were built to match the observed BPT diagrams \cite{1981PASP...93....5B}, where available, between redshifts 0 and 2.
A similar approach is adopted for [O{\sc ii}], inferred from H$\beta$ via an empirical relation where the mean and scatter are functions of metallicity.
For Ly$\alpha$, the line luminosity is chosen proportional to H$\alpha$, with a fixed proportionality constant, but is subsequently attenuated with a variable escape fraction.
This modeling produces a scatter in the various line luminosities, with a reasonable level of correlation, and is thus appropriate for our study.

By contrast, the modeling of the far IR lines is more simplified. 
For instance, the [C{\sc ii}] and [N{\sc ii}] luminosities scale with the total IR luminosity of the galaxy.
The CO and C{\sc i} lines scale with the gas mass in the galaxy.
While no scatter is introduced between the various CO transitions, some scatter in present between the CO, [C{\sc ii}] and [N{\sc ii}] luminosities.

To simulate the galaxy population seen by a LIM experiment, we run EGG with the following options:
\begin{center}
\texttt{egg-gencat verbose area=1 mmin=8 mmax=14 zmin=0.05 zmax=10}
\end{center}
This corresponds to a stellar mass limited sample, with $M_\star \in [10^8 M_\odot, 10^{14} M_\odot]$, over the redshift range $z \in [0.05, 10]$ and a sky area of $1\,{\rm deg}^2$. 
This produces about one million galaxies split between 109 redshift slices.

\subsection{LIM cross-spectra and decorrelation}
\label{sec:decorrelation}

One may want to cross-correlate different line lintensity maps, e.g.\ to suppress line interlopers.
In this case, knowing the amount of correlation between the two intensity maps is crucial.
This is quantified by the scale-dependent correlation coefficient:
\beq
r_{1,2}(k, \mu, z)
\equiv
\frac{P_{1,2}(k, \mu, z)}{\sqrt{P_{1,1}(k, \mu, z) P_{2,2}(k, \mu, z)}}
\eeq

On large scales where the 2-halo term dominates, this ratio is unity $r_{1,2}(k, \mu, z) = 1$ and the two intensity maps are perfectly correlated.
On smaller scales, where the 1-halo or shot noise term dominates, the power spectrum is of the form
\beq
P = \frac{\bar{I}_1 \bar{I}_2}{\bar{n}^{\rm eff}}
= \bar{I}_1 \bar{I}_2 \frac{1}{\bar{n}}
\left( 1 + \sigma_{1,2}^2 \right)
\eeq
where $\bar{n}^{\rm eff}$ and $\bar{n}$ are the effective and mean number density of halos or galaxies and $\sigma_{1,2}^2$ the halo or galaxy line noise contributions in the 1-halo and shot noise regimes, respectively.
As explained in \S\ref{sec:line_noise}, these quantities are ``cutoff dependent'' and must be defined with care.
If the line noise is negligible, $\sigma_{1,2}^2 \ll 1$, then the two intensity maps are again perfectly correlated: $r_{1,2}(k, \mu, z) = 1$.
However, if the noises are large, $\sigma_{1,1}^2, \sigma_{1,2}^2, \sigma_{2,2}^2 \gg 1$, then the correlation coefficient drops to
\beq
r_{1,2}(k, \mu, z)
= \frac{\sqrt{\bar{n}^{\rm eff}_{1,1}\ \bar{n}^{\rm eff}_{2,2}}}{\bar{n}^{\rm eff}_{1,2}}
\simeq \frac{\sigma_{1,2}^2}{\sigma_{1,1} \sigma_{2,2}},
\eeq
i.e.\ the correlation coefficient between the two intensity maps is equal to the correlation coefficient of the halo (galaxy) luminosities in lines 1 and 2, in the 1-halo (shot noise) regime.
Thus, a large scatter in the line luminosities of halos or galaxies can cause a decorrelation between two intensity maps.

In the 1-halo regime, assuming $L_i(m) \propto \text{SFR}^{\gamma_i}(m, z)$, the correlation coefficient between lines $1$ and $2$ becomes:
\beq
r_{1,2}(k\rightarrow 0, \mu, z)
=
\frac{\int dm \ n(m)\ \text{SFR}^{\gamma_1+\gamma_2}(m)}
{\sqrt{
\int dm \ n(m)\ \text{SFR}^{2\gamma_1}(m)
\int dm \ n(m)\ \text{SFR}^{2\gamma_2}(m)
}}.
\eeq
In the 1-halo regime, all lines with the same power law index $\gamma=1$ relating luminosity to SFR will thus be perfectly correlated. 
This is typically the case of the UV and optical lines.
For far IR lines like N{\sc ii} $122\,\mu$m, N{\sc iii} $58\,\mu$m, C{\sc ii} $158\,\mu$m, which scale as star formation rate to the power $0.78-1.02$ according to ref.~\cite{Fonseca17}, the correlation coefficient is $r>0.96$ out to redshift 5.
The decorrelation between these lines in the 1-halo regime is thus negligible.
However, CO has $\gamma=0.6$, leading to a significant decorrelation with the other lines considered in the 1-halo regime. 

In the shot noise regime, the joint luminosity function for H$\alpha$ and [O{\sc iii}] in ref.~\cite{Mehta15} predicts a correlation coefficient between these lines of $r=0.65$. 
In Fig.~\ref{fig:rij_goals_egg}, we show the correlation coefficients of line luminosities for an individual galaxy, as predicted by the EGG simulator and the as measured in the GOALS data. 
The correlation coefficients for line luminosities from individual galaxies are $r=0.088-1$ from EGG and $r=0.5-1$ from GOALS, depending on the lines considered.
Astrophysically, these correlation coefficients are a direct consequences of the physical processes inside galaxies responsible for the produceing the emission lines (Sec.~\ref{sec:physical_origin}).

\begin{figure}[h!]
\centering
\includegraphics[width=0.45\textwidth]{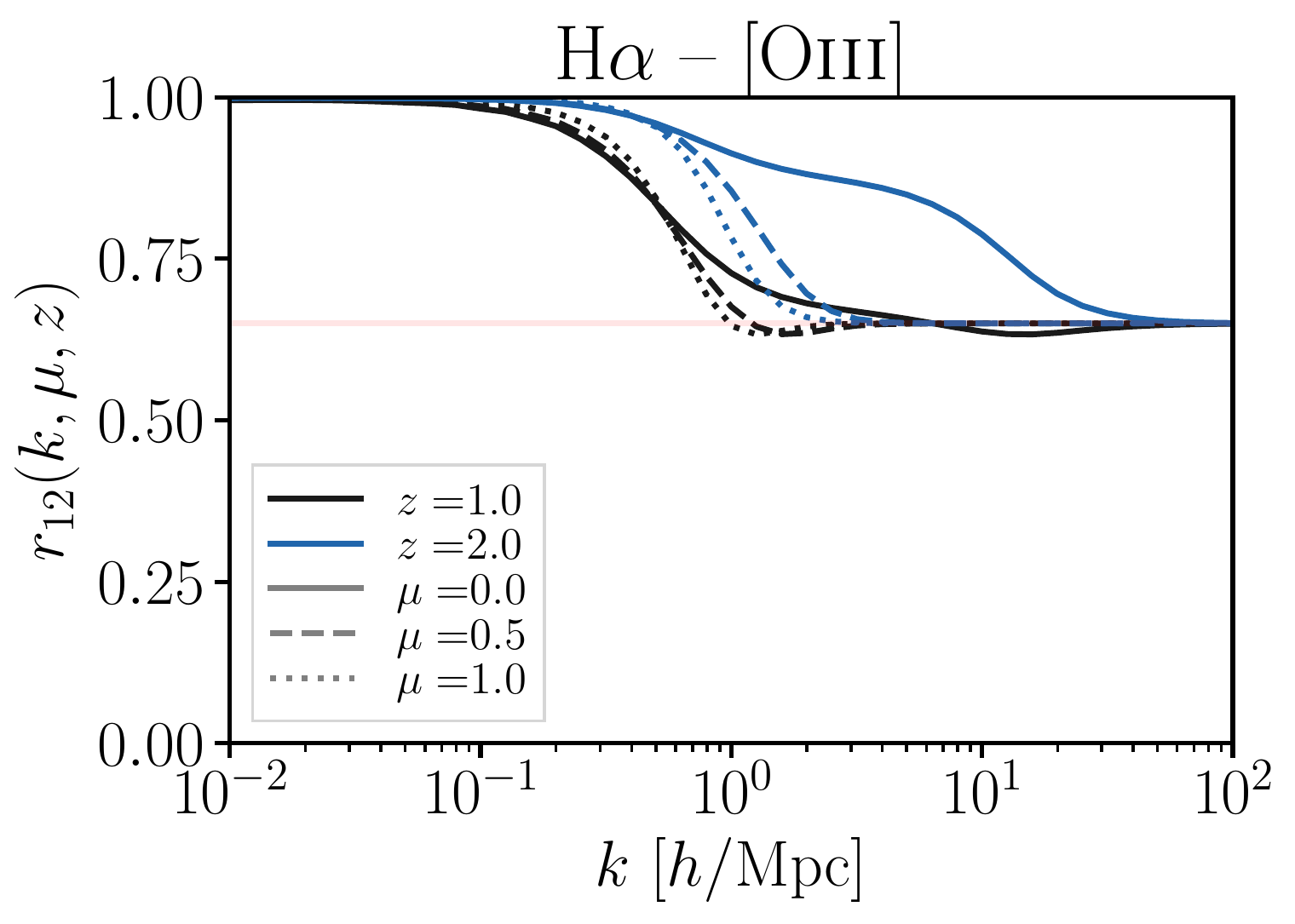}
\includegraphics[width=0.45\textwidth]{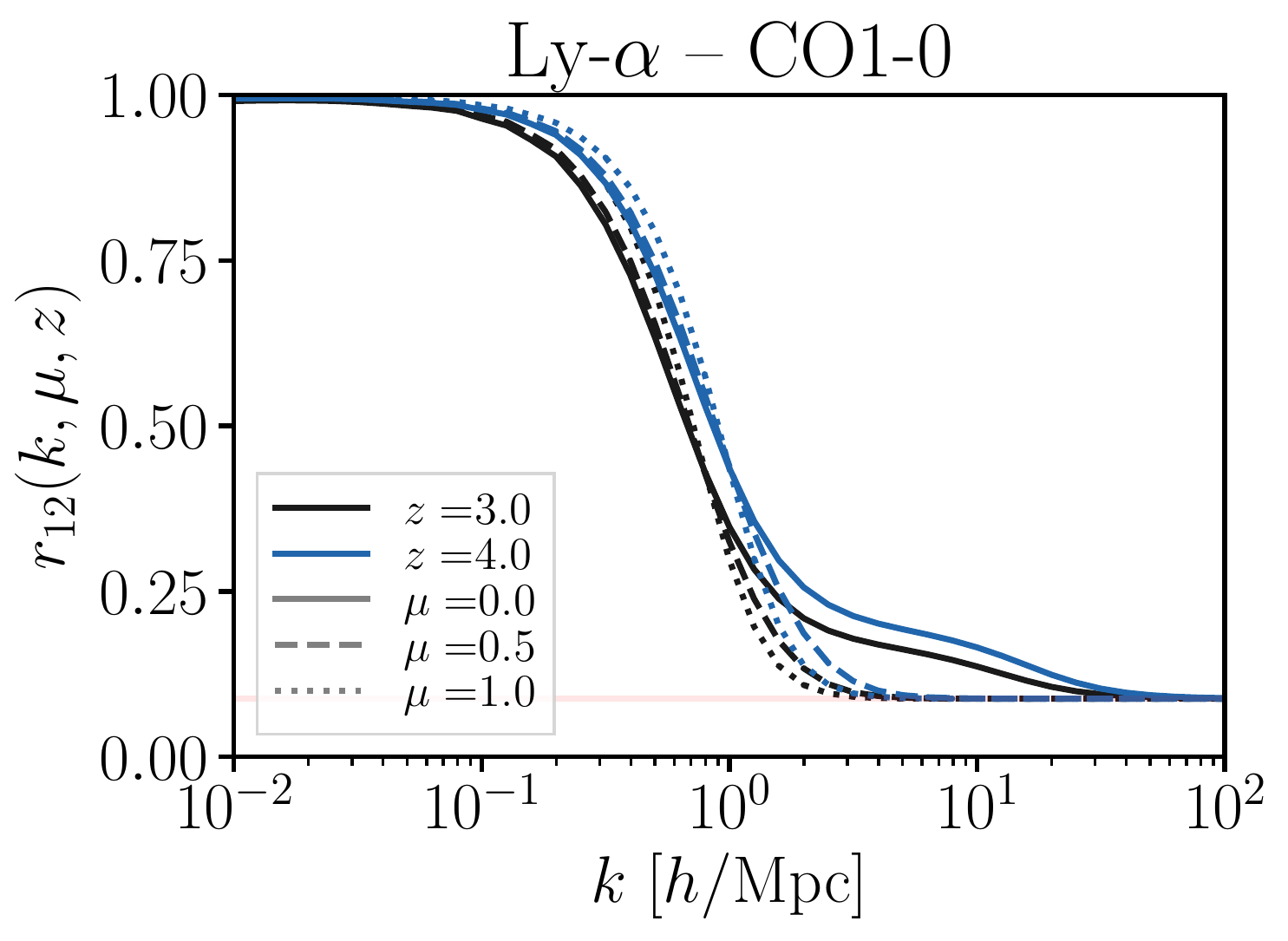}
\caption{
Generic evolution of the correlation coefficient between line intensity maps.
We consider the correlation betwewen H$\alpha$ \cite{Cochrane17} and [O{\sc iii}] \cite{Mehta15} (left panel),
and between Ly-$\alpha$ \cite{Cassata11} and CO \cite{Popping16} (right panel).
On large scales, where the 2-halo term dominates for both lines, the two maps are perfectly correlated.
On small scales, where both maps are shot noise dominated, we recover the line luminosity correlation coefficient (red line) $r=0.65$ (left, inferred from \cite{Mehta15}) or $r=0.088$ (right, from \cite{Schreiber17}).
On intermediate scales, if the 1-halo terms dominate, the correlation coefficient is close to unity. This occurs at $z=2$ for H$\alpha$ and [O{\sc iii}].
If instead the 1-halo and shot noise terms are comparable, e.g. at $z=1$ and $k=1 h/$Mpc for H$\alpha$ -- [O{\sc iii}] (left panel), the correlation coefficient of the sum of 1-halo and shot noise can actually be lower than that of the shot noise alone ($r=0.65$).
In both panels, the FOG effect suppresses the 1-halo term along the LOS, bringing the correlation coefficient down to the shot noise value.
This occurs at $k=1-20 h/$Mpc both for H$\alpha$ -- [O{\sc iii}] (left panel) and Ly-$\alpha$ -- CO1-0 (right panel).
}
\label{fig:r}
\end{figure}

Given the correlation coefficients at the level of individual galaxies, we can compute the scale-dependent correlation coefficients between any pair of intensity maps.
We show them for a pair of highly-correlated lines (H$\alpha$, [O{\sc iii}]) and for a pair of low-correlation lines (Ly-$\alpha$, CO) in Fig.~\ref{fig:r}.

\subsection{Implications}
\label{sec:decorrelation_applications}

The value of the correlation coefficient computed above has several implications:

\begin{enumerate}
\item \textbf{Estimating the auto-spectrum from cross-spectra only.}
Cross-spectra allow reduced foreground contamination. 
In ref.~\cite{Beane19}, cross-spectra of 21cm, [C{\sc ii}] and [O{\sc iii}] lines are therefore combined to estimate the 21cm power spectrum:
\beq
\hat{P}_{aa}
\equiv 
\frac{P_{ab} P_{ac}}{P_{bc}}
= 
P_{aa} \frac{r_{ab} r_{ac}}{r_{bc}}.
\label{eq:auto_from_crosses}
\eeq
Ref.~\cite{Beane19} models the line emission from simulated halos to estimate the level of decorrelation between these lines, and how it limits the method.
Here, we instead use an analytical halo model to predict the line correlation coefficients, including not only the effect of individual halos (1-halo term) but also that of individual galaxies (shot noise term).
We further include the effect of RSD on the correlation coefficient.
Additional noise or foregrounds, uncorrelated between intensity maps $a, b$ and $c$, will drop out of the cross-correlations.
As a result, the relevant correlation coefficients in Eq.~\eqref{eq:auto_from_crosses} are still the noiseless, foreground free ones.
On the other hand, any foreground component which contaminates two maps, e.g., $a$ and $b$, will bias the cross-correlation $P_{ab}$.
This will thus bias the estimated auto-spectrum $\hat{P}_{aa}$.
\item \textbf{Using a line at a different frequency as a foreground tracer.}
Consider a simple model where the measured line intensity at any frequency is the sum of a target ($t$) line and a foreground ($f$) line. We can produce intensity maps at two frequencies $\nu_1$ and $\nu_2$, such that the foreground emission in $\nu_1$ comes from the same redshift as the target emission in $\nu_2$.
In other words:
\beq
\left\{
\bal
&I_{\nu_1} = t(z_0) + f(z_1)\\
&I_{\nu_2} = t(z_1) + 
\underbrace{f(z_2)}_{\text{ignore}}\\
\eal
\right.
.
\eeq
How well can one subtract the foreground emission in $I_{\nu_1}$ using $I_{\nu_2}$?
We use a linear combination:
\beq
\bal
\hat{t}(z_0)
&= I_{\nu_1} - \alpha I_{\nu_2}\\
&= t(z_0) 
+ \underbrace{\left[ f(z_1) - \alpha t(z_1) \right]}_{\text{to be reduced}}\\
\eal
\eeq
where the coefficient $\alpha$ that minimizes the power spectrum of the square bracket above is
$\alpha = P_{12} / P_{22} = P_{tf}(z_1) / P_{tt}(z_1)$.
The residual power spectrum of the square bracket is then
$P_{ff}(z_1) \left[ 1 - r_{tf}^2(z_1) \right]$.
In other words, the maximum reduction of the foreground power spectrum achievable is entirely determined by $1 - r_{tf}^2(z_1)$.
Equivalently, the total power spectrum of the estimated target line is
$P_{\hat{t}} = P_{11} \left[ 1 - r^2_{12} \right]$,
where the maximum power reduction is determined by $\left[ 1 - r^2_{12} \right]$.
We show the corresponding reduction factors in Fig.~\ref{fig:r2}.
\begin{figure}[h!]
\centering
\includegraphics[width=0.45\textwidth]{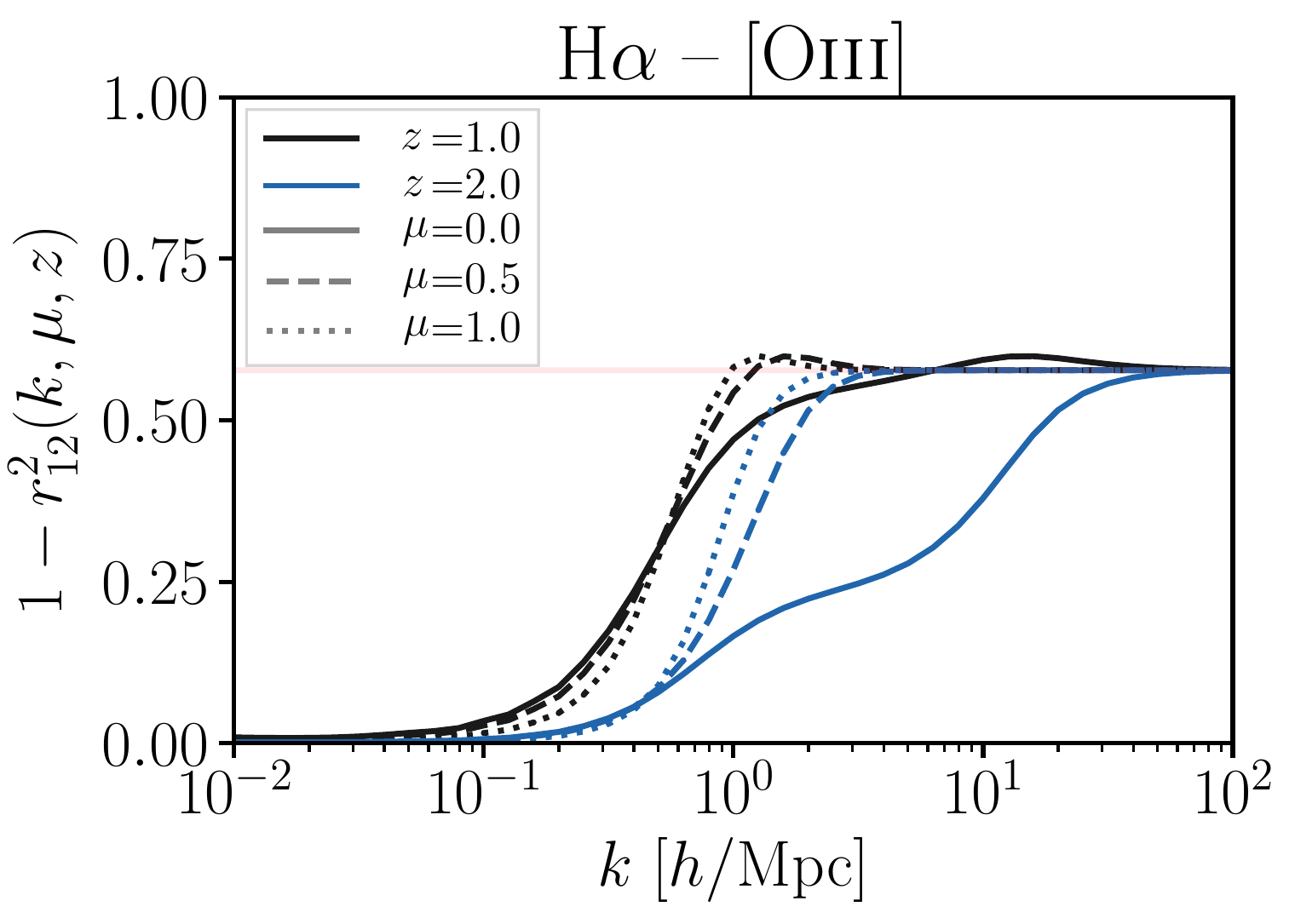}
\includegraphics[width=0.45\textwidth]{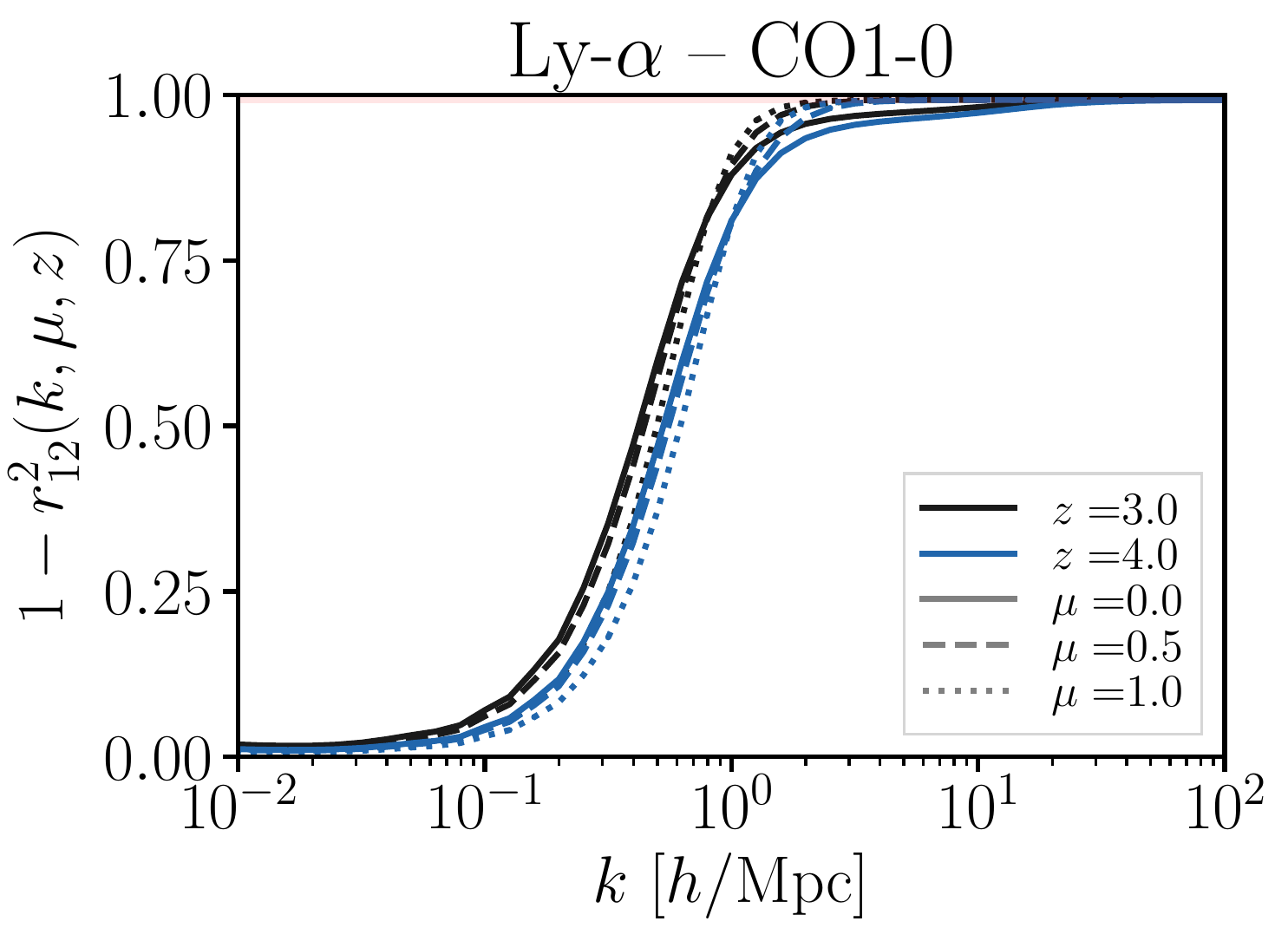}
\caption{
Factor $1-r^2$, which determines the ability to use one line as a tracer of the other, in order to subtract foregrounds/interlopers.  Values near zero indicate good subtraction while values near unity indicate ineffective subtraction.
}
\label{fig:r2}
\end{figure}
\item \textbf{Astrophysics of star formation}
The lines considered in this paper all trace the star formation rate in galaxies.
However, they are produced in different regions of the interstellar medium, with different densities, temperatures, ionization fractions and chemical compositions.
Thus, the correlation coefficient between two different line luminosities, for a given galaxy, informs us about the correlated abundance of these ISM regions, in a way that would be impossible by measuring each line separately.
This idea was explored in \cite{Breysse17b, Heneka17, Sun19}.
By connecting the correlation coefficient between various at the galaxy level to the correlation coefficient observable at the intensity map level, our formalism enables extracting this astrophysical information from LIM.
\end{enumerate}

\section{Conclusions}

Line intensity mapping (LIM) has become a new frontier for studying the Universe, with early observations giving way to dedicated surveys.  
In this paper, we have provided a new halo model formalism, consistently capturing the various sources of galaxy and halo luminosity fluctuations and their effects on the LIM power spectrum.

To this end we have generalized the conditional luminosity function (CLF) formalism of ref.~\cite{Yang03} to a multi-line CLF (\S\ref{sec:clf}).
This extension produces a new, general and flexible halo model, which predicts not only the mean intensity and auto-spectrum of a line intensity map, but also cross-spectra of different lines, and the cross-spectra of LIM with galaxy catalogs and galaxy and CMB lensing (Appendices.~\ref{app:halo_model} and \ref{app:angular_clustering}).
We have provided the corresponding expressions for the mean, 2-halo, 1-halo and shot noise terms, in 3D redshift-space and for 2D projected maps (Appendices~\ref{app:halo_model}-\ref{app:angular_clustering}).

The formalism highlights the similarity between the 1-halo and shot noise terms, both arising from the (approximately) Poissonian shot noise of halos and galaxies, respectively.
In galaxy clustering, they are simply given by the inverse number density of halos and galaxies.
In LIM, the CLF-based halo model shows that they are enhanced by a term we refer to as halo and galaxy ``line noises'', respectively (\S\ref{sec:line_noise}).
We showed that the halo line noise is entirely sourced by luminosity fluctuations across halo masses, and is insensitive to luminosity fluctuations at fixed halo mass.
The galaxy line noise is instead sourced by all galaxy luminosity fluctuations, i.e. $\langle L_\text{gal}^2\rangle - \langle L_\text{gal}\rangle^2$.
We thus defined LIM-relevant halo and galaxy ``effective number densities'', accounting for these effects.
By tracking the sources of luminosity scatter, we showed that the amplitude of the 1-halo term does not necessarily exceed that of the shot noise on large scales, for Poissonian halos and galaxies (\S\ref{sec:galaxy_halo_consistency}).
We showed examples of lines for either case.
This has implications for interloper rejection methods relying on the sparsity of the sources \cite{Cheng20} -- halos in the 1-halo regime and galaxies in the shot noise regime \cite{paper2}.
Because it encodes the imperfect correlation between different line luminosities from the same galaxy, the multi-line CLF allows us to quantify the level of decorrelation between two LIMs from the same redshift (\S\ref{sec:line_correlation}).
This is a crucial modeling improvement, since many LIM analyses rely on cross-spectra to subtract interlopers or foregrounds, or to estimate auto-spectra.
Furthermore, by connecting LIM cross-correlations to the galaxy-level properties, the multi-line CLF allows us to constrain properties of the galaxy interstellar media inaccessible with auto-spectra alone.

To evaluate the halo model, we have relied on observations and simulations (\S\ref{sec:observational}).
We used observed luminosity functions for H$\alpha$ \cite{Sobral13, Colbert13, Cochrane17}, [O{\sc iii}] \cite{Colbert13, Mehta15}, Lyman-$\alpha$ \cite{Cassata11}, CO \cite{Popping16} and [C{\sc ii}] \cite{Popping16}.
While most observations and simulations focus on the single-line LF or CLF, modeling LIM cross-spectra crucially requires knowledge of the multi-line CLF.
To estimate the correlation between different lines (\S\ref{sec:line_correlation}), we relied on measurements from \cite{Mehta15} for H$\alpha$ and [O{\sc iii}] and the GOALS survey \cite{Diaz-Santos17} for far infrared lines, as well as simulation results from \texttt{EGG} \cite{Schreiber17,SchreiberInPrep}.
We encourage semi-analytic models (SAMs), simulation and observational studies to output their multi-line CLFs whenever possible. Fitting functions for them, from simulations such as \cite{Yang20}, would be a valuable input for the halo model cross-spectra.
In return, these simulations could be used to check the halo model prediction.

Other applications of the CLF are possible. A valuable extension would be to extend this formalism to also predict higher-point correlators of LIM, or joint multi-line voxel intensity distributions \cite{Breysse17a}.

We have made our code publicly available at \url{https://github.com/EmmanuelSchaan/HaloGen/tree/LIM}.
This modular code computes the LIM mean intensity, auto- and cross-spectra in 3d redshift-space and 2d projections.
A key input to this code is the set of luminosity functions for the various lines. 
The uncertainty in these LFs at high redshift is very large at present, leading to a large uncertainty in the predicted power spectra.
Power spectra from this code should thus not be considered ground truth, and may differ significantly from predictions in the literature based on different input luminosity functions.
For this reason, the code was made highly modular, allowing the user to easily switch input (luminosity functions, star formation rate, etc.), making it a useful prediction tool for LIM forecasts.

\acknowledgments
We would like to thank Yun-Ting Cheng, Tzu-Ching Chang and Olivier Dor\'e for their kind hospitality and for all our useful discussions during the preparation of this manuscript.
We thank Corentin Schreiber for sharing with us details of the line luminosity implementation in his code \texttt{EGG}.
We thank the organizers and participants of the L2S2: Lines in the Large-scale Structure conference\footnote{\url{https://l2s2.sciencesconf.org/}} for many useful discussions.
We thank Patrick Breysse, Simone Ferraro, Adam Lidz, Abhishek Maniyar, Anthony Pullen, Uros Seljak and David Spergel for their helpful feedback on this paper.
We thank the anonymous referee for their useful comments.
E.S.~is supported by the Chamberlain fellowship at Lawrence Berkeley National Laboratory. 
M.W.~is supported by the DOE and the NSF.
This research has made use of NASA's Astrophysics Data System and the arXiv preprint server.
This research used resources of the National Energy Research Scientific Computing Center (NERSC), a U.S. Department of Energy Office of Science User Facility operated under Contract No. DE-AC02-05CH11231.

\appendix

\section{Halo model}
\label{app:halo_model}

\subsection{Pedagogical derivation of observables and a halo model}
\label{app:pedagogical_derivation}

Since there are many expressions and approximations made in different papers in the literature, in this appendix we present a pedagogical derivation of our basic observables and an implementation of the halo model tailored to LIM.  We have attempted to be clear about the approximations and assumptions made at each stage.  Our approach builds upon the extensive earlier work in this subject, including refs.\ \cite{White01,Gong20,Seljak01,Kang02,Tinker07,Kanemaru15,Cheng16,Sarkar18,Sarkar19} and the papers cited therein.

For a galaxy with line luminosity $L_1$ ([W]) in emission line 1, the observed line flux ([W m$^{-2}$]), integrated over the frequency width of the line, is
$L_1 / \left( 4\pi D_L^2 \right) = L_1 / \left( 4\pi \chi^2 (1+z)^2 \right)$, where have assumed a spatially flat Universe with comoving radial distance $\chi(z)$.

The sky specific intensity $I_\nu$ ([W m${}^{-2}$ sr${}^{-1}$ Hz${}^{-1}$]), measures the sky flux per spatial resolution element $d\Omega$ ([sr]) and frequency resolution element $d\nu$ ([Hz]) of the LIM experiment.
The narrow emission lines are typically not spectrally resolved, so the observed specific intensity at $\nu = \nu_1^0/(1+z)$ includes the full line luminosity $L_1$ from all galaxies at redshift $z$:
\beq
\bal
I_1
&= 
\underbrace{\frac{1}{\int d\nu d\Omega}
\int 
d\nu d\Omega}
_{\substack{\text{average in spatial and}\\ \text{spectral resolution element}}}
\underbrace{ \frac{d\chi}{d\nu} 
\int dL_1 \;
\frac{dN_\text{gal}}{d\Omega d\chi dL_1} }
_{\substack{\text{galaxy number per} \\ \text{resolution element} \\ \text{[sr$^{-1}$ Hz$^{-1}$]}}}\;
\underbrace{\frac{L_1}{4\pi D_L}}
_{\substack{\text{galaxy line flux} \\ \text{[W m$^{-2}$]}}}
\quad\quad\text{[W m${}^{-2}$ sr${}^{-1}$ Hz${}^{-1}$]}\\
&=
\frac{dV}{d\Omega d\chi}\;
\frac{d\chi}{d\nu}\;
\int dL_1\;
\frac{dN_\text{gal}}{dV dL_1}\; 
\frac{L_1}{4\pi D_L}\\
&=
\chi^2\;
\frac{c \left( 1 + z \right)^2}{\nu_1^0 H(z)} 
\int dm \;n(m)\;
\int dL_1\; \phi(L_1|m)\;
\frac{L_1}{4\pi \chi^2 (1+z)^2}\\
&=
\frac{1}{4\pi \nu_1^0}
\frac{c}{H(z)}
\int dm \;n(m)\;
\int dL_1\; \phi(L_1|m)\;L_1.\\
\eal
\eeq
where in the last two lines the first integrals are over halo mass, $m$, and the halo mass function, $n(m)$ -- the number of halos in mass range $(m;dm)$ per comoving volume -- and $\phi(L_1|m)$ is the conditional luminosity function defined in \S\ref{sec:clf}.
Hence the mean specific intensity is
\beq
\bar{I}_1
=
\frac{1}{4\pi \nu_1^0}
\frac{c}{H(z)}
\ \bar{n}_\text{gal} \bar{L}_1 
\qquad\text{[W m${}^{-2}$ sr${}^{-1}$ Hz${}^{-1}$]},
\eeq
with:
\beq
\left\{
\bal
&\bar{n}_\text{gal}
=
\int dm \;n(m)\;
\int dL_1\; \phi(L_1|m)
&&\text{Mean galaxy number density}
&&&[\text{cMpc}^{-3}]\\
&\bar{L}_1
=
\frac{1}{\bar{n}_\text{gal}}
\int dm \;n(m)\;
\int dL_1\; \phi(L_1|m)\;L_1
&&\text{Mean line $1$ luminosity}
&&&[\text{W}]\\
\eal
\right.
\eeq

In this halo model, the local number density of halos $n(m, x)$ is a biased, Poisson sampling of the underlying linear matter density field.
This produces the 2-halo and 1-halo terms in the power spectrum:
\beq
\langle 
\delta n(m, k)
\delta n(m', k')
\rangle
=
\left( 2\pi \right)^3 \delta^D(k+k')\,
\bar{n}(m)
\bar{n}(m')\,
\left[
\underbrace{b(m)b(m') P_\text{lin}(k)}
_\text{2-halo term}
+
\underbrace{\delta^D(m-m')
\frac{1}{\bar{n}(m)}}
_\text{1-halo term}
\right],
\label{eq:origin_1halo_2halo}
\eeq
where the 1-halo term is nothing but the shot noise of halos, and the 2-halo term is related to the linear matter density field through a deterministic, linear, scale independent and mass dependent halo bias.
This bias assumption is expected to be valid on large scales where the 2-halo term is significant.

We further assume that within a given halo, the number density of galaxies is a Poisson sampling of the (normalized) NFW profile \cite{NFW}, $u(k,m)$, of the halo.
We currently lack enough observational constraints on satellite vs.\ central galaxy line emission to include this distinction in our models, and we will therefore ignore this distinction here.
However, distinguishing satellites and centrals can be done easily in our formalism, as we show in Appendix~\ref{app:centrals_satellites}.
This Poisson process produces a galaxy shot noise term:
\beq
\bal
\langle 
\bar{n}(m) \delta\phi(L_1, L_2, k|m) \;
\bar{n}(m') \delta\phi(L_1', L_2', k'|m')
\rangle
&=
\left( 2\pi \right)^3 \delta^D(k+k')\\
&\times\delta^D(m-m')
\delta^D(L_1 - L_1')
\delta^D(L_2 - L_2') \\
&\times 
\; 
\underbrace{\bar{n}(m) \phi(L_1, L_2|m)}
_\text{galaxy shot noise}.
\eal
\label{eq:origin_shot_noise}
\eeq

Finally, we compute the specific intensity power spectrum in redshift space,
following \cite{White01,Gong20,Seljak01,Kang02,Tinker07,Kanemaru15,Cheng16,Sarkar18,Sarkar19}.
We take into account the effect of FOG on the halo profile by substituting:
\beq
u^{(s)}(k,\mu, m, z) 
=
u(k,m) e^{-k^2 \mu^2 \sigma^2(m) / 2}
\eeq
where the velocity dispersion $\sigma$ is possibly mass-dependent.
The Kaiser effect follows from the conservation of the luminosity in redshift-space:
\beq
\bal
\frac{\delta I_1^{(s)}(k)}{\bar{I}_1}
=
- \nabla\cdot \psi^{(s)}(k)
=
- \nabla\cdot \psi(k) + f\mu^2 \delta_m(k)
=
\frac{\delta I_1(k)}{\bar{I}_1} + f\mu^2 \delta_m(k).
\eal
\eeq
Here we introduced the Lagrangian displacement, $\psi$, which sources the matter density fluctuations 
($\delta_m + \nabla\cdot \psi = 0$)
and its redshift-space counterpart 
$\psi^{(s)} = \psi + f \left( \psi\cdot \hat{n} \right) \hat{n}$.

With all the ingredients above, we start with the 2-halo term, where the halo number density fluctuations trace the matter density field.
This 2-halo term corresponds to the first term in Eq.~\eqref{eq:origin_1halo_2halo},
and can be computed by taking the power spectrum of the ``2-halo'' field:
\beq
\bal
\delta I_1^\text{2h}
&=
\frac{1}{4\pi \nu_1^0}
\frac{c}{H(z)}\; \\
&\times
\left(
\int
dm\ n(m)\; 
\Big[ b(m) + 
\underbrace{f \mu^2}_\text{Kaiser} \Big] 
\;u(k,m)
\underbrace{e^{-k^2 \mu^2 \sigma^2(m) / 2}}
_\text{FOG}
\;
\int
dL_1 \phi(L_1|m)\; L_1
\right)
\delta_{\rm lin}(z)\\
&=
\bar{I}_1
\left[ b_1(k, \mu, z) + F(k, \mu, z) \mu^2 \right]
\delta_\text{lin}
,
\eal
\eeq
and therefore
\beq
P_{1,2}^\text{(s) 2-halo}(k,\mu, z)
=
\bar{I}_1 \bar{I}_2
\left[ b_1 + F \mu^2 \right]
\left[ b_2 + F \mu^2 \right]
P_\text{lin}
\eeq
where we have defined the relevant physical parameters:
\beq
\left\{
\bal
&b_j(k, \mu, z)
\equiv
\frac{1}{\bar{n}_\text{gal} \bar{L}_j}
\int
dm\ n(m)\;
N_\text{gal}(m) L_j(m)
\;
b(m) 
\;
u(k,m)
e^{-k^2 \mu^2\sigma^2 / 2}
\\
&F(k, \mu, z)
\equiv
f\;
\int dm \; n(m)\;
\left( \frac{m}{\bar{\rho}} \right)
u(k,m) 
e^{-k^2 \mu^2 \sigma^2/2}
\\
\eal
\right.
\eeq
On scales larger than $\sigma$, such that the FOG effect is negligible, and on scales larger than the halo size, such that the halo profile $u(k,m)\simeq 1$, the bias $b_j$ becomes the scale-independent large-scale bias of linear theory, and $F$ tends to the usual logarithmic growth rate $f$.
On smaller scales, both the bias $b_j$ and effective growth rate $F$ become scale-dependent and anisotropic.

Focusing now on the 1-halo term, due to the Poisson shot noise of halos, we use Eq.~\eqref{eq:origin_1halo_2halo} to obtain:
\beq
\bal
P_{1,2}^\text{(s) 1-halo}(k, \mu, z)
&=
\int dm \; n(m)
\left| u(k,m) \right|^2
e^{-k^2 \mu^2 \sigma^2(m)}
\left[ \int dL_1 \phi(L_1, m) L_1 \right]
\left[ \int dL_2 \phi(L_2, m) L_2 \right]\\
&=
\bar{I}_1 \bar{I}_2
\frac{U_{1,2}^2 (k,\mu, z)}
{\bar{n}_{1,2}^\text{h,eff}},\\
\eal
\eeq
where we further introduced the line-pair-weighted, squared halo profile
\beq
U_{1,2}^2(k, \mu, z)
\equiv
\frac{
\int dm \; n(m)
\left| u(k,m) \right|^2
e^{-k^2\mu^2\sigma^2}
L_1(m) L_2(m)
}
{
\int dm \; n(m)\ L_1(m)L_2(m)
}
\eeq
which tends to $1$ as $k\to 0$ and the total halo number density,
\beq
\bar{n}_{1,2}^{\rm h,eff}
\equiv
\frac{\left(\int dm \; n(m)L_1(m)\right)\left(\int dm \; n(m)L_2(m)\right)}
     {\int dm \; n(m)L_1(m)L_2(m)}
\qquad [\text{cMpc}^{-3}] \quad .
\eeq

Finally, the galaxy shot noise, coming from Poisson fluctuations in the number of galaxies within a halo, Eq.~\eqref{eq:origin_shot_noise}, becomes:
\beq
\bal
P_{1,2}^\text{shot}(z)
&=
\frac{1}{\left(4\pi\right)^2 \nu_1^0 \nu_2^0}
\left(\frac{c}{H(z)}\right)^2
\int dm\; n(m)
\int dL_1 dL_2\ \phi(L_1, L_2,m)\, L_1 L_2
\\
&=
\frac{\bar{I}_1 \bar{I}_2}
{\bar{n}_{1,2}^\text{gal,eff}} \quad ,\\
\eal
\eeq
with
\beq
\bar{n}^\text{gal eff}_{1,2}
=
\frac{
\left( \int dm \; n(m) L_1(m)  \right)
\left( \int dm \; n(m) L_2(m)  \right)
}
{
\int dm \; n(m)
\int dL_1 dL_2\ \phi(L_1, L_2|m) L_1 L_2
}
\qquad [\text{cMpc}^{-3}] \quad .
\eeq

\subsection{Halo occupation distribution: centrals and satellites}
\label{app:centrals_satellites}

In this appendix, we show that our halo model based on the CLF naturally describes the distinction between satellite and central galaxies.
As shown in \cite{Wolz19}, the particular choice of halo occupation distribution for the line-emitting galaxies can have a significant impact on the 1-halo LIM power spectrum.
The central galaxies are assumed to be at the centers of halos, with no velocity with respect to the halo, and therefore no FOG effect, whereas the satellites are assumed to sample the NFW matter density profile of the halo, and are alone responsible for the FOG.
We define their numbers $N_\text{cen}(m,z), N_\text{sat}(m,z)$ and mean luminosities $L_j^\text{cen}(m,z) L_j^\text{sat}(m, z)$ in a halo of mass $m$, at redshift $z$.
The mean intensity is then simply the sum of the satellite and central luminosities:
\beq
\bar{I}_j
=
\frac{1}{4\pi \nu_1^0}
\frac{c}{H(z)}
\ 
\left[
L_j^\text{cen}
N_\text{cen}
+
L_j^\text{sat}
N_\text{sat}
\right]
.
\eeq

The 2-halo term is given by:
\beq
P_{1,2}^\text{2-halo}(k, \mu, z)
=
\bar{I}_1 \bar{I}_2
\left( b_1 + F \mu^2 \right)
\left( b_2 + F \mu^2 \right)\; 
P_\text{lin},
\eeq
where the bias is
\beq
b_j(k, \mu, z)
\equiv \mathcal{L}_j^{-1}
\int
dm\ n(m)\;
b(m) 
\;
\left[
L_j^\text{cen}
N_\text{cen}
+
L_j^\text{sat}
N_\text{sat}
u(k,m,z)
e^{-k^2 \mu^2\sigma_d^2(m) / 2}
\right]
\eeq
and the effective growth rate of structure is
\beq
F(k, \mu, z)
\equiv
f\;
\int dm \; n(m)\;
\left( \frac{m}{\bar{\rho}} \right)
u(k,m) 
e^{-k^2 \mu^2 \sigma_d^2(m) /2}.
\eeq
The 1-halo term becomes:
\beq
\bal
P_{i,j}^\text{1-halo}(k, \mu, z)
=
\left(\frac{c}{4\pi H(z)}\right)^2 \frac{1}{\nu_i^0 \nu_j^0}
\int dm \; n(m)\
&\left[
L_j^{\text{sat}\ 2}
N_\text{sat}^2
\left| u(k,m,z) \right|^2
e^{-k^2 \mu^2\sigma_d^2(m)}\right.\\
&\left. + 2
N_\text{sat}N_\text{cen}
L_j^\text{sat}L_j^\text{cen}
e^{-k^2 \mu^2\sigma_d^2(m)/2}
\right]
.
\eal
\eeq

Finally, the shot noise term is unchanged, since it is not affected by the spatial distribution of the galaxies (centrals vs.\ satellites) nor by their peculiar velocities.

\subsection{Impact of luminosity fluctuations on the halo model}
\label{app:luminosity_scatter_halo_model}

The distribution of galaxy luminosities has a non-zero scatter around its mean.
How does this scatter affect the 2-halo, 1-halo and shot noise terms?
In this appendix, we show explicitly that the 2-halo and 1-halo terms depend only on the mean galaxy luminosity, whereas the shot noise term is enhanced by the scatter in galaxy luminosities. 

To better understand this, let us consider one pixel in configuration space.  For now, we consider only halos of a fixed mass, with number $N_h^\text{pix}$ within the observed pixel.  We want to compute the variance of the pixel luminosity $L$, to show how the 2-halo, 1-halo and galaxy shot noise terms arise, and to show that the fluctuations in galaxy luminosities only affect the galaxy shot noise term, not the 1-halo term.
The statistics of the number of halos in the pixel is assumed to be Poissonian:
\beq
\mathcal{P}(N_h^\text{pix}) \sim \text{Poiss}(N_h^\text{pix} | \bar{N}_h^\text{pix})
\eeq
Each halo contains a certain number $N_g$ of galaxies.
For a given halo, the number of galaxies in the halo is Poissonian:
\beq
\mathcal{P}(N_g) \sim \text{Poiss}(N_g| \bar{N}_g)
\eeq
The total number of galaxies $N_g^\text{pix}$ in the pixel is the sum of the number of galaxies in all of the halos in the pixel. 
A sum of a fixed number of Poisson variables is Poissonian.
However, the sum of a Poisson number of Poisson variables is not Poissonian!
In other words:
\beq
\bal
\mathcal{P}(N_g^\text{pix}) &\neq \text{Poiss}(N_g^\text{pix} | \bar{N}_h^\text{pix} \bar{N}_g)\\
\text{but}\quad \mathcal{P}(N_g^\text{pix} | N_h^\text{pix}) &\sim \text{Poiss}(N_g^\text{pix} | N_h^\text{pix} \bar{N}_g)\\
\eal
\eeq
Here the presence or absence of a bar over $N_h^\text{pix}$ matters.
Finally, each galaxy in a halo has a luminosity $L_g$ sampled independently from a probability distribution $\mathcal{P}_1(L_g)$. This probability distribution is the same object as $\phi$, but normalized to integrate to unity.
For the power spectrum or the variance, all we care about is that this distribution has a mean $\bar{L}_g$ and a variance which we write as $\bar{L}_g^2 \sigma_g^2$.

Finally, the total luminosity $L$ in the pixel is the sum of the luminosities $L_g$ of each of the $N_g^\text{pix}$ galaxies in the pixel.
Its probability distribution is complex.
However, at fixed $N_g^\text{pix}$, it is simply $\mathcal{P}_1^{* N_g^\text{pix}}$, i.e.\ the convolution of $\mathcal{P}_1$ with itself $N_g^\text{pix}$ times.
For the power spectrum, all we care about is that $\mathcal{P}_1^{* N_g^\text{pix}}$ has mean $N_g^\text{pix} \bar{L}_g$ and variance $N_g^\text{pix} \bar{L}_g^2 \sigma_g^2$.
Again, the absence of a bar on $N_g^\text{pix}$ matters here.
In other words:
\beq
\bal
\left\langle L^2 | N_g^\text{pix} \right\rangle
&=
\left\langle L | N_g^\text{pix}\right\rangle^2 + \text{var}[L | N_g^\text{pix}]\\
&=
\bar{L}_g^2 \left[ N_g^\text{pix 2} + N_g^\text{pix} \sigma_g^2 \right] 
\eal
\eeq

Putting all this together, we can compute the variance of the pixel luminosity $L$:
\beq
\bal
\left\langle L^2\right\rangle
&=
\sum_{N_g^\text{pix}} \mathcal{P}(N_g^\text{pix}) \left\langle L^2 | N_g^\text{pix}\right\rangle \\
&=
\bar{L}_g^2
\sum_{N_g^\text{pix}} \mathcal{P}(N_g^\text{pix}) \left[ N_g^{\text{pix } 2} + N_g^\text{pix} \sigma_g^2 \right]\\
&=
\bar{L}_g^2
\sum_{N_h^\text{pix}} \mathcal{P}(N_h^\text{pix})
\sum_{N_g^\text{pix}} \mathcal{P}(N_g^\text{pix} | N_h^\text{pix}) \left[ N_g^{\text{pix } 2} + N_g^\text{pix} \sigma_g^2 \right]\\
&=
\bar{L}_g^2
\sum_{N_h^\text{pix}} \mathcal{P}(N_h^\text{pix})
\sum_{N_g^\text{pix}} \text{Poiss}(N_g^\text{pix} | N_h^\text{pix}\bar{N}_g) \left[ N_g^{\text{pix } 2} + N_g^\text{pix} \sigma_g^2 \right]\\
&=
\bar{L}_g^2
\sum_{N_h^\text{pix}} \mathcal{P}(N_h^\text{pix})
\left[ (N_h^\text{pix} \bar{N}_g)^2 + N_h^\text{pix}\bar{N}_g + N_h^\text{pix}\bar{N}_g \sigma_g^2 \right]\\
&=
\bar{L}_g^2
\sum_{N_h^\text{pix}} \text{Poiss}(N_h^\text{pix} | \bar{N}_h^\text{pix})
\left[ (N_h^\text{pix} \bar{N}_g)^2 + N_h^\text{pix} \bar{N}_g + N_h^\text{pix} \bar{N}_g \sigma_g^2 \right]\\
&=
\bar{L}_g^2
\left[ (\bar{N}_h^\text{pix} \bar{N}_g)^2 + \bar{N}_h^\text{pix} \bar{N}_g^2
+ \bar{N}_h^\text{pix} \bar{N}_g + \bar{N}_h^\text{pix} \bar{N}_g \sigma_g^2 \right]\\
&=
\bar{L}_g^2
\left[ 
\underbrace{(\bar{N}_h^\text{pix} \bar{N}_g)^2}_\text{2-halo}
+ 
\underbrace{\bar{N}_h^\text{pix} \bar{N}_g^2}_\text{1-halo}
+ 
\underbrace{\bar{N}_h^\text{pix} \bar{N}_g (1 + \sigma_g^2)}_\text{shot noise + line noise}
\right]\\
\eal
\eeq
This confirms that the fluctuations in galaxy luminosities only appear in the galaxy shot noise term.
For the 1-halo and 2-halo terms, everything is as if each halo had the same luminosity $\bar{L}_g \bar{N}_g$, with no variation from halo to halo.


\subsection{Multi-line CLF: tracer auto-spectrum}
\label{app:tracer_auto_correlation}

In order to compare the value of galaxy number counts and LIM as tracers of the matter density field, we need to consistently predict their power spectra, using the same CLF halo model.
We assume that the galaxy sample is selected based on line luminosity, such that the number of satellites and centrals per halo are determined by the respective CLFs.
For example, for galaxies selected to have a line luminosity above a threshold $L_\text{min}$, we have
$N_\text{gal}(m,z) = \int_{L_\text{min}}^\infty dL\ \Phi(L| m, z)$.
Here, we simply review the expressions for the clustering power spectrum in this halo model.
For generality, we include the distinction between centrals and satellites.
The mean number density of galaxies is simply
$\bar{n}_g = \int dm\ n(m) \left[ N_\text{cen}(m) + N_\text{sat}(m) \right]$.
The 2-halo term in the power spectrum is given by:
\beq
P_g^\text{2-halo}(k, \mu, z)
=
\left( b_g + F \mu^2 \right)^2\; 
P_\text{lin},
\eeq
where the bias is
\beq
b_g(k, \mu, z)
\equiv
\int
\frac{1}{\bar{n}_g}
dm\ n(m)\;
b(m) 
\;
\left[
N_\text{cen}
+
N_\text{sat}
u(k,m,z)
e^{-k^2 \mu^2\sigma_d^2(m) / 2}
\right]
\eeq
and the effective growth rate of structure is
\beq
F(k, \mu, z)
\equiv
f\;
\int dm \; n(m)\;
\left( \frac{m}{\bar{\rho}} \right)
u(k,m) 
e^{-k^2 \mu^2 \sigma_d^2(m) /2}.
\eeq
The 1-halo term becomes:
\beq
\bal
P_g^\text{1-halo}(k, \mu, z)
=
\frac{1}{\bar{n}_g^2}
\int dm \; n(m)\
&\left[
N_\text{sat}^2
\left| u(k,m,z) \right|^2
e^{-k^2 \mu^2\sigma_d^2(m)}\right.\\
&\left. + 2
N_\text{sat}N_\text{cen}
e^{-k^2 \mu^2\sigma_d^2(m)/2}
\right]
.
\eal
\eeq
Finally, the shot noise power spectrum is simply 
$P^\text{shot}_g = \bar{n}_g^{-1}$.

\subsection{Multi-line CLF: cross-correlation with a tracer}
\label{app:tracer_cross_correlation}

One important observable in LIM surveys is the cross-correlation with a discrete tracer of the mass, e.g., individually detected quasars (QSO).
Previous papers model these cross-correlations (e.g., \cite{Sun20}). Here, we describe them in terms of the CLF.
To describe such cross-correlation, we restrict the multi-line CLF to count only sources of a given type.
In what follows, we focus on QSO for concreteness, but the formalism applies identically to any tracer of the matter density.
We thus introduce the QSO multi-line CLF $\phi (L_1, ..., L_n|m, \text{QSO})$ such that $\phi (L_1, ..., L_n|m, \text{QSO})dL_1...dL_n$ is the mean number of QSOs in a halo of mass $m$, with luminosity $L_i$ in each line $i$. 
In particular, the QSO CLF describes the full complexity of the QSO selection function.
If the QSO selection function is only based on fluxes in several bands, the QSO CLF can be simply obtained by adding these bands to the standard CLF, and multiplying it by the selection function.
As above, we can then define the mean number density of QSOs per halo or per unit comoving volume:
\beq
\left\{
\bal
&N_\text{QSO}(m)
=
\int dL_1\ \phi(L_1|m, \text{QSO})\\
&\bar{n}^\text{QSO}
=
\int dm \ n(m) N_\text{QSO}(m)\\
\eal
\right.
,
\eeq
and the mean QSO luminosity in line $j$:
\beq
\left\{
\bal
&L^\text{QSO}_j(m)
=
\int dL_j\ \phi(L_j|m, \text{QSO}) L_j\\
&\mathcal{L}^\text{QSO}_j
\equiv
\int dm \ n(m) L^\text{QSO}_j(m)\\
\eal
\right.
.
\eeq
The QSO selection function, described by the QSO CLF, typically includes a minimum luminosity in some bands, corresponding to the detection threshold of the survey considered.
This provides an effective lower luminosity (and halo mass) cutoff in the integrals above.

The cross-power spectrum between the intensity in line 1 $I_1$ and the number density of QSO $n_\text{QSO}$ is again the sum of three contributions:
\begin{equation}
    P_{1,\text{QSO}}(k,\mu,z) = P_{1,\text{QSO}}^\text{2-halo}(k, \mu, z) +
    P_{1,\text{QSO}}^\text{1-halo}(k, \mu, z) +
    P_{1,\text{QSO}}^\text{shot}(z),
\label{eqn:Pcross}
\end{equation}

The 2-halo term is given by
\beq
\boxed{
P_{1,\text{QSO}}^\text{2-halo}(k, \mu, z)
=
\bar{I}_1 \bar{n}^\text{QSO}
\left( b_1 + F \mu^2 \right)
\left( b_\text{QSO} + F \mu^2 \right)\; 
P_\text{lin},
}
\eeq
where the QSO bias term is
\beq
b_j(k, \mu, z)
\equiv \frac{1}{\bar{n}^\text{QSO}}
\int
dm\ n(m)\;
N_\text{QSO}(m)
\;
b(m) 
\;
u(k,m)
e^{-k^2 \mu^2\sigma^2 / 2}
.
\eeq

The 1-halo term becomes
\beq
\boxed{
P_{i,\text{QSO}}^\text{1-halo}(k, \mu, z)
=
\bar{I}_i \bar{n}^\text{QSO}
\frac{U_{i,\text{QSO}}^2(k, \mu, z)}{\bar{n}^\text{h eff}_{i,\text{QSO}}},
}
\eeq
where the effective halo profile and number densities are given by:
\beq
\left\{
\bal
&U_{i,\text{QSO}}^2(k, \mu, z)
\equiv
\frac{
\int dm \; n(m)
\left| u(k,m) \right|^2
e^{-k^2\mu^2\sigma^2}
L_i(m) N_\text{QSO}(m)
}
{
\int dm \; n(m)
L_i(m) N_\text{QSO}(m)
}\\
&\bar{n}^\text{h eff}_{i,\text{QSO}}
\equiv
\frac{
\left( \int dm \; n(m) L_i(m)  \right)
\left( \int dm \; n(m) N_\text{QSO}(m)  \right)
}
{\int dm \; n(m) L_i(m) N_\text{QSO}(m)}\\
\eal
\right.
\eeq

Finally, the shot noise cross-power spectrum is given by
\beq
\boxed{
P_{i,\text{QSO}}^\text{shot}(z)
=
\frac{\bar{I}_i \bar{n}^\text{QSO}}{\bar{n}^\text{gal eff}_{i,\text{QSO}}}
,
}
\eeq
with the effective galaxy number density
\beq
\bar{n}^\text{gal eff}_{i,j}
=
\frac{
\left( \int dm \; n(m) L_i(m)  \right)
\left( \int dm \; n(m) N_\text{QSO}(m)  \right)
}
{
\int dm \; n(m)
\int dL_i\ \phi(L_i|m, \text{QSO}) L_i
}.
\eeq

This formalism thus connects the observable cross-correlations between a tracer and intensity maps to the underlying CLF of the tracer.
By doing so, it provides astrophysical insights into the abundance of the prevalence of the interstellar medium regions responsible for each line inside the tracers.
Refs.~\cite{Wolz17, Breysse19} indeed explore these cross-correlations, in order to learn about the gas abundances in galaxies and the properties of feedback in star formation at high redshift.

\section{Projected angular power spectrum}
\label{app:angular_clustering}

One of the primary advantages of LIM over purely photometric surveys is enhanced redshift precision.  However there is also an intermediate regime, where the redshifts are not highly precise and redshift errors may have an unknown distribution, for which an analysis of angular clustering in multiple redshift bins is more appropriate.  Our formalism generalizes easily to this situation, and we include the relevant expressions here for completeness.

\begin{equation}
    I_X(\hat{n}) = \int d\chi\ W_X(\chi) I(\chi\hat{n})
\end{equation}
with $\chi$ the comoving radial distance and $W_X$ defining the shell.  The angular power spectrum of the cross-correlation between two shells is
\begin{equation}
    C_\ell^{XY} = \frac{2}{\pi}\int d\chi_1\,d\chi_2\ W_X(\chi_1)W_Y(\chi_2)
    \int k^2\,dk\ P_{XY}(k;\chi_1,\chi_2)j_\ell(k\chi_1)j_\ell(k\chi_2)
\label{eqn:cell_non_limber}
\end{equation}
with $j_\ell$ the spherical Bessel function of order $\ell$.  Except on very large scales, where linear theory is appropriate, we can approximate $P_{XY}(k;\chi_1,\chi_2)\simeq P_{XY}(k,\mu,z(\chi_1))$.

If we now decompose our power spectrum at $z$ into
\begin{equation}
    P(k,\mu,z) = P^{\rm 2-halo} + P^{\rm 1-halo} + P^{\rm shot}
\end{equation}
we find significant simplifications.  The shot-noise term is $k$-independent, and so can be pulled out of the $k$ integral.  The integral over $k$ then gives a $\delta$-function so that
\begin{equation}
    C_\ell^{\rm shot} =  \int\frac{d\chi}{\chi^2}\ W_X(\chi)W_Y(\chi)
    \frac{\bar{I}_\chi\bar{I}_\chi}{\bar{n}_{\rm gal,eff}}
\end{equation}
If the bin width is significantly larger than the velocity dispersion of halos dominating the emission we can set $\mu=0$ in the 1-halo term.  Since this term typically contributes mostly on small-scales, we can also use the Limber approximation (\cite{Limber53,Loverde08}; valid for $k\Delta\chi\gg 1$, with $\Delta\chi$ the width of the shell) to write
\begin{equation}
    C_\ell^{\rm 1-halo} = \int\frac{d\chi}{\chi^2}\ W_X(\chi)W_Y(\chi)
    \frac{\bar{I}_\chi\bar{I}_\chi}{\bar{n}_{\rm h,eff}}
    U^2\left(k=\frac{\ell+1/2}{\chi},\mu=0\right)
\end{equation}
This leaves only the 2-halo term.  In the Limber approximation
\begin{equation}
    C_\ell^{\rm 2-halo} = \int\frac{d\chi}{\chi^2}\ W_X(\chi)W_Y(\chi)
    \bar{I}_\chi\bar{I}_\chi
    b^2\left(k=\frac{\ell+1/2}{\chi},\mu=0\right)
    P_{\rm lin}\left(k=\frac{\ell+1/2}{\chi}\right)
\end{equation}
while at large scales one must perform the double integral in Eq.~\ref{eqn:cell_non_limber} using Eq.~\ref{eqn:P2halo}.

\subsection{Thin shell LIM}

Finally we comment upon a limit in which the results become particularly simple and intuitive.  If we take $W$ to be a top-hat centered at $\chi_0$ and of width $\Delta\chi\ll\chi_0$, normalized to unit integral, and we work in the Limber approximation,
\begin{equation}
    C_\ell \approx \mathcal{V}^{-1}\ P\left(k=\frac{\ell+1/2}{\chi_0},\mu=0,z\right)
\end{equation}
with $\mathcal{V}=\chi_0^2\,\Delta\chi$ the volume per steradian.  In this limit the 2D power spectrum equals the 3D power spectrum for modes of wavelength $\lambda=2\pi/k$ subtending an angle $\theta=2\pi/\ell$ at distance $\chi_0$ once accounting for the volume per solid angle.  

Similarly, for higher order $n$-point functions $P^{(n)}$, the Limber approximation
\beq
P^{(n)}_\ell
=
\int \frac{d\chi}{\chi^{2(n-1)}}\
W_X(\chi)^n\
P^{(n)} (\vk=\frac{\ell+1/2}{\chi})
\eeq
simplifies in the thin shell regime to:
\beq
P^{(n)}_\ell
=
\mathcal{V}^{-(n-1)}
P^{(n)}(\vk=\frac{\ell+1/2}{\chi}).
\eeq

In this limit non-overlapping redshift shells are uncorrelated, so all of the information is contained in the angular auto-spectra.  This makes it particularly easy to see what information is obtained from what parts of the clustering signal.

\subsection{Cross-correlation with CMB \& galaxy lensing}
\label{app:cross_lensing}

To compute the cross-power spectrum of LIM and galaxy or CMB lensing convergence $\kappa$,
we introduce the lensing kernel, for a source located at comoving radial distance $\chi_S$.
\beq
W^\kappa (\chi, \chi_S) = \frac{3}{2} \left( \frac{H_0}{c} \right)^2 \Omega_m^0 \frac{\chi}{a(\chi)} \left( 1 - \frac{\chi}{\chi_S} \right).
\eeq
For CMB lensing, the source distance is the surface of last scatter, at a distance $\chi_\text{SLS}$, corresponding to $z_\text{SLS} \sim 1100$:
\beq
W^{\kappa_\text{CMB}} (\chi) 
=
W^\kappa (\chi, \chi_\text{SLS}).
\eeq

For galaxy lensing, we average the lensing kernel $W^\kappa (\chi, \chi_S)$ over the redshift distribution $dn_S/d\chi_S$ of the source galaxies:
\beq
W^{\kappa_\text{gal}} (\chi) = 
\;\frac{1}{n_\text{S}} \int d\chi_S \ 
\frac{dn_\text{S}}{d\chi_S} \; 
W^\kappa(\chi, \chi_S), \text{\ \ \  with  }
n_S = \int d\chi \frac{dn_S}{d\chi}.
\eeq

In the Born and Limber approximations, we can thus predict the cross-spectrum of LIM with the (galaxy or CMB) lensing convergence.
The power spectrum is the sum of a 1-halo term
\beq
\bal
C_\ell^{\rm 1-halo} = 
\int\frac{d\chi}{\chi^2}\ 
W^\text{LIM} & (\chi)W^\kappa(\chi)
\left(\frac{c}{4\pi H(z)}\right) \frac{1}{\nu_\text{LIM}^0}\\
&\int dm \; n(m)
\left| u(k=\frac{\ell+1/2}{\chi},m) \right|^2
e^{-\left( \frac{\ell+1/2}{\chi} \right)^2\mu^2\sigma^2}
L_i(m) \frac{m}{\bar{\rho}_m}
\eal
\eeq
and a 2-halo term
\beq
\bal
C_\ell^{\rm 2-halo} = 
\int\frac{d\chi}{\chi^2}\ 
W^\text{LIM} & (\chi)W^\kappa(\chi)
\bar{I}_\text{LIM}
\left( b_\text{LIM} + F \mu^2 \right)
\left( b_m + F \mu^2 \right)\; 
P_\text{lin}.
\eal
\eeq
Here, $I_\text{LIM}$ and $b_\text{LIM}$ are the scale and redshift-dependent mean intensity and bias for LIM, used throughout this paper.
The effective growth rate of structure is the same as Eq.~\eqref{eq:effective_growth_rate}.
The quantity $b_m$ is the usual matter bias in the halo model:
\beq
b_m(k, \mu, z)
\equiv
\int
dm\ n(m)\; \frac{m}{\bar{\rho}_m}
\;
b(m) 
\;
u(k,m)
e^{-k^2 \mu^2\sigma_d^2(m) / 2}.
\eeq


\bibliographystyle{JHEP}
\bibliography{refs}

\end{document}